\documentclass[traditabstract]{aa}  
\usepackage{graphicx}
\usepackage{natbib}
\usepackage{lscape}
\usepackage{supertabular}

\newcommand{\tbd}[1]{{[TBD: {\bfseries #1}]}}
\newcommand{\comm}[2]{\textit{#1} \textbf{#2}}
\renewcommand{\tbd}[1]{}
\renewcommand{\comm}[2]{}

\usepackage{txfonts}

\begin{document}

\title{The CARMENES search for exoplanets around M dwarfs}
\subtitle{Wing asymmetries of H$\alpha$, \ion{Na}{i} D, and \ion{He}{i} lines}

\author{B. Fuhrmeister\inst{1}, S. Czesla\inst{1}, J. H. M. M. Schmitt\inst{1}
  \and  S.~V.~Jeffers\inst{2}
  \and  J.~A.~Caballero\inst{3}
  \and M.~Zechmeister\inst{2}
  \and   A.~Reiners\inst{2}
  \and   I.~Ribas\inst{4,5}
  \and P.~J.~Amado\inst{6}
  \and  A.~Quirrenbach\inst{7}
  \and  V.~J.~S.~B\'ejar\inst{8,13}
  \and  D.~Galad\'{\i}-Enr\'{\i}quez\inst{9}
  \and  E.~W.~Guenther\inst{10,8}
  \and  M.~K\"urster\inst{11}
  \and  D.~Montes\inst{12}
  \and  W.~Seifert\inst{7}}

\institute{Hamburger Sternwarte, Universit\"at Hamburg, Gojenbergsweg 112, 21029 Hamburg, Germany\\
  \email{bfuhrmeister@hs.uni-hamburg.de}
        \and
        Institut f\"ur Astrophysik, Friedrich-Hund-Platz 1, D-37077 G\"ottingen, Germany 
        \and
        Centro de Astrobiolog\'{\i}a (CSIC-INTA), Campus ESAC, Camino Bajo del Castillo s/n, E-28692 Villanueva de la Ca\~nada, Madrid, Spain 
        \and
        Institut de Ci\`encies de l'Espai (ICE, CSIC), Campus UAB, c/ de Can Magrans s/n, E-08193 Bellaterra, Barcelona, Spain
        \and
        Institut d'Estudis Espacials de Catalunya (IEEC), E-08034 Barcelona, Spain
           \and 
        Instituto de Astrof\'isica de Andaluc\'ia (CSIC), Glorieta de la Astronom\'ia s/n, E-18008 Granada, Spain 
        \and 
        Landessternwarte, Zentrum f\"ur Astronomie der Universit\"at Heidelberg, K\"onigstuhl 12, D-69117 Heidelberg, Germany 
        \and
        Instituto de Astrof\'{\i}sica de Canarias, c/ V\'{\i}a L\'actea s/n, E-38205 La Laguna, Tenerife, Spain
        \and
        Centro Astron\'omico Hispano-Alem\'an (MPG-CSIC), Observatorio Astron\'omico de Calar Alto, Sierra de los Filabres, E-04550 G\'ergal, Almer\'{\i}a, Spain 
        \and
        Th\"uringer Landessternwarte Tautenburg, Sternwarte 5, D-07778 Tautenburg, Germany 
        \and
        Max-Planck-Institut f\"ur Astronomie, K\"onigstuhl 17, D-69117 Heidelberg, Germany 
        \and
        Departamento de Astrof\'{\i}sica y Ciencias de la Atm\'osfera, Facultad de Ciencias F\'{\i}sicas, Universidad Complutense de Madrid, E-28040 Madrid, Spain 
        \and
        Departamento de Astrof\'{\i}sica, Universidad de La Laguna, E-38206 Tenerife, Spain 
}
        
\date{Received 31/10/2017; accepted dd/mm/2018}

\abstract
{Stellar activity is ubiquitously encountered in M~dwarfs and often characterised by
the H$\alpha$ line. In the most active M~dwarfs, H$\alpha$ is found in emission, sometimes with a complex line profile.  Previous studies have reported extended wings and asymmetries 
in the H$\alpha$ line during flares.
We used a total of 473 high-resolution spectra of 28 active M~dwarfs obtained by
the CARMENES (Calar Alto
high-Resolution search for M dwarfs with Exo-earths with Near-infrared and optical 
Echelle Spectrographs) spectrograph to study the occurrence  of broadened and
asymmetric H$\alpha$ line profiles and their association with flares, and 
examine possible physical explanations.
  We detected a total of 41 flares and 67 broad, potentially asymmetric, wings in
  H$\alpha$.
  The broadened H$\alpha$ lines
  display a variety of profiles with symmetric cases and both red and blue
  asymmetries.
  Although some of these line profiles are found during flares,
  the majority are at least not obviously associated with flaring.
  We propose a mechanism
  similar to coronal rain  or chromospheric downward condensations as a cause for the observed red asymmetries; the
  symmetric cases may also be caused by Stark broadening.
  We suggest that blue asymmetries are associated with rising material, and
  our results are consistent with a prevalence of blue asymmetries during the
  flare onset. Besides
  the H$\alpha$ asymmetries, we find some cases of additional line asymmetries in \ion{He}{i} D$_{3}$, 
  \ion{Na}{i}~D lines, and the \ion{He}{i} line at 10830\,\AA\, taken all simultaneously thanks to
  the large wavelength coverage of CARMENES.
  Our study shows that
  asymmetric H$\alpha$ lines are a rather common phenomenon in M~dwarfs and need to
  be studied in more detail to obtain a better understanding of the atmospheric dynamics
 in these objects.}

\titlerunning{Wing asymmetries of emission lines in M dwarfs' CARMENES spectra}
\authorrunning{B. Fuhrmeister et~al.}
\maketitle


\section{Introduction}
Stellar activity is a common phenomenon in M dwarfs. In particular, the
H$\alpha$ line is often used as an activity indicator for these stars because it
is more easily accessible than more traditional activity tracers used for solar-like stars such as the \ion{Ca}{ii} H\&K lines. In all but the earliest M dwarfs, 
the photosphere is too cool to show the H$\alpha$
line in absorption. Therefore, H$\alpha$ emission and absorption must
stem from the chromosphere.  As the level of activity increases, 
the H$\alpha$ line first appears as a deepening absorption line, then fills in, and, finally, goes
into emission for the most active stars \citep{Mullan}. These latter stars are called
active ``dMe'', while M dwarfs with H$\alpha$ in absorption are usually
called inactive. M dwarfs with neither emission nor absorption in the
H$\alpha$ line may either be very inactive or medium active;
a different activity indicator is needed for a discrimination in this case
\citep[see e.\,g.][]{Walkowicz}.

\citet{Newton} found a threshold in the
mass--rotation-period plane, separating active and inactive M~dwarfs.
This is in line with \citet{Jeffers2017}, who reported that the detection of H$\alpha$
active stars is consistent with the saturation
rotational velocity observed in X-rays by \citet{Reiners2014}. The
            saturation period increases from 10 d for spectral
            types < 2.5 \citep{Jeffers2017}, to 30--40 d
            for  M3.0 to M4.5 \citep{Jeffers2017,Newton}, 
            and 80--90 d in
            late-M stars (> 5.0) \citep{West2015,Newton}.
 
For the more slowly rotating active M dwarfs, where
the emission profile is not dominated by rotation, the H$\alpha$ line can
exhibit a complex double peaked profile, which is reproduced by non-local thermal equilibrium
chromospheric models \citep{Short1}. 
Observations additionally often show an asymmetric pattern for which static
models cannot account \citep[e.g.][]{Cheng}.

The surface of the Sun can be resolved in great detail, and thus
numerous reports of solar observations accompanied by broadened and asymmetric
chromospheric line profiles exist
(e.\,g. \citet{Schmieder1987, Zarro1988, Cho2016}; see also the review
by \citet{Berlicki}).
While symmetric line broadening can be caused by the Stark effect or turbulence
\citep{Svestka, Kowalski}, asymmetric line broadening is sometimes attributed to
bulk motion, for example in the form of so-called coronal rain
\citep[e.\,g.][]{Antolin12, Antolin, Lacatus}
or chromospheric condensations (\citet{Ichimoto, Canfield}; see also
Sect.~\ref{scenario}).
Following previous studies, we use the terms red and blue asymmetry to refer to
line profiles with excess emission in each respective part of the line profile.

While red-shifted excess emission is frequently observed on the Sun, blue
asymmetries are less often observed in solar flares. In fact,
\citet{Svestka1962} found that of 92 studied flares only $23$\,\% showed a blue
asymmetry, while $80$\,\% exhibited a red asymmetry. A similar result was
derived by \citet{Tang1983}, who reported an even lower fraction of $5$\,\% of
flares with blue asymmetries.

Other stars are far less intensively monitored with high-resolution
spectroscopy than the Sun, so that spectral time series of stellar flares often
result from chance observations \citep[e.g.][]{Klocova2017}.
Also, flares on other stars are generally only observed in
disk-integrated light. Nonetheless, also during
stellar flares, H$\alpha$ line profiles with broad wings and asymmetric profiles have
been observed \citep[e.g.][]{LHS2034, CNLeoflare}.

In a systematic flare study of the active M3 dwarf AD Leo, \citet{Crespo}
detected a total of 14 flares, out of which the strongest three showed Balmer
line broadening and two 
also showed red wing asymmetries, with the more pronounced asymmetry for the
larger flare. Additionally, the same authors detected weaker red asymmetries during
the quiescent state, which they tentatively attributed to persistent coronal rain.
Notably, \citet{Crespo} did not find any asymmetry or broad wings for \ion{Ca}{ii}
H\&K.

During a flare on the M dwarf Proxima~Centauri,
\citet{proxcen} found blue and red asymmetries in the Balmer lines,
the \ion{Ca}{ii} K line, and one \ion{He}{i} line. A similar observation was
made during a mega-flare on CN~Leo, where
\citet{CNLeoflare} detected asymmetries in the Balmer lines, in two \ion{He}{i}
lines, and the \ion{Ca}{ii} infrared triplet lines. Both studies reported a blue
asymmetry during flare onset and a red asymmetry during decay,
along with temporal evolution in the asymmetry pattern on the scale of minutes.

\citet{Montes} studied the H$\alpha$ line of highly active binary stars
and weak-line T\,Tauri stars and reported the
detection of numerous broad and asymmetric wings in H$\alpha$, possibly related to
atmospheric mass motion.
In a study of two long-lasting ($>$ 1~day)
  flares on the RS Canum Venaticorum binary II Peg, \citet{Berdyugina}
  found broad, blue-shifted components in the \ion{He}{i} D$_{3}$ line as well
  as many other strong absorption lines. The narrow components
  of these lines exhibited a red shift. The authors attributed these
  phenomena to up- and down flows.
A blue wing asymmetry
during a long duration flare on the active K2 dwarf LQ Hydrae was reported
by \citet{Soriano}. The blue excess emission persisted for two days after
the flare and was interpreted as a stellar coronal mass ejection.
\citet{Gizis} observed a broadened H$\alpha$ line during a strong flare
  on the field L1 ultra-cool dwarf WISEP J190648.47+401106.8 accompanying amplified continuum
and He emission. Therefore, also very late-type objects undergo
these type of event.
As an example of line broadening on a more inactive M dwarf, we point
out the observations of a flare on Barnard's star by
  \citet{Paulson}. During the event, 
  the Balmer series went into emission up to H13 and showed broadening, 
  which the authors attributed to the Stark effect.
  A marginal asymmetry may also have been present.

All these examples of searches on individual stars or serendipitous flare
observations show that
M~dwarfs regularly exhibit broadened and asymmetric Balmer lines though
  even strong flares do not need to produce broad components or asymmetries 
  \citep[e.\,g.][]{Worden}. We are not aware of any systematic study of such
  effects using high-resolution spectroscopy in a large sample of active M dwarf stars.
Therefore, we utilise the stellar sample
monitored by the CARMENES (Calar Alto
high-Resolution search for M dwarfs with Exo-earths with Near-infrared and optical 
Echelle Spectrographs) consortium for planet search around M dwarfs
\citep{CARMENES1} to systematically study chromospheric emission-line
asymmetries.
The CARMENES sample spans the whole M dwarf regime from M0 to M9 and comprises
inactive as well as active stars. Since the most active M dwarfs are known to
flare frequently, we focus on a sub-sample of active objects, which we consider
the most promising sample for a systematic study of chromospheric line profile
asymmetries.
Specifically, we want to study how often such asymmetries
occur, if they are always associated with flares, and if there are physical
parameters such as  fast rotation that favour their occurrence.

Our paper is structured as follows: in Sect. 2 we describe the CARMENES observations and data reduction.
In Sect. 3 we give an overview of our stellar sample selection, and we describe our
flare search criterion in Sect. 4.
In Sect. 5 we present the detected wing asymmetries and discuss them in Sect. 6.
Finally, we provide our conclusions in Sect. 7.


\section{Observations and data reduction}

All spectra discussed in this paper were taken with CARMENES \citep{CARMENES1} . The main scientific objective of 
CARMENES is the search for low-mass planets (i.e. super- and exo-earths) 
orbiting mid to late M  dwarfs in their habitable zone.
For this purpose, the CARMENES consortium is conducting a 750-night exo-planet
survey, targeting $\sim$300 M~dwarfs \citep{Alonso-Floriano, Reiners2017}.
The CARMENES spectrograph, developed by a consortium of eleven Spanish and
German institutions, is mounted at the 3.5\,m Calar Alto telescope, located in
the Sierra de Los Filabres, Almer\'ia, in southern Spain, at a height of about
2200\,m.
The two-channel, fibre-fed spectrograph covers the wavelength range from 0.52\,$\mu$m to 0.96\,$\mu$m 
in the visual channel (VIS) and from 0.96\,$\mu$m to 1.71\,$\mu$m in the near infrared
channel (NIR) with a spectral resolution of R $\sim$ 94\,600 in VIS and R $\sim$ 80\,400
in NIR. 

The spectra were reduced using the CARMENES reduction pipeline
and are taken from the CARMENES archive \citep{Caballero2, Zechmeister2017}.  
For the analysis, all stellar spectra were corrected for barycentric and stellar
radial
velocity shifts. Since we were not interested in high-precision radial velocity
measurements, we used the mean radial velocity for each star computed from
the pipeline-measured absolute radial velocities given in the fits headers of
the spectra;
the wavelengths are given in vacuum.
We did not correct for telluric lines since they are
normally weak and rather narrow compared to the H$\alpha$ line, which is the prime
focus of our study.
The exposure times varied from below one hundred seconds for the brightest stars
to a few hundred seconds for most of the observations, with the maximum exposure time
being 1800\,s for the faintest stars in the CARMENES sample.

\section{Stellar sample}\label{sec:sample}

Since line wing asymmetries in H$\alpha$ have so far been reported primarily during
flares, we selected only active stars with H$\alpha$ in emission. 
For the selection we used the CARMENES input catalogue Carmencita 
\citep{carmencita}, which lists H$\alpha$ pseudo-equivalent widths, pEW(H$\alpha$), from various
sources (see Table \ref{stars}). The term pseudo-equivalent width is used
here since around H$\alpha$ there is no true continuum in M dwarfs, but numerous
molecular lines, defining a pseudo-continuum. We took into consideration all
stars with a negative pEW(H$\alpha$) in Carmencita and ten or more observations listed in the archive as of 28~February~2017.
Since pEW(H$\alpha$) values higher than $-1.0$~\AA\, need not correspond
to true emission lines, but are sometimes due to flat or even shallow
absorption lines, we visually inspected all such H$\alpha$ lines in the CARMENES spectra
and excluded
stars showing absorption lines or flat spectra. Moreover, we rejected one
known binary star. There is another star, Barta 161 12,  which was
listed by \citet{Malo} as a very young double-line
spectroscopic binary candidate based on observations with ESPaDOnS, and
which indeed shows suspicious changes in the H$\alpha$ line but not in the
photospheric lines in CARMENES spectra.
Since its H$\alpha$ profile shows in many cases a more complex behaviour than the other stars 
(see Sect. \ref{absorption}) we excluded it from the general analysis but list it for reference in the tables and the appendix.

We did not exclude stars
with an H$\alpha$ line profile consisting of a central absorption component
and an emission component outside the line centre, since we interpret these
stars to be intermediate cases with H$\alpha$ between absorption and emission;
the corresponding stars are V2689~Ori, BD--02~2198, GJ~362, HH~And, and DS~Leo.
These selection criteria left us with a sample of 30 stars listed in Table~\ref{stars},
where we specify the Carmencita identifier, the common name of the star, its spectral type,
the pEW(H$\alpha$), the projected rotation velocity \mbox{($v\sin
i$)}, and a rotation period, all taken from Carmencita with
the corresponding references. 

\begin{table*}
\caption{\label{stars} Basic information on the investigated stars. }
\footnotesize
\begin{tabular}[h!]{llll clccl}
\hline
\hline
\noalign{\smallskip}
Karmn & Name  & SpT  & Ref. & pEW(H$\alpha$) & Ref. & $v \sin{i}$\,$^{a}$ & 
$P$ & Ref.  \\
&       &       & (SpT)     & [\AA]  &pEW(H$\alpha$)    & [km\,s$^{-1}$]
& [d] & ($P$)   \\
\noalign{\smallskip}
\hline
\noalign{\smallskip}
J01033+623  &V388 Cas    & M5.0\,V   &AF15   &-10.1  &AF15  &10.5 &    ...& ...             \\ 
J01125-169  &YZ Cet      & M4.5\,V   &PMSU   &-1.15  &PMSU  &<2.0 & 69.2& SM16      \\               
J01339-176  &LP 768-113  & M4.0\,V   &Sch05  &-1.68  &Gai14 &<2.0 &...  &...  \\                       
J01352-072  &Barta 161 12& M4.0\,V   &Ria06  &-5.36  & Shk10&51.4  &  ...&...\\
J02002+130  &TZ Ari      & M3.5\,V   &AF15   &-2.34  &Jeff17& <2.0 &   ...&...   \\          
J04153-076  &GJ 166~C   & M4.5\,V   &AF15   &-4.16  &Jeff17&<2.0 &   ...&...   \\            
J05365+113  &V2689 Ori   & M0.0\,V   &Lep13  &-0.25  &Jeff17&3.8 &   ...&...   \\            
J06000+027  &G 099-049    & M4.0\,V   &PMSU   &-2.31  &Jeff17&4.8 &   ...&...     \\         
J07361-031  &BD--02\,2198   & M1.0\,V   &AF15   &-0.58  &Jeff17& 3.1& 12.16 & Kira12\\     
J07446+035  &YZ CMi      &M4.5\,V   &PMSU   &-7.31  &Jeff17 & 3.9 &2.78& Chu74       \\
J09428+700  &GJ 362      &M3.0\,V   &PMSU   &-1.08  &PMSU   &<2.0 &  ...&...    \\             
J10564+070  &CN Leo      &M6.0\,V   &AF15   &-9.06  &Jeff17 &<2.0&  ...&...   \\              
J11026+219  &DS Leo      &M2.0\,V   &Mon01  &-0.14  &Jeff17 &2.6& 15.71& FH00     \\   
J11055+435  &WX UMa      &M5.5\,V   &AF15   &-10.2  &AF15   &8.3 &   ...&...   \\               
J12156+526  &StKM 2-809   &M4.0\,V   &Lep13  &-7.18  &Jeff17 &35.8&  ...&...   \\              
J12189+111  &GL Vir      &M5.0\,V   &PMSU   &-6.2   &Jeff17 &15.5& 0.49 & Irw11    \\  
J15218+209  &OT Ser      &M1.5\,V   &PMSU   &-2.22  &Jeff17 &4.3 & 3.38 & Nor07   \\   
J16313+408  &G 180-060    &M5.0\,V   &PMSU   &-7.7   &Jeff17 &7.3& 0.51& Har11    \\    
J16555-083  &vB 8        &M7.0\,V   &AF15   &-6.0   &AF15   &5.3&...&...    \\                        
J18022+642  &LP 071-082  &M5.0\,V   &AF15   &-4.67  &Jeff17 &11.3& 0.28 & New16a     \\
J18174+483  &TYC 3529-1437-1  &M2.0\,V    &Ria06  & ...  &  ...     &<2.0 &  ...& ...     \\                  
J18356+329  &LSR J1835+3259  &M8.5\,V   &Schm07 &-3.2   &RB10   &48.9&   ...&...   \\           
J18363+136  &Ross 149    &M4.0\,V  &PMSU   &-1.61  &Jeff17 &<2.0&   ...&...    \\            
J18482+076  &G 141-036    &M5.0\,V   &AF15   &-4.42  &Jeff17 &2.4&  2.756 & New16a  \\   
J19169+051S &vB 10       &M8.0\,V   &AF15   &-9.5   &AF15   & 2.7&...& ...     \\                     
J22012+283  &V374 Peg    &M4.0\,V   &PMSU   &-5.48  &Jeff17 &36.0 &  0.45 & Kor10    \\
J22468+443  &EV Lac      &M3.5\,V   &PMSU   &-4.54    &New   &3.5 &   4.376 & Pet92   \\  
J22518+317  &GT Peg      &M3.0\,V   &PMSU   &-3.26    &Lep13   &13.4 &   1.64& Nor07     \\ 
J23419+441  &HH And      &M5.0\,V   &AF15   &-0.6   &AF15   &<2.0 & ...& ...  \\              
J23548+385  &RX J2354.8+3831&M4.0\,V   &Lep13  &-0.31  &Lep13  &3.4&  4.76 & KS13    \\
\noalign{\smallskip}
\hline

\end{tabular}
\tablebib{
AF15:~\citet{Alonso-Floriano};  Chu74:~\citet{Chu74};  FH00:~\citet{FH00}; Gai14:~\citet{Gai14}; Har11:~\citet{Har11};
Irw11:~\citet{Irw11};  
Kira12:~\citet{Kira12}; Kor10:~\citet{Kor10}; KS13:~citet{KS13}; Lep13:~\citet{Lepine};
Mon01:~\citet{Mon01}; New:\citet{Newton}; ~New16a:~\citet{New16a}; Nor07:~\citet{Nor07}; 
PMSU:~\citet{PMSU};  Ria06:~\citet{Ria06}; 
Sch05:~\citet{Sch05}; Jeff17:~\citet{Jeffers2017};  Schm07:~\citet{Schmidt-M7flare}; 
Shk10:~\citet{Shk10}; SM16:~\citet{SM16}; Pet92:~\citet{Tes04}. \\
$^{a}$ All rotational velocities are measured by  \citet{Reiners2017} and also included in the catalogue
  of \citet{Jeffers2017}.}
\normalsize
\end{table*}

For technical
reasons, such as too low signal-to-noise ratio in the region of H$\alpha$, we had
to exclude a number of spectra from the analysis. Unfortunately, this affected
all spectra of the latest object in our sample, namely LSR~J1835+2839, reducing
the number of stars in our sample to 28 with a total of 473 usable spectra (not
counting Barta 161 12).
The final sample then comprised stars with spectral types ranging from M0 through M8,
among which the mid M spectral types are the most numerous.

\section{Characterisation of the activity state}

To assess the activity state of the star for each spectrum we define an H$\alpha$ index following
 the method of \citet{Robertson}. In particular, we use the mean flux density in a
 spectral band centred on the spectral feature (the so-called line band) and
 divide by the sum of the average flux density in two reference bands. We use the
 definition
\begin{equation}
I_{\rm line}=w \, \left(1-\frac{\overline{F_{\rm line}}}{\overline{F_{\rm ref1}}+\overline{F_{\rm
ref2}}}\right) \; ,
\end{equation}
where $w$ is the width of the line band, and
$F_{\rm line}$, $F_{\rm ref1}$, and $F_{\rm ref2}$ are the mean flux densities in the
line and the two reference bands. Our definition slightly differs from that of \citet{Robertson}.
For equidistant wavelength grids, which are an accurate approximation
for the spectra used in our study, the index $I_{\rm{H}\alpha}$ can
easily be converted into pEW(H$\alpha$) by the relation pEW(H$\alpha$) = 2$I_{\rm{H}\alpha}
- w$ and, therefore, carries the same information content.
In $I_{\rm{H}\alpha}$, the transition between absorption and
emission occurs at $0.5\,w$. According to its definition, changes in
$I_{\rm{H}\alpha}$ can be caused by variations of the line, a shift in the continuum level, or both.

For the definition of the width $w$ of the line band we follow
\citet{Robertson} and \citet{GomesdaSilva}, adopting
1.6~\AA\ for the (full) width of the line band.
 We note that during flares the line
 width can easily amount to 5~\AA\, or even more. In that case, line flux is left
 unaccounted for and the true value of $I_{\rm line}$ is underestimated.
However, using a larger bandwidth reduces the sensitivity of the index, and
 since we are not interested in a detailed study of the index evolution during flares, but
 only want to identify more active phases, we refrain from adapting the width of
 the line band.
 We use the red reference band of the H$\alpha$ line to locally normalise our
 spectra for graphical display. The two reference bands are centred on 6552.68 and
 6582.13 \AA\, and are 10.5 and 8.5 \AA\, wide, respectively.
 
Because line asymmetries have been reported to occur during flares,
we coarsely categorise the spectra as flare or quiescent spectra according to
the H$\alpha$ index I$_{\rm H\alpha}$.   Since the sequencing of the CARMENES observations 
is driven by the planet search \citep{Garcia-Piquer}, we lack a continuous time series, which 
precludes a search for the characteristic flare temporal behaviour with a short rise and a longer
decay phase. Finding flares based on a snapshot spectrum and especially on H$\alpha$ emission
  line strength alone is difficult, since there is intrinsic non-flare variation.
  Other studies \citep[e.g.][]{Hilton} used H$\beta$ (4861.28 \AA) additionally in
  flare searches, but this is not covered by our spectra. We therefore define a threshold for I$_{\rm H\alpha}$
  below which we define the star to be in a flaring state. Since especially the more active stars
display a considerable spread in the values of the I$_{\rm H\alpha}$ also during their
bona-fide quiescent state, we had to empirically define this threshold. In
particular, we mark all spectra with an I$_{\rm H\alpha}$ lower than 1.5
times the median I$_{\rm H\alpha}$ as flaring. This flare definition is
thus compatible with the 30\% variability during quiescence found by \citet{Lee}
and adopted also by \citet{Hilton}. We are aware that this method is
insensitive to small
flares, and also to the very beginning and end of stronger flares where the lines show
relatively little excess flux. 
To identify these cases one
would need a continuous time series and shorter exposure times.

\begin{figure}
\begin{center}
\includegraphics[width=0.5\textwidth, clip]{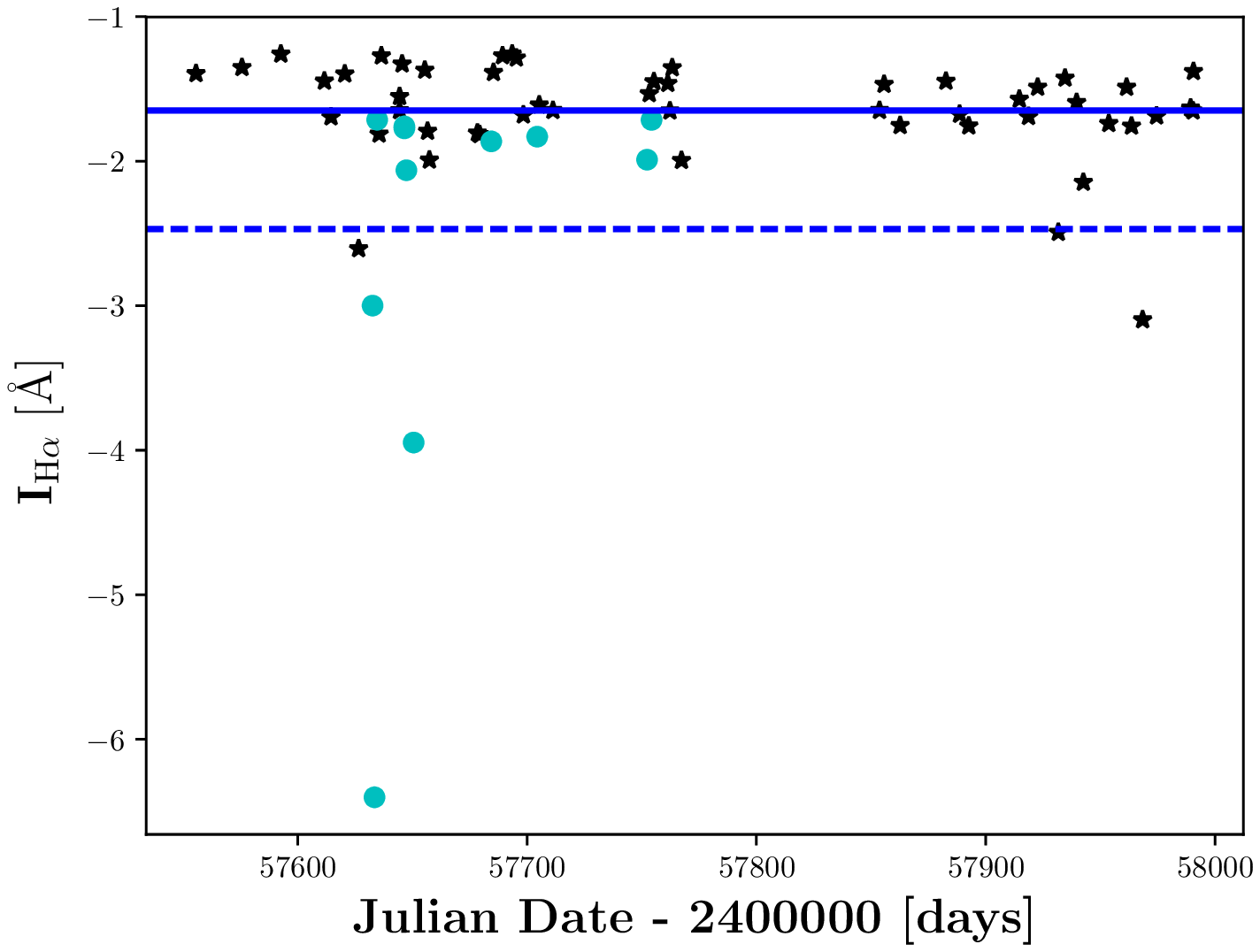}

\includegraphics[width=0.5\textwidth, clip]{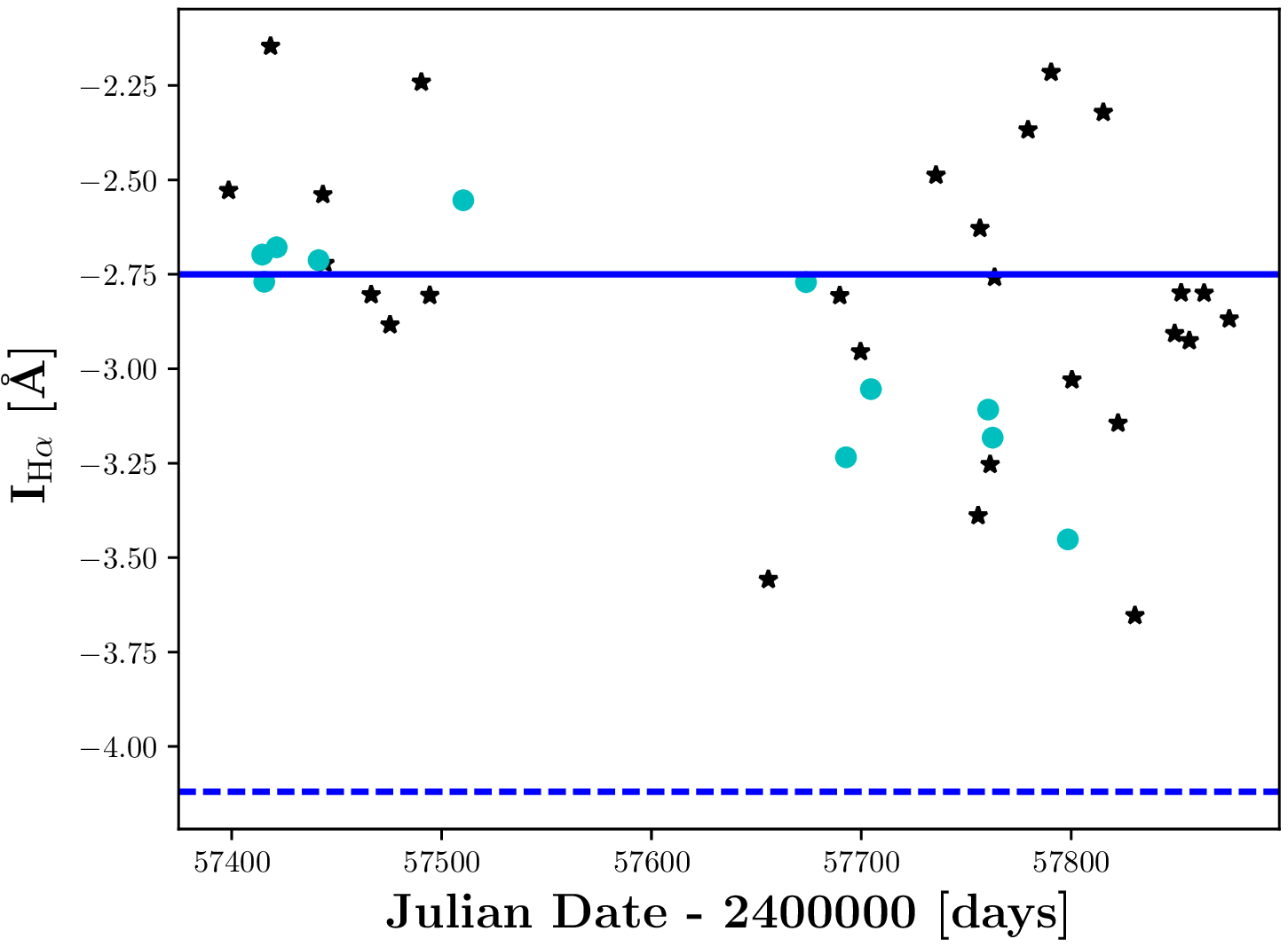}
\caption{\label{lightcurve} Time series of the H$\alpha$ index I$_{\rm H\alpha}$ for the
  stars EV~Lac (top) and YZ~CMi (bottom) as black asterisks. 
  Spectra with detected asymmetries are marked by cyan circles.
  The solid (blue) line marks the median index and the dashed line marks
  our flare threshold.}
\end{center}
\end{figure}

In Fig.~\ref{lightcurve} we show, as an example, the temporal behaviour 
of the I$_{\rm H\alpha}$ time series of EV~Lac and YZ~CMi. Both stars show continuous
variability in their I$_{\rm H\alpha}$ also during the quiescent state, which
is, in fact, more reminiscent of the quiescent flickering observed in the
X-ray emission of active M~dwarfs \citep{Robrade2005}. Nonetheless, we continue to
refer to this state as quiescent. 
While for EV~Lac the four flares in the I$_{\rm H\alpha}$ time series are 
set apart from the quiescent state, the I$_{\rm H\alpha}$ light curve of YZ CMi
reveals the difficulty of finding small flares without continuous observations.
Following our flare definition, there are no flares in the YZ CMi light curve.
Yet the star seems to be more variable in the second half of the light
curve, indicating some transition in its state of activity.

\begin{figure}
\begin{center}
\includegraphics[width=0.5\textwidth, clip]{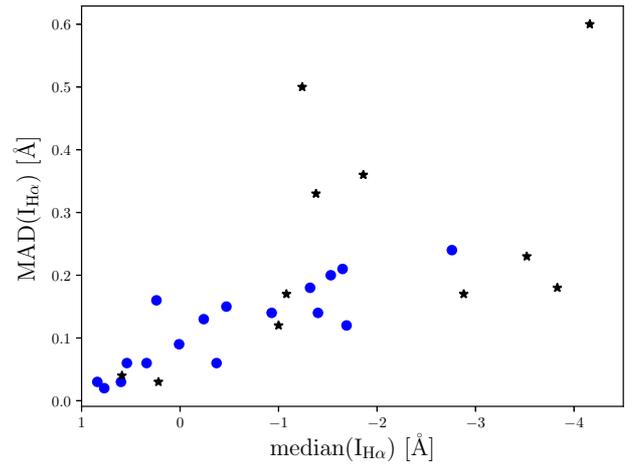}
\caption{\label{madvsmedian} Variability (MAD) of I$_{\rm H\alpha}$ as a
  function of its median value for spectral type M0.0 -- M4.0 (blue circles) and M5.0
  and later (black asterisks).}
\end{center}
\end{figure}

In Table~\ref{stats} we provide the minimum and maximum of the 
measured I$_{\rm H\alpha}$, as well as its median and median absolute deviation 
(MAD) for all our sample stars.
While for the stars EV~Lac and YZ~CMi the spread in I$_{\rm H\alpha}$ 
during quiescence as measured by MAD is about the same  ($\approx 0.22$~\AA;
cf. Table~\ref{stats}), the sample stars show a considerable
range with DS~Leo and CN~Leo marking the lower and upper bounds with MADs of
$0.02$~\AA\ and $0.6$~\AA.

Both the median(I$_{\rm H\alpha}$) and its dispersion as measured by the MAD
describe the stellar activity level, and we
compare both values in Fig.~\ref{madvsmedian}. A correlation seems to be
present, indicating lower variability levels for less active stars.
This is consistent with the results reported by \citet{Bell} and
\citet{Gizis2002}, who also found larger scatter for more
active stars. For our sample,
Fig.~\ref{madvsmedian} suggests a flattening of the variability distribution for
higher activity levels, which was found neither by \citet{Bell} nor \citet{Gizis2002}. We therefore
  consider this to be a spurious result, attributable to the low number
  statistics for high-activity stars in our sample or to the underestimation of I$_{\rm H\alpha}$
  for broadened profiles in the stars with the highest activity level.

\section{Line wing asymmetries}
\subsection{Overview}

\begin{figure}
\begin{center}
\includegraphics[width=0.5\textwidth, clip]{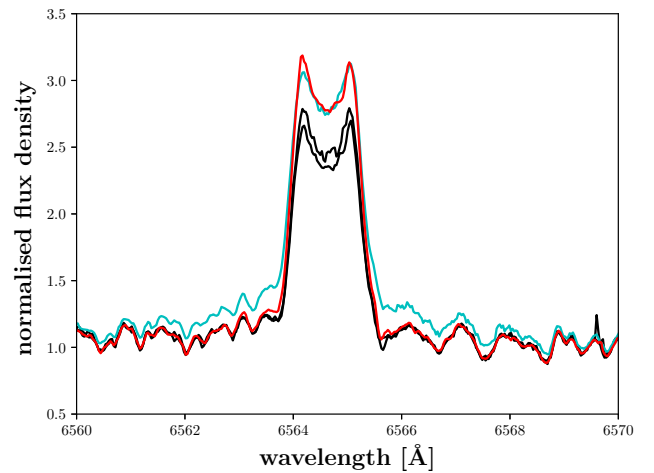}
\caption{\label{OTSer} Four normalised H$\alpha$ line profiles of
OT~Ser. The two black spectra represent the
quiescent state of the star, while the red and the cyan spectrum correspond to
an enhanced level of activity.}
\end{center}
\end{figure}

In our M dwarf CARMENES spectra we find diverse examples of H$\alpha$ lines
with broad line wings with both relatively symmetric or highly asymmetric profiles. 
An example of a symmetric profile is given in Fig.~\ref{OTSer}, which shows four out of the
17 spectra of OT~Ser considered in this study. In particular, we show two
spectra representing the typical quiescent state of OT Ser (in black), while
the spectra plotted in red and cyan represent a more active state.  As is 
shown in Fig.~\ref{OTSer}, the states are nearly identical in the line core, yet
the cyan spectrum exhibits rather broad line wings, reaching about 4 \AA\ beyond 
the emission core.

To quantify the profile and its asymmetry,
we follow \citet{CNLeoflare}, \citet{Soriano}, and other studies and first
obtain a residual spectrum by subtracting the spectrum
with the highest (i.e. least active) I$_{\rm H\alpha}$. By subtraction of the spectrum
representing the quiescent state, we infer the additional (potentially
flaring) flux density and how it is distributed between the narrow line core and the
broad line component. We then fit the residual flux density
with a narrow and a broad Gaussian profile, representing the line core and wing.
The fit with this two-component model
results in a reasonably good description of the line profile even in the most
asymmetric cases. We nevertheless caution that the fit
may be not unique.
For example this may happen when one Gaussian flank of the broad component is
hidden in the narrow component (see the example in the next section and Fig.
\ref{CNLeofit}).

\subsection{Red and blue line asymmetries:  CN~Leo and vB~8}

\begin{figure}
\begin{center}
\includegraphics[width=0.5\textwidth, clip]{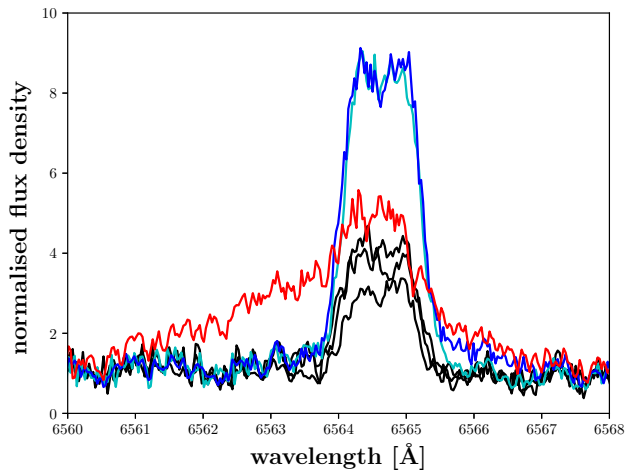}
\caption{\label{vb8} Normalised H$\alpha$ line profiles spectra of vB~8.
The three black spectra are typical for the quiescent state, while the
red, blue, and cyan spectra show profiles corresponding to enhanced
activity levels.}
\end{center}
\end{figure}

A rather striking example of line asymmetries can be found in the CARMENES 
spectra of vB~8 shown in Fig.~\ref{vb8} (and with two-Gaussian fit  in Fig. \ref{fit3}).
Three example spectra show the typical variation during the quiescent state.
One spectrum shows an extreme blue line asymmetry with only marginal emission
in the line core.
Two more example spectra both exhibit nearly identical line cores with large
amplitudes, but only one of them shows an additional red
asymmetry.
This  demonstrates the variety of profiles encountered in our analysis.

\begin{figure}
\begin{center}
\includegraphics[width=0.5\textwidth, clip]{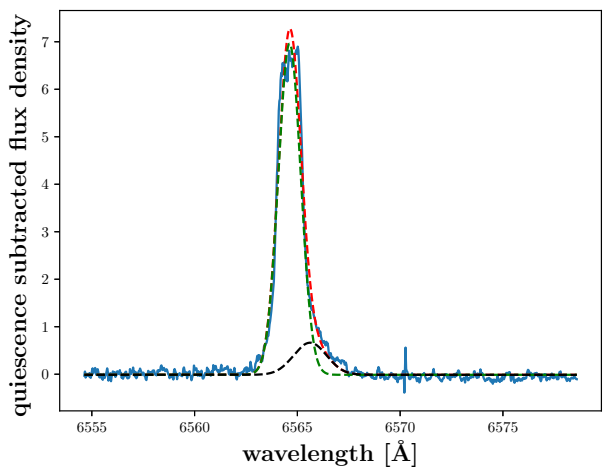}
\caption{\label{CNLeofit} Residual H$\alpha$
  line profile of CN Leo (blue) along with best-fit model with two Gaussian
  components. The dashed red line represents the best-fit two-component
  Gaussian model, and the dashed black and green lines represent the individual
  Gaussian components.}
\end{center}
\end{figure}

To illustrate our fits of red and blue line asymmetries, we consider the
case of CN~Leo.  In
Fig.~\ref{CNLeofit} we plot the residual H$\alpha$ line profile 
(i.e. the profile after subtraction of the quiescent state) with a
rather moderate red asymmetry along with a two-component fit; in this
case we require both the Gaussian component representing the line core 
and the one representing the wing  to be rather narrow.
The profile of the emission line core is not entirely Gaussian,
especially at the top. Nonetheless, we consider the resulting characterisation
of the width and area of the line core sufficiently accurate for our purposes.
The problem of the non-uniqueness of the fit is
apparent: the width of the wing component cannot tightly be constrained 
because the blue flank of this component is hidden in the core component.
Fortunately, this problem occurs only rarely and the parameters derived for most
of the broad components are trustworthy. To further illustrate the quality
  of our fits, we show the strongest
  and weakest asymmetry with its corresponding fit for blue and red asymmetries and for
  more symmetric profiles in Appendix~\ref{appendixd} in Figs. \ref{fit1} to \ref{fit6}.

\subsection{CARMENES sample of broad and asymmetric H$\alpha$ line profiles}

In the fashion described above we analysed all H$\alpha$ line profiles. 
In Table~\ref{asyms} we list all detected asymmetries for all sample stars,
number them for reference, and provide the best-fit parameters for
each of the two Gaussian components: the area in \AA, the shift of the
central wavelength with respect to the nominal wavelength of
the H$\alpha$ line $v_{\rm central}$,
and the standard deviation $\sigma$ in \AA. Moreover, we
list the full width at half maximum (FWHM), the ratio
of the Gaussian area $A$ determined for the broad and
total components, the epoch of each observation as Julian Date-2\,400\,000 in days, and
the exposure time in seconds.
In  Appendix~\ref{appendix} we show all
H$\alpha$ line profiles with detected asymmetries, and in
Table~\ref{stats} we summarise statistical aspects of our findings.
Specifically, we give the number of detected flares and broad components (categorised into
red, blue, and symmetric) and how
often both coincide. 





\normalsize

\section{Discussion}
\subsection{Overview: statistics}

In the 473 CARMENES spectra selected for our study we find a total of 41 flares and 67 line wing
asymmetries.  Only in eight out of the 28 sample stars, we do not find any
H$\alpha$ line observations with
broad wings.
This indicates a
mean duty cycle of $9$\,\% and $15$\,\% for flares (following  our definition) and
H$\alpha$ line asymmetries averaged over the studied sample, respectively.
Therefore, broad and asymmetric H$\alpha$ line profiles appear to be a rather
common phenomenon among active M~stars.  Also, line asymmetries are not necessarily
linked to flares; in fact, we observe flares without line asymmetries and line
asymmetries without flares.
Red asymmetries occur more often than blue asymmetries,
and symmetric broadening is detected least frequently.

\subsection{Limits of the line-profile analysis}
\label{sec:Timing}

The CARMENES observations provide a comprehensive and
temporally unbiased sample of M~dwarf line profiles, the nature of which
imposes, however, some limitations on the analysis. As we
lack continuous spectral time series, the course of spectral evolution cannot be
resolved, and because
integration times vary between less than $100$\,s and up to $1800$\,s for the
faintest objects, variability on short timescales remains elusive. In the specific case of flares,
we remain ignorant of the phase covered by the spectrum.
Nonetheless, the spectra allow both
a statistical evaluation of the sample properties and meaningful analyses of
individual profiles. In particular, any sufficiently strong or long-lived
structure will be preserved in the individual spectra.

\subsection{Physical scenario for line broadening with potential asymmetry}
\label{scenario}


\subsubsection{Symmetrically broadened lines}\label{sec:symm}
Symmetric line broadening can be caused by the Stark effect or
non-thermal turbulence. For the Sun, \citet{Svestka} argued that symmetric Balmer line broadening
is caused by Stark broadening and not Doppler broadening.
More recently, \citet{Kowalski} introduced a new method to handle the Balmer
line broadening caused by the elevated charge density during flares to better
reconcile modelling and observations. 

Another explanation for the more symmetric lines is the long
exposure time, possibly blurring phases with blue and red shifts into
relatively symmetric, broad lines. Indeed, we see a tendency 
for broader lines to be more symmetric in our data, but that may be caused by
Stark broadening as well (see Sect. \ref{linewidth}).

\subsubsection{Red line asymmetries}

As to line asymmetries, \citet{Lacatus}  presented solar \ion{Mg}{ii}~h and~k line profiles
with a striking red asymmetry, observed with the \textit{Interface Region
Imaging Spectrograph} (\textit{IRIS}) following an
X2.1-class flare; the relevant data and fits are shown in their Fig. 2.
The authors ascribed the observed line profiles to coronal rain, which is cool plasma
condensations travelling down magnetic field lines from a few Mm above the solar
surface. 

This coronal rain is thought to
result from catastrophic cooling and is not only observed in the context of
flares, but appears to be a rather ubiquitous coronal phenomenon \citep{Antolin12,
Antolin}. In fact, \citet{Antolin12} supposed that coronal rain is an
important mechanism in the balance of chromospheric and coronal material and
estimated the fraction of the solar corona with coronal rain to be between
$10$\,\% and $40$\,\%.


Based on the \ion{Mg}{ii} line profiles, \citet{Lacatus} derived a
net red shift of $60$\,km\,s$^{-1}$, which they
attributed to bulk rain motion.  Velocities on the order of $100$\,km\,s$^{-1}$
are, indeed, rather typical for such condensations \citep[e.g.][]{Schrijver2001,
deGroof2005, Antolin12, Verwichte2017b}, which is also the maximum shift
we observed.

Along with the red shift,
\citet{Lacatus} observed broad wings with a Doppler width of $\approx
100$\,km\,s$^{-1}$, appearing about six minutes after the impulsive flare phase.
Their lines are far broader than the net red shift, which is also true for the
majority of our measurements. The authors argue that these widths are
best explained by unresolved Alfv\'enic fluctuations.
Broadening by Alfv\'enic wave turbulence was first suggested by
\citet{Matthews}, based on observations of a solar flare associated with a
Sun-quake in the
\ion{Mg}{ii} h \& k lines (see also \citet{Judge2014}).

In a study of spatially resolved H$\alpha$ emission during a solar
flare, \citet{Johns-Krull} found quiescence subtracted spectra exhibiting weak
asymmetries with slightly more flux in the red wing. We completely lack spatial
resolution, and the amount of asymmetry reported by \citet{Johns-Krull} is so
small that we would most likely not have detected it in our spectra.
\citet{Johns-Krull} interpreted the red wing asymmetry as being caused by rising material absorbing
at the blue side of the line. 

Besides coronal rain, also chromospheric condensations produce
red asymmetries on the Sun. Such
condensations normally occur during the impulsive phases of flares driven
by non-thermal downward electron beams
  \citep[e.\,g.][]{Ichimoto, Canfield}. These so-called red
  streaks reach velocities between 40 to 100\,km\,s$^{-1}$ with the velocities decreasing on short timescales and the emission ceasing before H$\alpha$ emission
  reaches its maximum. \citet{Canfield} measured only velocities up to 60\,km\,s$^{-1}$ and
  ascribed this low maximum velocity to their
  comparably low time resolution of 15\,s. Also \citet{Graham} found lower
  velocities and very short decay times of only about one minute, which
  even partially preceded the also observed up-flows.

For the \ion{Si}{iv} and \ion{C}{ii} transition region lines, 
\citet{Reep} found red emission asymmetries lasting more than 30\,min and
attributed these to the successive
energisation of multiple threads in an arcade-like structure. The authors
argue against coronal rain because no blue shifts were detected at the
beginning of the flare. Instead, the red shifts started simultaneously with a
hard X-ray burst. \citet{Reep} further argue that a loop 
needs some time to destabilise for coronal rain to
start after the heating ceases.
\citet{Rubiodacosta} and \citet{rubiodacosta1} modelled solar flare
observations in H$\alpha$, \ion{Ca}{ii} 8542~\AA, and \ion{Mg}{ii} h \& k with
multi-threaded down-flows but still had trouble  reproducing the observed
wing flux in their models.

Since we lack information about when 
during the flare the emission responsible for the
broad components occurred (and if at all in some cases), we cannot properly distinguish between the coronal
rain and chromospheric condensation scenario. While coronal rain is expected to
take place later in the decay phase, chromospheric condensations are
thought to be associated with the impulsive phase.
We consider it improbable though that transient events lasting only about a minute can account
  for the strongest asymmetric signatures in our up to 30-minutes-long exposures.
  Moreover, the blue asymmetries found in our sample are never associated with
  large amplitudes in the narrow main component, suggesting coverage of the flare onset only
    (as expected; see Sect.
  \ref{sec:blue} and \ref{sec:symm}). In contrast, elevated emission in the
  narrow component was detected for some lines with red
  asymmetries, which indicates that at least the onset of the decay
  phase was covered; however, the impulsive
  phase may be covered, too.
  While it appears to us that the coronal rain scenario is more easily
  compatible with the majority of our observations, we cannot ultimately
  discriminate between the scenarios on
  the basis of our data, and both phenomena may actually have
  occurred.

\subsubsection{Blue line asymmetries}\label{sec:blue}

On the Sun, blue asymmetries are also observed. For instance, \citet{Cho2016}
reported the appearance of a blue asymmetry in the
H$\alpha$ line associated with the start of a rising filament preceding a solar
flare. The authors attribute the asymmetry to a mixture of material moving at
velocities between $-138$\,km\,s$^{-1}$ and $38$\,km\,s$^{-1}$. While
\citet{Cho2016} reported an excess absorption in H$\alpha$, blue shifted emission
has also been observed on the Sun \citep{Svestka1962, Tang1983}. In fact, the flare observations of the mid-M~dwarfs CN~Leo and
Proxima~Centauri, which also cover the flare onset,
show both blue asymmetries during flare onset, followed by
red asymmetries later on during the decay phase \citep{CNLeoflare, proxcen}.
Accordingly, we 
ascribe the blue shifted components to chromospheric evaporation.

\subsection{Strength of the asymmetries}
\begin{figure}
\begin{center}
\includegraphics[width=0.5\textwidth, clip]{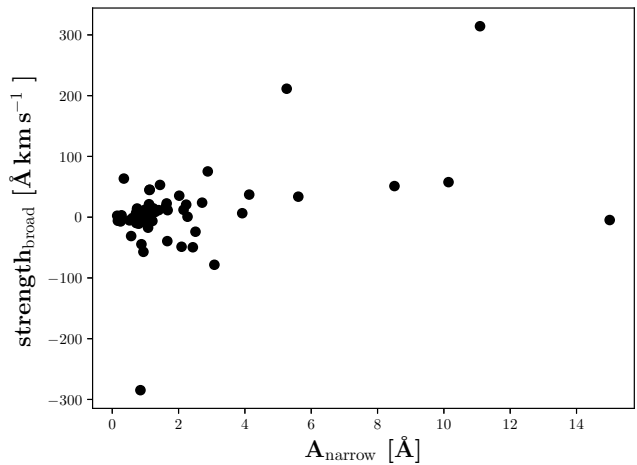}\\
\caption{\label{strength} Strength of the asymmetries as a function of the narrow component amplitude.}
\end{center}
\end{figure}

As a way to measure the strength of the asymmetry, we compute the area of
the broad Gaussian multiplied by the shift of its central wavelength. We show this
asymmetry strength in Fig.~\ref{strength}.

Based on this measure, we find three outstandingly strong
asymmetries. The one with the blue shift belongs to vB~8 and the corresponding
spectrum is shown in Fig.~\ref{vb8}. The remaining
two correspond to asymmetries no.~50 and 55 in Table~\ref{asyms}, are red
shifted, and stem from EV Lac. The associated spectra are shown in
Fig.~\ref{asymEVLac} (and with Gaussian fit in Fig. \ref{fit1}). The first of these asymmetries
(no. 50) also shows the second highest flux level in the narrow line
along with broad wings in the \ion{Na}{i}~D lines. It is clearly associated with
a large flare on EV~Lac.

\subsection{Connection of asymmetries to flares}


As evidenced in Table~\ref{stats}, broad wings in H$\alpha$
with more or less pronounced asymmetries of the broad component appear to be quite common among
M stars. In contrast to the literature, where wings and asymmetries therein are
normally connected to flares, many asymmetries found in our study do not
coincide with flares (using our flare definition).
Assuming that stellar flares show a sufficiently uniform
pattern of chromospheric line evolution, this finding shows that: (i)
line asymmetries may not in all cases be connected to flares; (ii)
they may be found also in smaller flares, (iii) they
may persist up to or even beyond the end of the decay phase of stronger flares, 
or, (iv) they occur during
the impulsive phase, where the chromospheric lines just start to react to the
flare. We cannot distinguish these latter three scenarios since we do not have
time series, but only snapshots. The last possibility is rather unlikely, since
the impulsive phase of flares typically lasts only a few tens of seconds, while
the typical exposure times of our sample stars is a few hundred seconds, so that in
most cases also the flare peak should be covered by the exposure.

Despite the lack of a clear association between asymmetries and flares,
asymmetries generally seem to be accompanied by somewhat enhanced activity
states:
only in six cases, where an asymmetry was detected, did the narrow Gaussian
component area remain
small (<0.5 \AA), indicating that the stellar activity state was only
marginally higher than the minimum state that we subtracted. In all other
cases, we found indications for enhanced emission in the narrow component, albeit not
sufficiently strong to be qualified as a flare.

\begin{figure}
\begin{center}
\includegraphics[width=0.5\textwidth, clip]{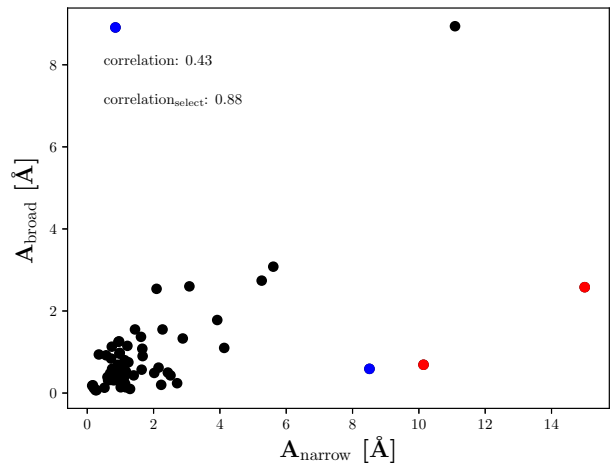}
\caption{\label{ftotalfbroad} H$\alpha$ line residual pEW(H$\alpha$) in the broad component versus the narrow
  residual pEW(H$\alpha$) both measured by the Gaussian area, A, of the
  line fit. Red dots correspond to the two values
  for CN~Leo and blue dots to the two values for vB~8. Correlation
  coefficients correspond to the whole sample and the sample excluding
  CN~Leo and vB~8.}
\end{center}
\end{figure}

Moreover, we find that the pEW(H$\alpha$) of the broad
  component increases  as the
  strength of the narrow component of the H$\alpha$ line increases (see
  Fig.~\ref{ftotalfbroad}).
  A similar
  correlation between the stellar activity state and
  the strength of the broad component was found by \citet{Montes}. Formally, we
obtain a Pearson correlation coefficient of 0.43, taking all
spectra into account where we found a broad component.
Excluding the four values pertaining to CN~Leo and vB~8 (see
Fig.~\ref{ftotalfbroad}),
we compute a correlation coefficient of 0.88. CN~Leo and vB~8 are the only two
stars in the sample with spectral type M6 or later for which we found asymmetries, but we are not
aware of any physical reason why these two stars should behave differently,
though they may have higher contrast to photospheric emission in this wavelength region.

\subsection{Occurrence rate of asymmetries and activity state or spectral type}

According to our analysis the four fastest rotators in our sample (V374~Peg,
LSR~J1835+3259, StKM 2-809, and GL~Vir) with \mbox{$v\sin i$} ranging from
21\,km\,s$^{-1}$ to 43\,km\,s$^{-1}$ all show asymmetries.
For the fastest rotator, V374~Peg, a
member of the roughly 300~Myr old Castor moving group 
(\citet{Caballero2010}, but see \citet{Mamajek}), we also detect the highest
fraction of asymmetries. On the other end of the activity range, we find that
three out of the eight stars without any detected asymmetry
are less active, showing weak H$\alpha$
emission (DS Leo, HH And, and BD--02\,2198). Both hint
at a relation
between activity state and the frequency of asymmetries.

\begin{figure}
\begin{center}
\includegraphics[width=0.5\textwidth, clip]{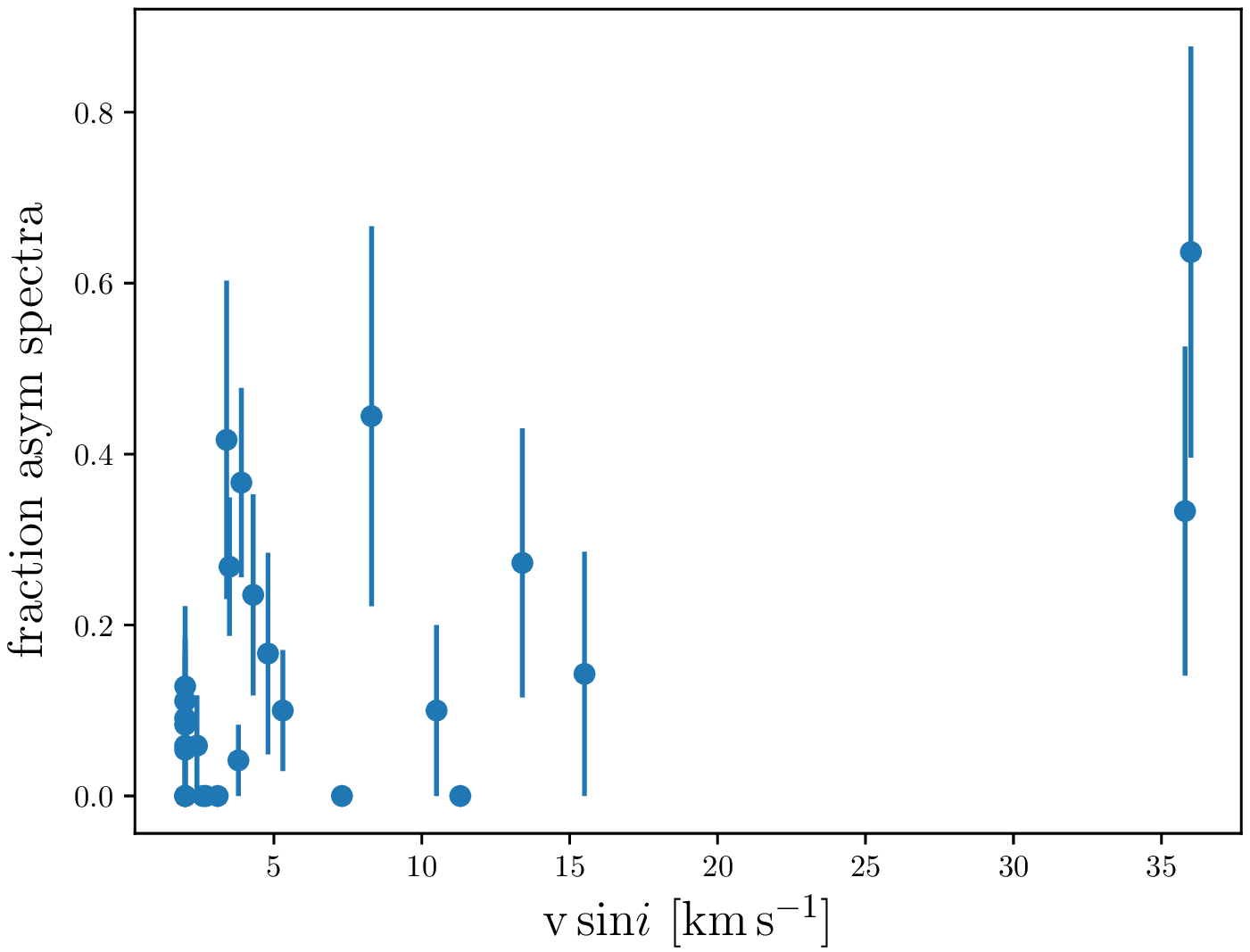}

\includegraphics[width=0.5\textwidth, clip]{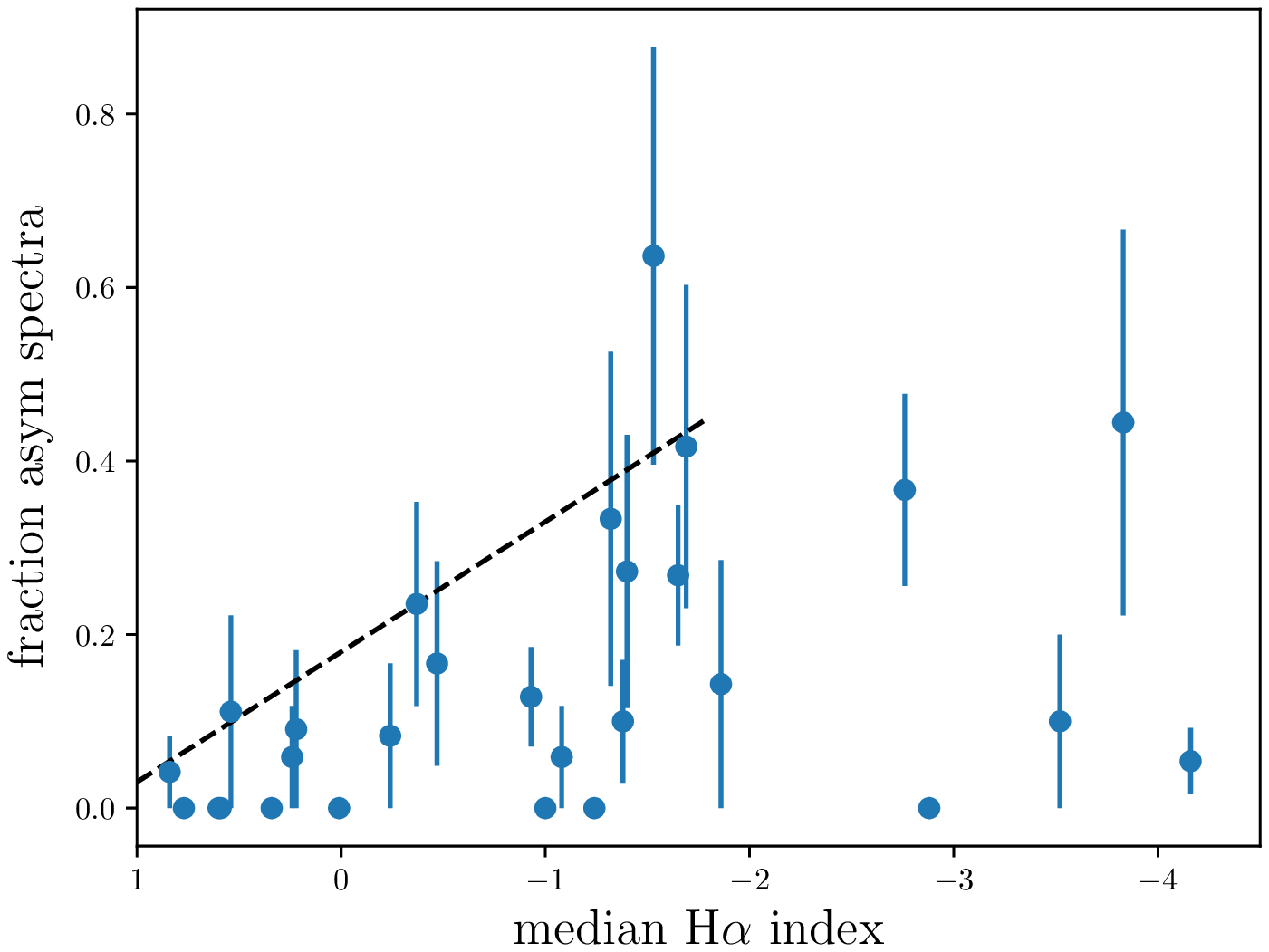}
\caption{\label{fracvsvsini} Fraction of spectra with 
  asymmetries (blue circles) as a function of rotational velocity (top) and
  median(I$_{\rm H\alpha}$) (bottom) with a
  tentative upper envelope (dashed black line).}
\end{center}
\end{figure}

In Fig.~\ref{fracvsvsini}
we show the fraction of spectra exhibiting asymmetries
as a function of (projected) rotational velocity (top) and median(I$_{\rm H\alpha}$)
(bottom). With a mean fraction of 5\,\%, the probability of
showing an asymmetry is smaller for slowly rotating stars ($< 3.5$ km\,s$^{-1}$)
than for fast rotators, where the mean fraction is 23\,\%. A
t-test shows a probability of about 1\,\% that the two means are equal.

The picture is similar for the rate of asymmetries as a function of
the median(I$_{\rm H\alpha}$) (Fig.~\ref{fracvsvsini}, bottom panel).
We find the highest fractions of asymmetric line profiles
among the more active stars. In fact, the distribution suggests the existence
of an upper envelope  indicated by the dashed line
in Fig.~\ref{fracvsvsini}. While we cannot determine any formal
justification, the presence of an upper envelope appears to be
plausible, if the asymmetries are related to the release of magnetic energy into
the stellar atmosphere. At lower activity levels such releases would then
be expected to occur less frequently than at higher activity levels, where they
can -- but apparently do not need to -- lead to detectable line profile
asymmetries.

\begin{figure}
\begin{center}
\includegraphics[width=0.5\textwidth, clip]{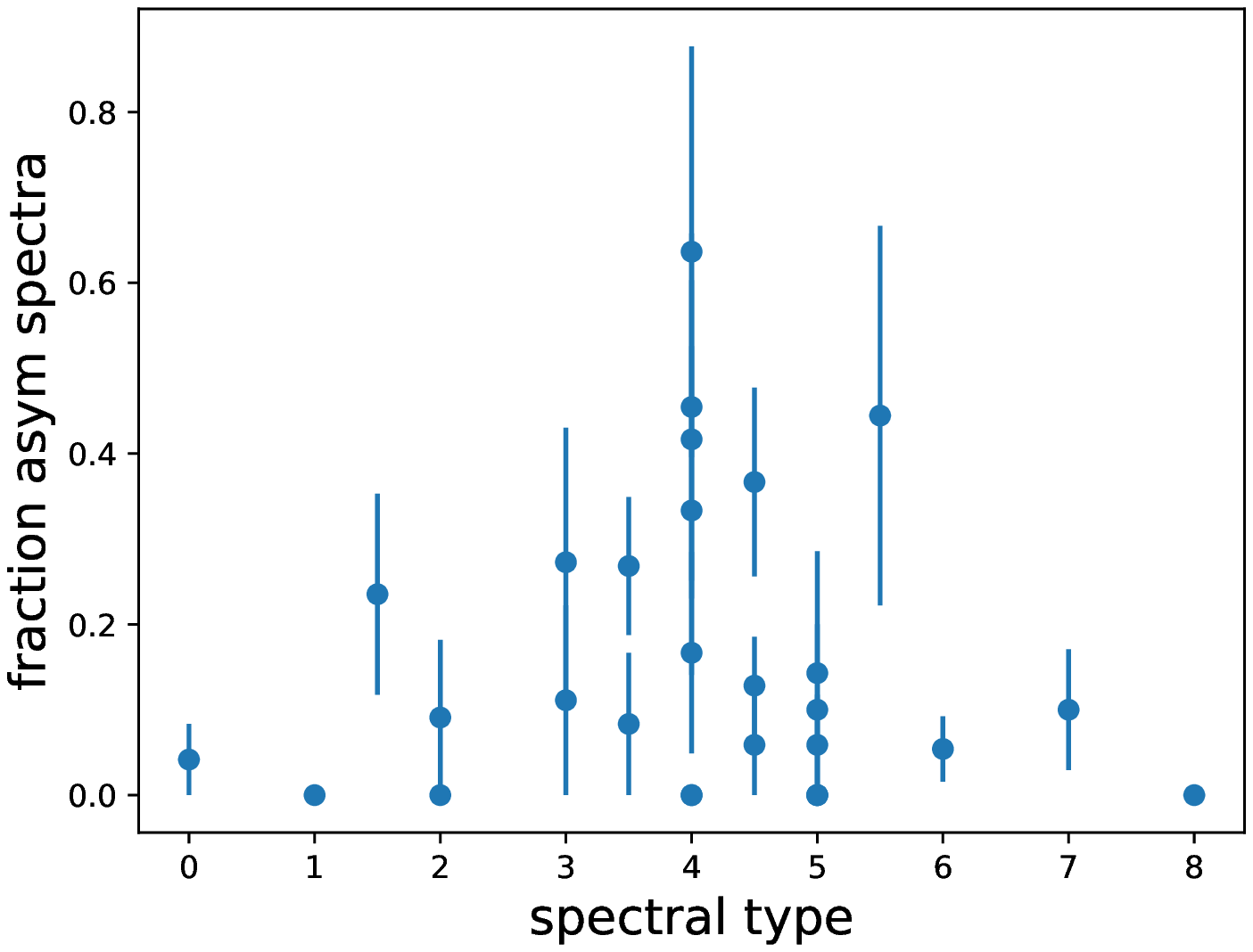}
\caption{\label{fracvstype} Fraction of spectra with 
  asymmetries (blue circles) as a function of spectral type.}
\end{center}
\end{figure}

Another parameter that may relate to the occurrence of asymmetries is spectral type.
  We plot the distribution of the fraction
of asymmetries as a function of spectral type in Fig.~\ref{fracvstype}.
There seems to be a concentration of a high occurrence rate of asymmetries around the mid-M spectral types.
The mean occurrence rate for early spectral types (M0 -- M2.5\,V) is 0.07, while for mid spectral type
(M3 -- M5.0\,V) it is 0.17, and for even later spectral types it is 0.14. We caution though, that this may be caused
by selection effects, since for example all our early stars are slow rotators.

\begin{table*}
\caption{\label{stats} Statistics of detected asymmetries and I$_{\rm H\alpha}$ measurements.}
\begin{tabular}[h!]{lccccccccccc}
\hline
\hline
\noalign{\smallskip}
Star       & Archived& Used & $N_{\rm flare}$ & $N_{\rm asymm red}$& $N_{\rm asymm blue}$& $N_{\rm symm}$& $N_{\rm flare,asymm}$& \multicolumn{4}{c}{I$_{\rm H\alpha}$}\\
 &  & & &   &  &  &  &max & min &median & MAD\\
\noalign{\smallskip}
\hline
\noalign{\smallskip}
V388 Cas   & 11  &  10 &  0  & 1  & 0 & 0 & 0  & -2.93  & -5.06  &  -3.52 & 0.23\\
YZ Cet     & 17  &  17 &  4  & 1  & 0 & 0 & 1  &  0.48  & -0.90  &  0.24  & 0.16\\
LP 768-113 & 10  &  10 &  5  & 0  & 0 & 0 & 0  &  0.17  & -0.23  &  0.01 & 0.09\\
Barta 161 12&11  &  (11) &  (1)  & (1)  & (3) & (1)  & (0)  &-0.93   & -2.05  & -1.32 & 0.34\\
TZ Ari     & 14  &  12 &  4  & 0  & 1  & 0 & 1  & -0.08  & -0.57  &  -0.24 & 0.13 \\
GJ 166 C  & 49  &  39 &  1  & 2   & 1  & 2  & 0  & -0.65  & -1.53  &  -0.93 & 0.14 \\
V2689 Ori  & 25  &  24 &  0  & 1  & 0 & 0 & 0  &  0.89  &  0.75  &  0.84  & 0.03\\
G099-049   & 13  &  12 &  2  & 1  & 0 & 1 & 1  & -0.24  & -0.79  &  -0.47 & 0.15\\
BD-022198  & 11  &   8 &  1  & 0  & 0 & 0 & -  &  0.65  &  0.37  &  0.60 & 0.03\\
YZ CMi     & 32  &  29 &  0  & 5  & 4 & 2 & 0  & -2.14  & -4.41  & -2.76 & 0.24\\
GJ 362     & 10  &   9 &  0  & 0 & 0  & 1  & 0  &  0.63  &  0.47  & 0.54 & 0.06\\
CN Leo     & 43  &  37 &  3  & 1 & 0 & 1  & 2  & -2.17  &-10.60  & -4.16 & 0.60 \\
DS Leo     & 27  &  26 &  0  & 0 & 0 & 0 & 0  &  0.81  &  0.70  & 0.77  & 0.02\\
WX UMa     & 12  &   9 &  0  & 1 & 2 & 1 & 0  & -2.83  & -4.05  & -3.83 & 0.18 \\
StKM 2-809  & 11  &   9 &  0  & 1 & 1 & 1  & 0  & -0.98  & -1.73  & -1.32 & 0.18\\
GL Vir     &  10 &    7 &  0  & 1 & 0 & 0  & 0  & -1.28  & -2.25 & -1.86 & 0.36 \\
OT Ser     &  17 &   17 &  2  & 1 & 2 & 1 & 2  & -0.23  & -0.64 & -0.37 & 0.06 \\
G180-060   &  11 &    6 &  0  & 0 & 0 &0  & 0  & -2.52  & -3.22 & -2.88 & 0.17 \\
vB 8       &  64 &   20 &  5 & 1 & 1 & 0   &2  &-0.32   &-4.62   & -1.38 & 0.33\\
LP 071-082 &  12 &   11 &  1  & 0 & 0 & 0  & -  & -0.87  & -1.67 & -1.00 & 0.12\\
J18174+483     &  11 &   11 &  0  & 0 & 1 &0 & 0  &  0.34  &  0.18 & 0.22 & 0.03\\
Ross 149   &  11 &   11 &  2  & 0  &0 & 0& -  &  0.42  & -0.01 &  0.34 & 0.06\\
G141-036   &  19 &   17 &  1  & 1  & 0 & 0& 1  & -0.65  & -3.57 &  -1.08  & 0.17\\
vB 10      &  22 &   16 &  0  & 0  & 0 & 0 & -  &  0.83  & -3.96 &  -1.24  & 0.50\\
V374 Peg   &  12 &   11 &  1  & 3 & 2 & 2  & 1  & -1.14  & -2.93 &  -1.53 & 0.20\\
EV Lac     &  45 &   41 &  4  & 6 & 3 & 2  & 3  & -1.26  & -6.40 &  -1.65 & 0.21 \\
GT Peg     &  12 &   12 &  1  & 2 & 0 &1  & 1  & -1.04  & -2.50 &  -1.40 & 0.14 \\
HH And     &  30 &   30 &  3  & 0 & 0 & 0 & 0  & 0.74   & 0.01 & 0.59  & 0.04\\
RX J2354.8+3831 &  13 &   12 &  1  & 3 & 2 & 0  & 1  & -1.31  & -2.67 & -1.69 & 0.12 \\
\noalign{\smallskip}
\hline
\noalign{\smallskip}
Total      &     & 473 & 41 & 32 & 20 & 15 &  16 \\         
\noalign{\smallskip}
\hline
\end{tabular}
Note: The spectra of Barta 161 12 are not counted in the total.
\end{table*}

\subsection{The narrow Gaussian component}
\label{sec:narrowshift}
The mean width of the narrow Gaussian component as
measured by its FWHM found in our analysis is
$65$\,km\,s$^{-1}$ with a standard deviation of $10$\,km\,s$^{-1}$.
The mean absolute shift of the narrow Gaussian core components is
$3.9$\,km\,s$^{-1}$, which is significantly lower than its mean width
and about an order of magnitude less than the mean
absolute shift of $35$\,km\,s$^{-1}$ found for the broad component.
For only nine spectra we measure a shift exceeding 10\,km\,s$^{-1}$. Of these,
five originate from the fast rotator V374~Peg, for which we suspect a rotational
contribution to the shift (see also Sect.~\ref{absorption}).
The four remaining cases are due to TZ~Ari, OT~Ser, vB~8, and WX~UMa, where the
shift cannot be explained by rotation.
In the specific case of OT~Ser, the
spectrum also appears to suffer from some spectral defects that may influence
the fit.

We find only small shifts in the
position of the core component of the H$\alpha$ line as measured by the narrow
Gaussian component, in particular when compared with the line width. The shift
in the position of the narrow Gaussian component is uncorrelated with its
area. The distribution is shown in Fig.~\ref{shift}.
At any rate, the two-component Gaussian model only provides an approximate
representation of the true line profile, which is generally more complex,
so that some residual shifts are expected.
The scatter appears to decrease for stronger
emission line cores, which may be due to a better constrained fit in these
cases. Therefore, we consider the measurements consistent with a stationary
emission line core.

\subsection{Shift of the broad Gaussian component}\label{broadshift}

The measured shift of the broad component varies between about $-100$ and
$100$\,km\,s$^{-1}$.
There are 15 (22\,\%) spectra showing a shift of less than
$\pm 10$~km\,s$^{-1}$ in the broad component. In analogy with the distribution
of shifts in the narrow component (see Sect.~\ref{sec:narrowshift}),
we consider these shifts insignificant and the profiles symmetric. 
In our sample, we also find a surplus of red shifted broad wings. Only considering
broad components with an absolute shift larger than $10$~km\,s$^{-1}$, we find
20 blue shifted ($30$\,\%) and 32 red shifted ($48$\,\%) components, while
for absolute shifts larger than 50~km\,s$^{-1}$, there are four (6\,\%) blue shifted
and 13 (19\,\%) red shifted components.  An excess of red asymmetries is also
consistent with observations of solar flares \citep{Svestka1962, Tang1983}.



In the context of our envisaged scenario, the asymmetric line
profiles do require atmospheric motions.  We note that 
all measured velocities are far below the escape velocity of 500 to
600\,km\,s$^{-1}$ for mid to early M~dwarfs.
Theoretically, a (radial) mass ejection at the limb would be
observed at much lower velocity due to projection effects,
but we consider improbable the possibility of observing  ejections only at the stellar limb
and not at the centre.
Assuming a
strictly radial motion and a true velocity of $500$\,km\,s$^{-1}$, only $20$\,\%
of all directions yield projected velocities of $\pm 100$\,km\,s$^{-1}$ or less.
Thus, we consider it unlikely that the observed blue shifts are exclusively due
to limb ejections, although we cannot rule out the possibility
in individual cases. However, even the extreme case
of a blue asymmetry in vB~8 (Fig.~\ref{vb8}) does not require any material
near the escape velocity because the largest shifts in the line flank
at about 6561\,\AA\
correspond to velocities of only 160\,km\,s$^{-1}$.

Turning now to red shifts, the observed red shifts are comparable with the speeds measured for
solar coronal rain, and we consider a phenomenon similar to coronal rain a
viable explanation for the observed line shifts. An alternative
explanation could be chromospheric condensations. 

\begin{figure*}
\begin{center}
\includegraphics[width=0.5\textwidth, clip]{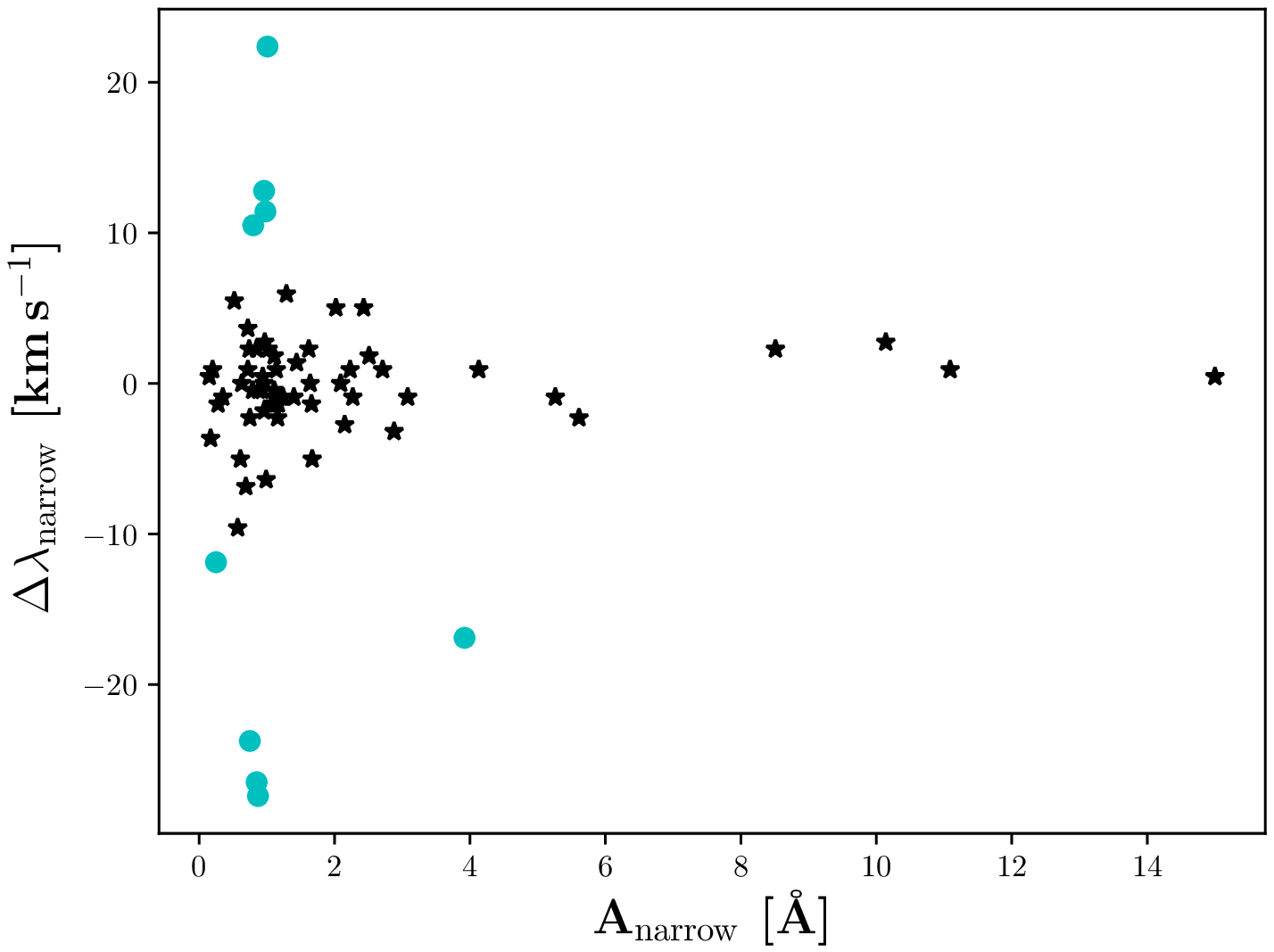}
\includegraphics[width=0.5\textwidth, clip]{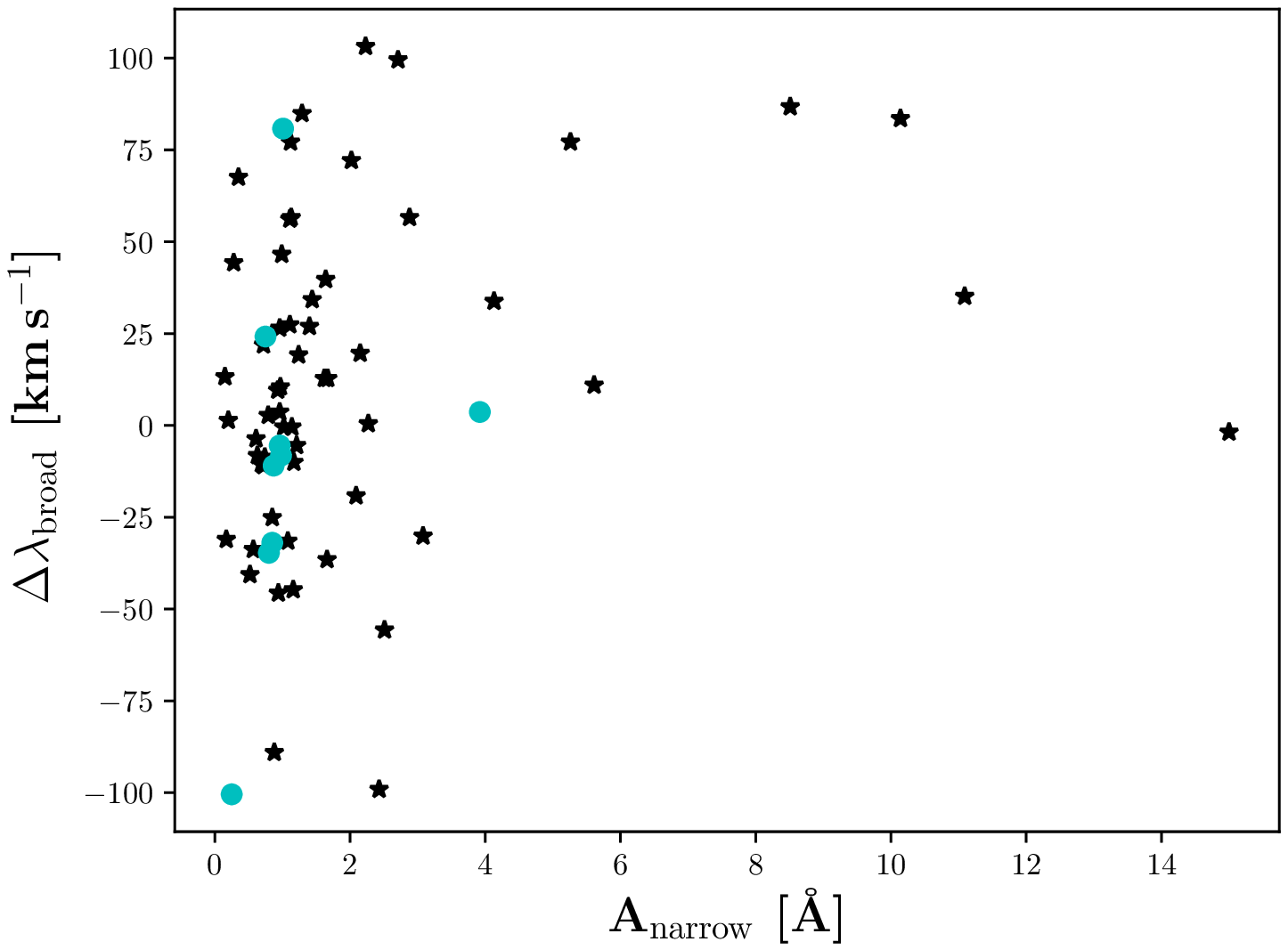}
\caption{\label{shift} Shift of narrow (left panel) and broad (right panel) Gaussian
  component as a function of their continuum-normalised residual flux density (measured by the Gaussian area A
  of the residual line fit).
Cyan dots mark outliers showing absolute shifts exceeding $10$\,km\,s$^{-1}$
in the narrow component.}
\end{center}
\end{figure*}

In the context of our scenario
(see Sect.~\ref{scenario}), blue asymmetries would be associated with rising
material, primarily in the early phases of flares.
In Fig.~\ref{shift} we show the shift of the broad component as a function of
continuum-normalised flux, which shows a
striking absence of blue shifts for fluxes in the
narrow component higher than about 3\,\AA. A single exception
corresponds to a spectrum of CN~Leo, showing a rather marginal blue shift
(which anyway makes it a symmetric case). 
For weaker narrow components, the distribution is more balanced. 
We argue that spectra showing strong blue shifted asymmetries with weak
narrow components cover primarily the onset of a flare, where the line
cores do not yet show a strong response.
The case of CN~Leo may be explained by the exposure covering both the flare
onset, peak, and potentially a fraction of the decay phase. Then,
both blue and red asymmetries would mix and give the appearance of a
broad line with only marginal net shift and substantial flux in the
narrow component.

While the observed shifts can, therefore, be reconciled with our scenario,
 red and blue shifts are caused by
completely different physical processes. In light of this, it appears remarkable
that the maximum velocities of the observed blue and red shifts are so
similar and we currently lack a concise physical explanation for this. 

%

\subsection{Width of the broad component}\label{linewidth}

\begin{figure}
\begin{center}
\includegraphics[width=0.5\textwidth, clip]{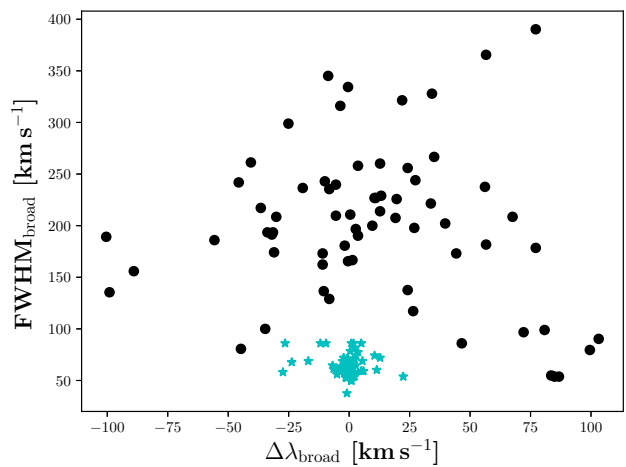}
\caption{\label{widthvsshift} Width of the broad (black) and narrow (cyan)
components as a function of
their shift. }
\end{center}
\end{figure}

The fitted widths of the broad
component vary between 0.5 and 3.6\,\AA\ with a mean of 1.72\,\AA,
corresponding to a velocity of $184.9$\,km\,s$^{-1}$ as FWHM.
In Fig.~\ref{widthvsshift}
we show the distribution of the width of the broad and narrow Gaussian
components as a function of their shift; the numbers are given in
Table~\ref{asyms}. Typical values range between $100$ and $200$\,km\,s$^{-1}$
with a maximum of $390$\,km\,s$^{-1}$.
Visual inspection of the fits shows, however, that the top two widths
of $390$ and $365$\,km\,s$^{-1}$, corresponding to
flares on EV~Lac and GT~Peg, are probably overestimated.

The distribution shown in Fig.~\ref{widthvsshift} suggests a decrease in the
width as the shift increases either to the red or the blue side. This impression
is corroborated by a value of $-0.54$ for
Pearson's correlation coefficient between
the width and the absolute value of the shift (excluding the two
outliers); the p-value for a chance result is $3\times 10^{-6}$.
A stronger contribution of Stark broadening at smaller shifts is
conceivable. Also the coarse time resolution may play a role, because
most exposures that cover the short flare onset should also cover the
flare peak and the start of the decay phase.
In our sample, we do not find strong blue asymmetries in combination with strong
narrow components. We argue that this indicates that evaporation was
accompanied by red asymmetries in all cases studied here. 
This could indicate that
evaporated material increases the density in the coronal loops,
leading to more efficient cooling, followed by H$\alpha$
rain. A red coronal rain asymmetry
from the flare peak then adds to the broadening,
making the profile more symmetric.


When the shift of the broad component is sufficiently small for the H$\alpha$
line to be considered symmetric, we cannot distinguish if the lines are
broadened by high charge density, turbulence, or downward and upward moving
material, whose distinct signatures are smeared out due to the long integration time.
In fact, the latter effect may account for widths
in excess of $300$\,km\,s$^{-1}$, which appear high for an interpretation in
terms of turbulence. 
However, the majority of the broadened lines show substantial asymmetry
in the broad component for which Stark broadening does not appear to be a
viable explanation. 
In their modelling of a chromospheric
condensation computed to represent a solar flare event, \citet{Kowalski1} find that Stark broadening
does play a significant role for the Balmer lines but not for other lines. The authors 
find red asymmetries with
  the line width from instantaneous model snapshots to be actually lower than
  that observed. The inclusion of moderate micro-turbulence, together with
  the velocity evolution of the shifts in the model accumulated over the
  exposure time, leads to a realistic broadening in their study.
The determined widths of the broad component are comparable to those observed in
the \ion{Mg}{ii} lines by \citet{Lacatus} on the Sun. By analogy, we consider magnetic turbulence a
viable broadening mechanism to explain the observed widths of the H$\alpha$
lines.

\subsection{H$\alpha$ line profiles of the fast rotators Barta 161 12, StKM 2-809, and V374~Peg}
\label{absorption}
\begin{figure}[h]
  \includegraphics[width=0.49\textwidth]{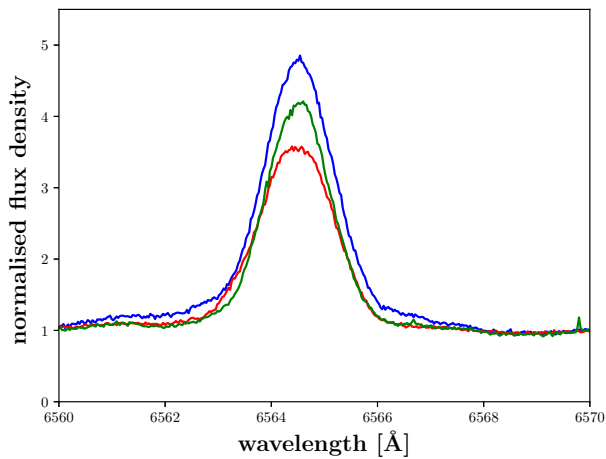}
  \caption{Normalised strongest (blue), weakest (red), and
  medium-activity (green) H$\alpha$ line profiles of StKM 2-809 observed
  by CARMENES.
  \label{fig:stkm2_3profiles}}
\end{figure}

The three fastest rotators Barta 161 12, StKM 2-809, and V374~Peg show complex
features in the H$\alpha$ line profiles observed by CARMENES.
For all three stars the H$\alpha$ line
profile continuously varies.
In Fig.~\ref{fig:stkm2_3profiles} we show three H$\alpha$ line profiles (prior
to subtraction) of StKM 2-809, namely, those with the weakest and strongest
detected H$\alpha$ lines and a spectrum with intermediate line strength.  
The lowest-activity H$\alpha$ line profile shows a more pronounced blue
wing than the intermediate-activity spectrum shown here. Subtracting the
lowest-activity state spectrum produces relative absorption features in
the residual spectrum in the blue flank of the H$\alpha$ line.
For StKM 2-809, we find such features in two out of three
spectra with asymmetry, but some spectra without marked asymmetry show
also complex H$\alpha$ line profiles. In this case, the absorption features
  may influence our fits but not to a degree that would call the presence of
  additional shifted emission into question. Therefore, we include the star
  in our normal analysis since we believe that the main effect is similar to
  that in slower rotators.

For V374~Peg  
 only one spectrum shows a significant relative
absorption feature along with an asymmetry (see Fig.~\ref{asymV374Peg},
fourth from bottom). Therefore, we believe that also for V374 Peg the
  asymmetries are mainly caused by the same mechanism, although some profiles
  also show plateaus and one profile is double-peaked (second from bottom in Fig.~\ref{asymV374Peg}).

Barta~161~12 shows the most pronounced variety of H$\alpha$ profiles, which in many cases need more
than two Gaussians for fitting (Fig.~\ref{asymBarta})
representing a different case of asymmetries. Only one spectrum of this star
seems to fall into the category of asymmetries discussed in this paper. Therefore, we
exclude it from our analysis as already mentioned in Sect. \ref{sec:sample}. For
the other two fast rotators the situation is less extreme.
All three stars show significant shift in the narrow H$\alpha$ line component
but never higher than their rotational velocity.


Whether or not the relative absorption features seen in the subtracted spectra
of the three fast rotators are due to true absorption, for example caused by a prominence, or are the results of
excess emission in the lowest-activity spectrum used in the subtraction cannot
be determined on the basis of our data. We also want to note the possibility that
  the plateaus and multi-peaked spectra of especially Barta~161~12 (Fig.~\ref{asymBarta},
  top two spectra) may originate
in distinct active regions, rotating along with the stellar surface.

\subsection{Broad components and asymmetries in other lines}


\subsubsection{The optical \ion{Na}{i} D and \ion{He}{i} D$_{3}$ lines}
\label{NaDlines}
The optical arm of CARMENES covers further known activity indicators beyond
the H$\alpha$ line. In particular, we investigated
the \ion{Na}{i}~D and \ion{He}{i}~D$_{3}$ lines and, indeed, found
asymmetric components in a
number of spectra of EV~Lac, OT~Ser, and StKM 2-809.

\begin{figure}
\begin{center}
  \includegraphics[width=0.5\textwidth, clip]{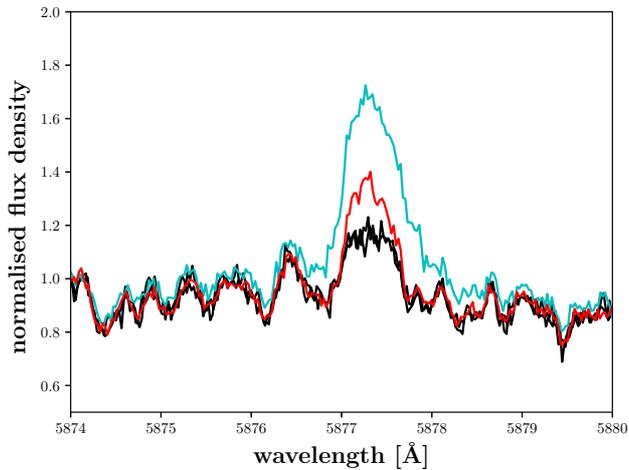}

  \includegraphics[width=0.5\textwidth, clip]{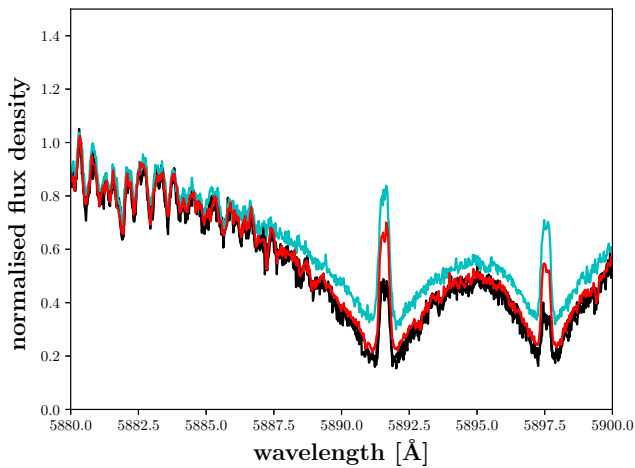}
  \caption{\label{otsernad} Line asymmetries in spectra of OT~Ser.
  Normalised \ion{He}{i}~D$_{3}$ (top) and \ion{Na}{i}~D line profiles (bottom). 
    In both cases, we show two typical quiescent spectra in black along with
    two spectra of enhanced activity in cyan and red (same spectra and
    colour coding as in Fig.~\ref{OTSer}).}
\end{center}
\end{figure}

In Fig.~\ref{otsernad} we show an example for the star OT~Ser; the corresponding
H$\alpha$ line profiles are presented in Fig.~\ref{OTSer} with the same colour
coding. While two enhanced spectra in the H$\alpha$ line are nearly identical, this
is not true for the corresponding \ion{He}{i}~D$_{3}$ and \ion{Na}{i}~D lines, indicating a different state of the chromosphere at the two times.

Because the broad components of the two \ion{Na}{i}~D lines overlap,
we fit them simultaneously with a total of four
Gaussian components, representing two narrow and two broad components. 
We fix the offset in wavelength of the narrow and broad components to the
known doublet separation and also couple their respective widths.
The best-fit values for the three spectra with broad wings in \ion{Na}{i}~D and
\ion{He}{i}~D$_{3}$ are given in Table~\ref{asymsnad}; the spectra are also
indicated in the notes column of Table~\ref{asyms}.

The narrow Gaussian component is in all cases only slightly shifted, and
its width stays about the same, with the
\ion{He}{i}~D$_{3}$ line being slightly broader than the \ion{Na}{i}~D lines.
Both lines are significantly narrower than the narrow component of the
H$\alpha$ line, which is broader because it is optically thick for our sample stars.
For the broad component, the picture is more complicated with most of the
fit results for the individual lines -- also compared to the broad component of
the H$\alpha$ line -- differing significantly.

These differences may be caused by the different height at which the lines
originate in the atmosphere;
H$\alpha$ is thought to be formed in the upper chromosphere, while the
\ion{Na}{i}~D lines originate in the lower chromosphere.
Although we consider the fits satisfactory
(see Figs.~\ref{asymnad1}--\ref{asymnad3}),
part of the differences may also be attributable to problems in the fit itself
such as the approximation by Gaussian profiles.
However, Figs.~\ref{asymnad1}--\ref{asymnad3}  demonstrate that
the \ion{He}{i}~D$_{3}$ line shows a red asymmetry in StKM 2-809, while the
\ion{Na}{i}~D lines are more symmetric, exhibiting emission in the blue wing as
well. Thus we are confident that a physical reason underlies the line profile
differences, which may be related
to distinct sites of line formation and radiative transfer. Again, however, we
caution that the phenomena may actually not have occurred simultaneously, but
arise from time averaging during the exposure (see Sect.~\ref{sec:Timing}). 

The spectrum of OT~Ser showing broadening in the \ion{Na}{i}~D lines also
exhibits relatively symmetric, broad wings in the H$\alpha$ line. While the
latter may be caused by Stark broadening, the \ion{Na}{i}~D lines are not
thought to be affected by this. Therefore, the broadened \ion{Na}{i}~D lines
suggest a different broadening mechanism, probably turbulence, at least in this
individual case.

\begin{table*}
\caption{\label{asymsnad} Detected asymmetries in \ion{Na}{i} D (5891.58 and 5897.56 \AA) and \ion{He}{i} D$_{3}$ (5877.24 \AA). }
\footnotesize
\begin{tabular}[h!]{llcccccccccc}
\hline
\hline
\noalign{\smallskip}
Star  & Line &\multicolumn{5}{c}{Gauss 1} & \multicolumn{5}{c}{Gauss 2}
\\
          & &Area   & $\lambda_{\rm{central}}$ &   $v_{\rm{central}}$ &$\sigma$& FWHM & Area  &
           $\lambda_{\rm{central}}$ & $v_{\rm{central}}$&  $\sigma$&FWHM \\
          &  &     &  [\AA]& [km\,s$^{-1}$]& [\AA]  & [km\,s$^{-1}$]&     
          & [\AA]& [km\,s$^{-1}$] & [\AA]&[km\,s$^{-1}$]  \\ 
\noalign{\smallskip}
\hline
\noalign{\smallskip}
OT Ser &  \ion{Na}{i} D$_{1}$ & 0.12 & 5897.54 & $-1.0$ & 0.20 & 24.0& 0.27 &
5897.64 &4.1& 1.97 & 236.\\
OT Ser &  \ion{Na}{i} D$_{2}$ & 0.13 & 5891.56 & $-1.0$ & 0.20 & 24.0& 0.34 &
5891.66 &4.1& 1.97 & 236.\\
  OT Ser & \ion{He}{i} D$_{3}$& 0.37 & 5877.37 &6.6& 0.34 &40.8 & 0.35 & 5878.47 &62.7& 2.74&329.0\\
  EV Lac & \ion{Na}{i} D$_{1}$ & 0.70& 5897.59 &1.5& 0.17 &20.4 & 0.29 & 5897.59 &1.5& 0.80 & 95.8\\
    EV Lac & \ion{Na}{i} D$_{2}$ & 0.72 & 5891.61 &1.5& 0.17 &20.4 & 0.51  & 5891.61 &1.5& 0.80 & 95.8\\
  EV Lac & \ion{He}{i} D$_{3}$& 1.60 & 5877.34 &5.1& 0.30 &36.0 & 1.5 & 5877.6
  &18.4& 1.11 &133.3\\
  StKM 2-809&\ion{Na}{i} D$_{1}$  & 0.19 & 5897.55 &0.&  0.40 &47.9 & 0.65 &
  5897.04 & $-26.5$ & 2.44 & 292.3\\
  StKM 2-809&\ion{Na}{i} D$_{2}$  & 0.19 & 5891.58 &0.&  0.40 &47.9 & 0.79 &
  5891.06 & $-26.5$& 2.44 & 292.3\\
  StKM 2-809&\ion{He}{i} D$_{3}$& 0.47 & 5877.45 &10.7& 0.34 &40.8& 0.23 & 5878.49 &63.7& 0.85& 102.1\\
\noalign{\smallskip}
\hline
\end{tabular}
\end{table*}

\subsubsection{The \ion{He}{i} and Pa$\beta$ near-infrared lines}\label{IRlines}
The infrared arm of CARMENES also covers known chromospherically-sensitive lines
such as the Pa$\beta$ line and the \ion{He}{i} line at $10833.31$\,\AA,
the latter of which has  primarily been used in the solar context \citep[e.g.][]{Andretta}.
We searched all 72 spectra with asymmetries in H$\alpha$ also for asymmetries in
\ion{He}{i} 10833\,\AA\ and Pa$\beta$.
Unfortunately, not all available optical
spectra have infrared counterparts (cf., Table~\ref{asyms}).
For the analysis, 40 spectra were available of which we discarded four
because of the telluric contamination was too severe.
Of the 36 usable spectra, 19 show no excess in the \ion{He}{i} line, two show a
weak excess, and 15 a clear excess. Unfortunately,
the low excess flux in the Pa$\beta$ lines precludes a meaningful
analysis of its symmetry here.

We show the 15 excess spectra in Appendix \ref{appendixc}
in Figs.~\ref{He10830V388} to \ref{He10830GTPeg}
along with the corresponding Pa$\beta$ excess fluxes.
Also for the \ion{He}{i} 10833\,\AA\ line, the degree of asymmetry is hard to
assess in most cases because the red side of the line is contaminated by strong
air glow or telluric lines. 

\begin{figure}
\begin{center}
  \includegraphics[width=0.5\textwidth, clip]{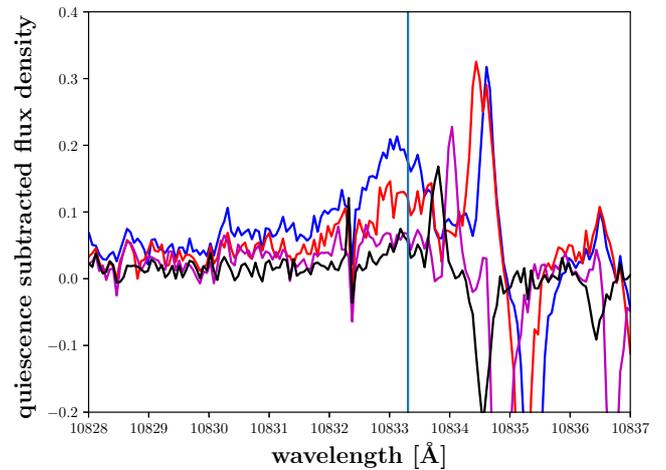}

  \includegraphics[width=0.5\textwidth, clip]{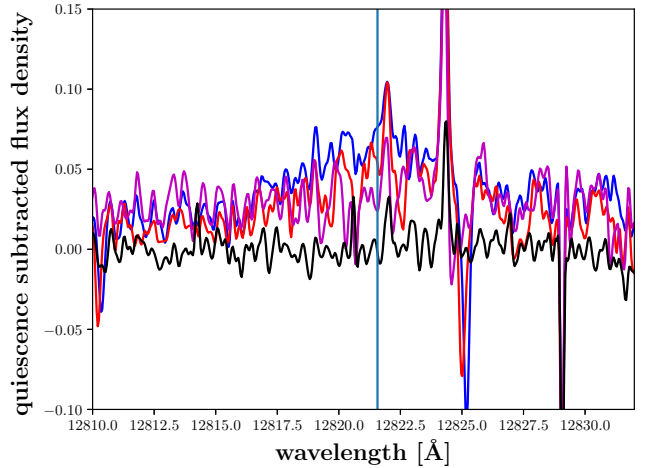}
  \caption{\label{evlacir} Line asymmetries in infrared spectra of EV Lac. Top: The quiescence subtracted \ion{He}{i}
    10833 \AA\, line profile.
    Bottom: The quiescence subtracted Pa$\beta$ line profiles smoothed over eight bins for reduction of noise.
    The colours indicate different active spectra of EV~Lac but correspond to the same spectra for both lines.
  In both cases the reference central wavelength is marked by the blue vertical line.}
\end{center}
\end{figure}

In Fig.~\ref{evlacir}, we show the \ion{He}{i} and Pa$\beta$ lines of EV~Lac
(the corresponding spectra are also marked by `IR' in Table~\ref{asyms}).
The infrared regime is more heavily affected by
telluric contamination and air glow lines \citep{Rousselot}. Because a number of
strong lines are located in the direct vicinity of the \ion{He}{i} line
(cf., Fig.~\ref{evlacir}), and an appropriate correction is very challenging,
we limit ourselves to a qualitative analysis.
In Fig.~\ref{evlacir}, we show four spectra with enhanced emission in
the quiescent subtracted flux density. While three spectra do not
show clear signs of an asymmetry, one spectrum does exhibit a strong
blue asymmetry (taken at JD 2\,457\,632.6289 d, cf., Table~\ref{asyms}) and also the corresponding
H$\alpha$ line shows a blue asymmetry in this case.
In the Pa$\beta$ line, we find enhancements in the line flux for the three
flare spectra as can be seen in the bottom panel of Fig.~\ref{evlacir}. 
Like the spectrum described above, this one also suggests a blue asymmetry. 

The two infrared lines studied here clearly react to different activity
states. In our spectra, emission in the \ion{He}{i} 10833\,\AA\ line is clearly
detected in $34$\,\% of the available infrared spectra also showing an H$\alpha$ asymmetry,
while the Pa$\beta$ line appears to be less sensitive to chromospheric states
causing broad wings. Asymmetries are difficult to assess in the Pa$\beta$ and \ion{He}{i} 10833\,\AA\
line. Only in three cases, namely two spectra of EV~Lac (first two) and the last
spectrum of StKM 2-809, we are confident that a blue asymmetry is present in the
\ion{He}{i} 10833\,\AA\ line.
These correspond also to blue shifts in H$\alpha$. 



\section{Conclusions}
We present the most comprehensive high-resolution spectral survey of
wings and asymmetries in the profiles of H$\alpha$ and other chromospheric
lines in M~dwarfs to date. We found 67 asymmetries in 473 spectra of 28 emission-line
M~dwarf stars.
Line wing asymmetries in the H$\alpha$ line, therefore, appear to be a wide-spread
phenomenon in active M~dwarfs, which we find in about 15\,\% of the
analysed spectra. Only eight stars do not show
any wings or asymmetries.

In contrast to that, asymmetries in other spectral lines seem to occur less frequently. In particular, only 4\,\% of the spectra with detected H$\alpha$
asymmetry show also an asymmetry in the optical \ion{Na}{i}~D and
\ion{He}{i}~D$_{3}$ lines.
In the \ion{He}{i}~10833\,\AA\ line, we find broad wings for 34\,\% of the
cases.
Excess emission in the Pa$\beta$ line was found only in a few cases, but
we considered a meaningful analysis of its asymmetry unfeasible.

While the occurrence rate of the asymmetries varies widely between stars, we
find a relation with the stellar activity level. In particular, more
active stars (as measured by rotational velocity or median(I$_{\rm H\alpha}$))
also show a higher fraction of spectra with asymmetric H$\alpha$ line profiles.
Nevertheless, we find that only 24\,\% of the asymmetries are associated
with a flare, as defined by our flare criterion, and we conjecture
that either asymmetries are not necessarily coupled to (large) flares or they can
persist to the very end of the decay phase.

We find our results to be consistent with a scenario in which red asymmetries
are associated with coronal rain or chromospheric condensations, two 
phenomena well known from the Sun. Unfortunately, they cannot be distinguished
on the basis of the occurring velocities or line width, and we lack the timing
information of when (and sometimes if) during a flare the asymmetry occurs.
\citet{Lacatus} explain the width by Alfv\'{e}nic
turbulence, which may also cause the broadening in our case. Because of
the relatively long exposure times for M~dwarfs, however, we cannot rule out
that the width is due to an evolution in the velocity field.

For the blue wing enhancements, we propose chromospheric evaporation as the mechanism responsible, which explains why we do not find substantially blue shifted
broad components in combination with strong narrow core components.
Supposedly, the chromospheric evaporation occurs during the early impulsive stage of a flare,
while the largest amplitudes in chromospheric lines are typically observed at
flare peak or early decay phase. While also symmetric line profiles may 
be explained by coronal rain with low (projected) velocities or time average effects, we cannot
exclude Stark broadening as an alternative explanation. One exception here is the case of
OT~Ser, which exhibits a symmetrically broadened H$\alpha$ line along with
broadened \ion{Na}{i}~D lines, which is not expected for the Stark effect;
in fact, Stark broadening affects higher Balmer lines more strongly than
H$\alpha$ itself. While CARMENES does not cover these lines, a dedicated study
of the effect may take advantage of this fact.

While the data presented here are well suited for a statistical study of
broadening and asymmetry in the H$\alpha$ line, there is a
clear need for further observations. Any detailed modelling attempt would
greatly benefit from dedicated observations that clarify whether asymmetries
are coupled to flares in all cases and if so, which sort of asymmetry (red or blue) is associated
to which phase of the flare.
Therefore, to better understand the phenomenon of asymmetric line profiles,
continuous spectral time series with a higher temporal cadence (or simultaneous photometry) and spectral
coverage of the end of the Balmer series are definitely needed.

\begin{acknowledgements}
  B.~F. acknowledges funding by the DFG under Cz 222/1-1 and thanks
E. N. Johnson and L. Tal-Or for helpful remarks.
  CARMENES is an instrument of the Centro Astron\'omico Hispano-Alem\'an de
  Calar Alto (CAHA, Almer\'{\i}a, Spain). 
  CARMENES is funded by the German Max-Planck-Gesellschaft (MPG), 
  the Spanish Consejo Superior de Investigaciones Cient\'{\i}ficas (CSIC),
  the European Union through FEDER/ERF FICTS-2011-02 funds, 
  and the members of the CARMENES Consortium 
  (Max-Planck-Institut f\"ur Astronomie,
  Instituto de Astrof\'{\i}sica de Andaluc\'{\i}a,
  Landessternwarte K\"onigstuhl,
  Institut de Ci\`encies de l'Espai,
  Insitut f\"ur Astrophysik G\"ottingen,
  Universidad Complutense de Madrid,
  Th\"uringer Landessternwarte Tautenburg,
  Instituto de Astrof\'{\i}sica de Canarias,
  Hamburger Sternwarte,
  Centro de Astrobiolog\'{\i}a and
  Centro Astron\'omico Hispano-Alem\'an), 
  with additional contributions by the Spanish Ministry of Economy, 
  the German Science Foundation through the Major Research Instrumentation 
    Programme and DFG Research Unit FOR2544 ``Blue Planets around Red Stars'', 
  the Klaus Tschira Stiftung, 
  the states of Baden-W\"urttemberg and Niedersachsen, 
  and by the Junta de Andaluc\'{\i}a.

\end{acknowledgements}

\bibliographystyle{aa}
\bibliography{papers}

\begin{thebibliography}{80}
\expandafter\ifx\csname natexlab\endcsname\relax\def\natexlab#1{#1}\fi

\bibitem[{{Alonso-Floriano} {et~al.}(2015){Alonso-Floriano}, {Morales},
  {Caballero}, {Montes}, {Klutsch}, {Mundt}, {Cort{\'e}s-Contreras}, {Ribas},
  {Reiners}, {Amado}, {Quirrenbach}, \& {Jeffers}}]{Alonso-Floriano}
{Alonso-Floriano}, F.~J., {Morales}, J.~C., {Caballero}, J.~A., {et~al.} 2015,
  \aap, 577, A128

\bibitem[{{Andretta} {et~al.}(2017){Andretta}, {Giampapa}, {Covino}, {Reiners},
  \& {Beeck}}]{Andretta}
{Andretta}, V., {Giampapa}, M.~S., {Covino}, E., {Reiners}, A., \& {Beeck}, B.
  2017, \apj, 839, 97

\bibitem[{{Antolin} \& {Rouppe van der Voort}(2012)}]{Antolin12}
{Antolin}, P. \& {Rouppe van der Voort}, L. 2012, \apj, 745, 152

\bibitem[{{Antolin} {et~al.}(2015){Antolin}, {Vissers}, {Pereira}, {Rouppe van
  der Voort}, \& {Scullion}}]{Antolin}
{Antolin}, P., {Vissers}, G., {Pereira}, T.~M.~D., {Rouppe van der Voort}, L.,
  \& {Scullion}, E. 2015, \apj, 806, 81

\bibitem[{{Bell} {et~al.}(2012){Bell}, {Hilton}, {Davenport}, {Hawley}, {West},
  \& {Rogel}}]{Bell}
{Bell}, K.~J., {Hilton}, E.~J., {Davenport}, J.~R.~A., {et~al.} 2012, \pasp,
  124, 14

\bibitem[{{Berdyugina} {et~al.}(1999){Berdyugina}, {Ilyin}, \&
  {Tuominen}}]{Berdyugina}
{Berdyugina}, S.~V., {Ilyin}, I., \& {Tuominen}, I. 1999, \aap, 349, 863

\bibitem[{{Berlicki}(2007)}]{Berlicki}
{Berlicki}, A. 2007, in Astronomical Society of the Pacific Conference Series,
  Vol. 368, The Physics of Chromospheric Plasmas, ed. P.~{Heinzel},
  I.~{Dorotovi{\v c}}, \& R.~J. {Rutten}, Heinzel

\bibitem[{{Caballero}(2010)}]{Caballero2010}
{Caballero}, J.~A. 2010, \aap, 514, A98

\bibitem[{{Caballero} {et~al.}(2016{\natexlab{a}}){Caballero},
  {Cort{\'e}s-Contreras}, {Alonso-Floriano}, {Montes}, {Quirrenbach}, {Amado},
  {Ribas}, {Reiners}, {Abellan}, {B{\'e}jar}, {Brinkm{\"o}ller}, {Czesla},
  {Dorda}, {Gallardo}, {Gonz{\'a}lez-{\'A}lvarez}, {Hidalgo}, {Holgado},
  {Jeffers}, {Kim}, {Klutsch}, {Lamert}, {Llamas}, {L{\'o}pez-Santiago},
  {Mart{\'{\i}}nez-Rodr{\'{\i}}guez}, {Morales}, {Mundt}, {Passegger},
  {Sch{\"o}fer}, {Seifert}, \& {Zechmeister}}]{carmencita}
{Caballero}, J.~A., {Cort{\'e}s-Contreras}, M., {Alonso-Floriano}, F.~J.,
  {et~al.} 2016{\natexlab{a}}, in 19th Cambridge Workshop on Cool Stars,
  Stellar Systems, and the Sun (CS19), 148

\bibitem[{{Caballero} {et~al.}(2016{\natexlab{b}}){Caballero}, {Gu{\`a}rdia},
  {L{\'o}pez del Fresno}, {Zechmeister}, {de Juan}, {Alonso-Floriano}, {Amado},
  {Colom{\'e}}, {Cort{\'e}s-Contreras}, {Garc{\'{\i}}a-Piquer}, {Gesa}, {de
  Guindos}, {Hagen}, {Helmling}, {Hern{\'a}ndez Casta{\~n}o}, {K{\"u}rster},
  {L{\'o}pez-Santiago}, {Montes}, {Morales Mu{\~n}oz}, {Pavlov}, {Quirrenbach},
  {Reiners}, {Ribas}, {Seifert}, \& {Solano}}]{Caballero2}
{Caballero}, J.~A., {Gu{\`a}rdia}, J., {L{\'o}pez del Fresno}, M., {et~al.}
  2016{\natexlab{b}}, in \procspie, Vol. 9910, Observatory Operations:
  Strategies, Processes, and Systems VI, 99100E

\bibitem[{{Canfield} {et~al.}(1990){Canfield}, {Penn}, {Wulser}, \&
  {Kiplinger}}]{Canfield}
{Canfield}, R.~C., {Penn}, M.~J., {Wulser}, J.-P., \& {Kiplinger}, A.~L. 1990,
  \apj, 363, 318

\bibitem[{{Cheng} {et~al.}(2006){Cheng}, {Ding}, \& {Li}}]{Cheng}
{Cheng}, J.~X., {Ding}, M.~D., \& {Li}, J.~P. 2006, \apj, 653, 733

\bibitem[{{Cho} {et~al.}(2016){Cho}, {Lee}, {Chae}, {Wang}, {Ahn}, {Yang},
  {Lim}, \& {Maurya}}]{Cho2016}
{Cho}, K., {Lee}, J., {Chae}, J., {et~al.} 2016, \solphys, 291, 2391

\bibitem[{{Chugainov}(1974)}]{Chu74}
{Chugainov}, P.~F. 1974, Izvestiya Ordena Trudovogo Krasnogo Znameni Krymskoj
  Astrofizicheskoj Observatorii, 52, 3

\bibitem[{{Cram} \& {Mullan}(1979)}]{Mullan}
{Cram}, L.~E. \& {Mullan}, D.~J. 1979, \apj, 234, 579

\bibitem[{{Crespo-Chac{\'o}n} {et~al.}(2006){Crespo-Chac{\'o}n}, {Montes},
  {Garc{\'{\i}}a-Alvarez}, {Fern{\'a}ndez-Figueroa}, {L{\'o}pez-Santiago}, \&
  {Foing}}]{Crespo}
{Crespo-Chac{\'o}n}, I., {Montes}, D., {Garc{\'{\i}}a-Alvarez}, D., {et~al.}
  2006, \aap, 452, 987

\bibitem[{{de Groof} {et~al.}(2005){de Groof}, {Bastiaensen}, {M{\"u}ller},
  {Berghmans}, \& {Poedts}}]{deGroof2005}
{de Groof}, A., {Bastiaensen}, C., {M{\"u}ller}, D.~A.~N., {Berghmans}, D., \&
  {Poedts}, S. 2005, \aap, 443, 319

\bibitem[{{Fekel} \& {Henry}(2000)}]{FH00}
{Fekel}, F.~C. \& {Henry}, G.~W. 2000, \aj, 120, 3265

\bibitem[{{Flores Soriano} \& {Strassmeier}(2017)}]{Soriano}
{Flores Soriano}, M. \& {Strassmeier}, K.~G. 2017, \aap, 597, A101

\bibitem[{{Fuhrmeister} {et~al.}(2011){Fuhrmeister}, {Lalitha}, {Poppenhaeger},
  {Rudolf}, {Liefke}, {Reiners}, {Schmitt}, \& {Ness}}]{proxcen}
{Fuhrmeister}, B., {Lalitha}, S., {Poppenhaeger}, K., {et~al.} 2011, \aap, 534,
  A133

\bibitem[{{Fuhrmeister} {et~al.}(2008){Fuhrmeister}, {Liefke}, {Schmitt}, \&
  {Reiners}}]{CNLeoflare}
{Fuhrmeister}, B., {Liefke}, C., {Schmitt}, J.~H.~M.~M., \& {Reiners}, A. 2008,
  \aap, 487, 293

\bibitem[{{Fuhrmeister} {et~al.}(2005){Fuhrmeister}, {Schmitt}, \&
  {Hauschildt}}]{LHS2034}
{Fuhrmeister}, B., {Schmitt}, J.~H.~M.~M., \& {Hauschildt}, P.~H. 2005, \aap,
  436, 677

\bibitem[{{Gaidos} {et~al.}(2014){Gaidos}, {Mann}, {L{\'e}pine}, {Buccino},
  {James}, {Ansdell}, {Petrucci}, {Mauas}, \& {Hilton}}]{Gai14}
{Gaidos}, E., {Mann}, A.~W., {L{\'e}pine}, S., {et~al.} 2014, \mnras, 443, 2561

\bibitem[{{Garcia-Piquer} {et~al.}(2017){Garcia-Piquer}, {Morales}, {Ribas},
  {Colom{\'e}}, {Gu{\`a}rdia}, {Perger}, {Caballero}, {Cort{\'e}s-Contreras},
  {Jeffers}, {Reiners}, {Amado}, {Quirrenbach}, \& {Seifert}}]{Garcia-Piquer}
{Garcia-Piquer}, A., {Morales}, J.~C., {Ribas}, I., {et~al.} 2017, \aap, 604,
  A87

\bibitem[{{Gizis} {et~al.}(2013){Gizis}, {Burgasser}, {Berger}, {Williams},
  {Vrba}, {Cruz}, \& {Metchev}}]{Gizis}
{Gizis}, J.~E., {Burgasser}, A.~J., {Berger}, E., {et~al.} 2013, \apj, 779, 172

\bibitem[{{Gizis} {et~al.}(2002){Gizis}, {Reid}, \& {Hawley}}]{Gizis2002}
{Gizis}, J.~E., {Reid}, I.~N., \& {Hawley}, S.~L. 2002, \aj, 123, 3356

\bibitem[{{Gomes da Silva} {et~al.}(2011){Gomes da Silva}, {Santos}, {Bonfils},
  {Delfosse}, {Forveille}, \& {Udry}}]{GomesdaSilva}
{Gomes da Silva}, J., {Santos}, N.~C., {Bonfils}, X., {et~al.} 2011, \aap, 534,
  A30

\bibitem[{{Graham} \& {Cauzzi}(2015)}]{Graham}
{Graham}, D.~R. \& {Cauzzi}, G. 2015, \apjl, 807, L22

\bibitem[{{Hartman} {et~al.}(2011){Hartman}, {Bakos}, {Noyes}, {Sip{\H o}cz},
  {Kov{\'a}cs}, {Mazeh}, {Shporer}, \& {P{\'a}l}}]{Har11}
{Hartman}, J.~D., {Bakos}, G.~{\'A}., {Noyes}, R.~W., {et~al.} 2011, \aj, 141,
  166

\bibitem[{{Hilton} {et~al.}(2010){Hilton}, {West}, {Hawley}, \&
  {Kowalski}}]{Hilton}
{Hilton}, E.~J., {West}, A.~A., {Hawley}, S.~L., \& {Kowalski}, A.~F. 2010,
  \aj, 140, 1402

\bibitem[{{Ichimoto} \& {Kurokawa}(1984)}]{Ichimoto}
{Ichimoto}, K. \& {Kurokawa}, H. 1984, \solphys, 93, 105

\bibitem[{{Irwin} {et~al.}(2011){Irwin}, {Berta}, {Burke}, {Charbonneau},
  {Nutzman}, {West}, \& {Falco}}]{Irw11}
{Irwin}, J., {Berta}, Z.~K., {Burke}, C.~J., {et~al.} 2011, \apj, 727, 56

\bibitem[{{Jeffers} {et~al.}(2018){Jeffers}, {Sch\"ofer}, {Lamert}, \&
  {Reiners}}]{Jeffers2017}
{Jeffers}, S.~V., {Sch\"ofer}, P., {Lamert}, A., \& {Reiners}, A. 2018, A\&A,
  accepted

\bibitem[{{Johns-Krull} {et~al.}(1997){Johns-Krull}, {Hawley}, {Basri}, \&
  {Valenti}}]{Johns-Krull}
{Johns-Krull}, C.~M., {Hawley}, S.~L., {Basri}, G., \& {Valenti}, J.~A. 1997,
  \apjs, 112, 221

\bibitem[{{Judge} {et~al.}(2014){Judge}, {Kleint}, {Donea}, {Sainz Dalda}, \&
  {Fletcher}}]{Judge2014}
{Judge}, P.~G., {Kleint}, L., {Donea}, A., {Sainz Dalda}, A., \& {Fletcher}, L.
  2014, \apj, 796, 85

\bibitem[{{Kiraga}(2012)}]{Kira12}
{Kiraga}, M. 2012, \actaa, 62, 67

\bibitem[{{Klocov{\'a}} {et~al.}(2017){Klocov{\'a}}, {Czesla}, {Khalafinejad},
  {Wolter}, \& {Schmitt}}]{Klocova2017}
{Klocov{\'a}}, T., {Czesla}, S., {Khalafinejad}, S., {Wolter}, U., \&
  {Schmitt}, J.~H.~M.~M. 2017, \aap, 607, A66

\bibitem[{{Korhonen} {et~al.}(2010){Korhonen}, {Vida}, {Husarik}, {Mahajan},
  {Szczygie{\l}}, \& {Ol{\'a}h}}]{Kor10}
{Korhonen}, H., {Vida}, K., {Husarik}, M., {et~al.} 2010, Astronomische
  Nachrichten, 331, 772

\bibitem[{{Kowalski} {et~al.}(2017{\natexlab{a}}){Kowalski}, {Allred}, {Daw},
  {Cauzzi}, \& {Carlsson}}]{Kowalski1}
{Kowalski}, A.~F., {Allred}, J.~C., {Daw}, A., {Cauzzi}, G., \& {Carlsson}, M.
  2017{\natexlab{a}}, \apj, 836, 12

\bibitem[{{Kowalski} {et~al.}(2017{\natexlab{b}}){Kowalski}, {Allred},
  {Uitenbroek}, {Tremblay}, {Brown}, {Carlsson}, {Osten}, {Wisniewski}, \&
  {Hawley}}]{Kowalski}
{Kowalski}, A.~F., {Allred}, J.~C., {Uitenbroek}, H., {et~al.}
  2017{\natexlab{b}}, \apj, 837, 125

\bibitem[{{Lacatus} {et~al.}(2017){Lacatus}, {Judge}, \& {Donea}}]{Lacatus}
{Lacatus}, D.~A., {Judge}, P.~G., \& {Donea}, A. 2017, \apj, 842, 15

\bibitem[{{Lee} {et~al.}(2010){Lee}, {Berger}, \& {Knapp}}]{Lee}
{Lee}, K.-G., {Berger}, E., \& {Knapp}, G.~R. 2010, \apj, 708, 1482

\bibitem[{{L{\'e}pine} {et~al.}(2013){L{\'e}pine}, {Hilton}, {Mann}, {Wilde},
  {Rojas-Ayala}, {Cruz}, \& {Gaidos}}]{Lepine}
{L{\'e}pine}, S., {Hilton}, E.~J., {Mann}, A.~W., {et~al.} 2013, \aj, 145, 102

\bibitem[{{Malo} {et~al.}(2014){Malo}, {Artigau}, {Doyon}, {Lafreni{\`e}re},
  {Albert}, \& {Gagn{\'e}}}]{Malo}
{Malo}, L., {Artigau}, {\'E}., {Doyon}, R., {et~al.} 2014, \apj, 788, 81

\bibitem[{{Mamajek} {et~al.}(2013){Mamajek}, {Bartlett}, {Seifahrt}, {Henry},
  {Dieterich}, {Lurie}, {Kenworthy}, {Jao}, {Riedel}, {Subasavage}, {Winters},
  {Finch}, {Ianna}, \& {Bean}}]{Mamajek}
{Mamajek}, E.~E., {Bartlett}, J.~L., {Seifahrt}, A., {et~al.} 2013, \aj, 146,
  154

\bibitem[{{Matthews} {et~al.}(2015){Matthews}, {Harra}, {Zharkov}, \&
  {Green}}]{Matthews}
{Matthews}, S.~A., {Harra}, L.~K., {Zharkov}, S., \& {Green}, L.~M. 2015, \apj,
  812, 35

\bibitem[{{Montes} {et~al.}(1998){Montes}, {Fern\'andez-Figueroa}, {de Castro},
  {Cornide}, {Poncet}, \& {Sanz-Forcada}}]{Montes}
{Montes}, D., {Fern\'andez-Figueroa}, M.~J., {de Castro}, E., {et~al.} 1998, in
  Astronomical Society of the Pacific Conference Series, Vol. 154, Cool Stars,
  Stellar Systems, and the Sun, ed. R.~A. {Donahue} \& J.~A. {Bookbinder}, 1516

\bibitem[{{Montes} {et~al.}(2001){Montes}, {L{\'o}pez-Santiago}, {G{\'a}lvez},
  {Fern{\'a}ndez-Figueroa}, {De Castro}, \& {Cornide}}]{Mon01}
{Montes}, D., {L{\'o}pez-Santiago}, J., {G{\'a}lvez}, M.~C., {et~al.} 2001,
  \mnras, 328, 45

\bibitem[{{Newton} {et~al.}(2017){Newton}, {Irwin}, {Charbonneau}, {Berlind},
  {Calkins}, \& {Mink}}]{Newton}
{Newton}, E.~R., {Irwin}, J., {Charbonneau}, D., {et~al.} 2017, \apj, 834, 85

\bibitem[{{Newton} {et~al.}(2016){Newton}, {Irwin}, {Charbonneau},
  {Berta-Thompson}, {Dittmann}, \& {West}}]{New16a}
{Newton}, E.~R., {Irwin}, J., {Charbonneau}, D., {et~al.} 2016, \apj, 821, 93

\bibitem[{{Norton} {et~al.}(2007){Norton}, {Wheatley}, {West}, {Haswell},
  {Street}, {Collier Cameron}, {Christian}, {Clarkson}, {Enoch}, {Gallaway},
  {Hellier}, {Horne}, {Irwin}, {Kane}, {Lister}, {Nicholas}, {Parley},
  {Pollacco}, {Ryans}, {Skillen}, \& {Wilson}}]{Nor07}
{Norton}, A.~J., {Wheatley}, P.~J., {West}, R.~G., {et~al.} 2007, \aap, 467,
  785

\bibitem[{{Paulson} {et~al.}(2006){Paulson}, {Allred}, {Anderson}, {Hawley},
  {Cochran}, \& {Yelda}}]{Paulson}
{Paulson}, D.~B., {Allred}, J.~C., {Anderson}, R.~B., {et~al.} 2006, \pasp,
  118, 227

\bibitem[{{Pettersen} {et~al.}(1992){Pettersen}, {Olah}, \& {Sandmann}}]{Tes04}
{Pettersen}, B.~R., {Olah}, K., \& {Sandmann}, W.~H. 1992, \aaps, 96, 497

\bibitem[{{Quirrenbach} {et~al.}(2016){Quirrenbach}, {Amado}, {Caballero},
  {Mundt}, {Reiners}, {Ribas}, {Seifert}, {Abril}, {Aceituno},
  {Alonso-Floriano}, {Anwand-Heerwart}, {Azzaro}, {Bauer}, {Barrado},
  {Becerril}, {Bejar}, {Benitez}, {Berdinas}, {Brinkm{\"o}ller}, {Cardenas},
  {Casal}, {Claret}, {Colom{\'e}}, {Cortes-Contreras}, {Czesla}, {Doellinger},
  {Dreizler}, {Feiz}, {Fernandez}, {Ferro}, {Fuhrmeister}, {Galadi},
  {Gallardo}, {G{\'a}lvez-Ortiz}, {Garcia-Piquer}, {Garrido}, {Gesa},
  {G{\'o}mez Galera}, {Gonz{\'a}lez Hern{\'a}ndez}, {Gonzalez Peinado},
  {Gr{\"o}zinger}, {Gu{\`a}rdia}, {Guenther}, {de Guindos}, {Hagen}, {Hatzes},
  {Hauschildt}, {Helmling}, {Henning}, {Hermann}, {Hern{\'a}ndez Arabi},
  {Hern{\'a}ndez Casta{\~n}o}, {Hern{\'a}ndez Hernando}, {Herrero}, {Huber},
  {Huber}, {Huke}, {Jeffers}, {de Juan}, {Kaminski}, {Kehr}, {Kim}, {Klein},
  {Kl{\"u}ter}, {K{\"u}rster}, {Lafarga}, {Lara}, {Lamert}, {Laun},
  {Launhardt}, {Lemke}, {Lenzen}, {Llamas}, {Lopez del Fresno},
  {L{\'o}pez-Puertas}, {L{\'o}pez-Santiago}, {Lopez Salas}, {Magan
  Madinabeitia}, {Mall}, {Mandel}, {Mancini}, {Marin Molina}, {Maroto
  Fern{\'a}ndez}, {Mart{\'{\i}}n}, {Mart{\'{\i}}n-Ruiz}, {Marvin}, {Mathar},
  {Mirabet}, {Montes}, {Morales}, {Morales Mu{\~n}oz}, {Nagel}, {Naranjo},
  {Nowak}, {Palle}, {Panduro}, {Passegger}, {Pavlov}, {Pedraz}, {Perez},
  {P{\'e}rez-Medialdea}, {Perger}, {Pluto}, {Ram{\'o}n}, {Rebolo}, {Redondo},
  {Reffert}, {Reinhart}, {Rhode}, {Rix}, {Rodler}, {Rodr{\'{\i}}guez},
  {Rodr{\'{\i}}guez L{\'o}pez}, {Rohloff}, {Rosich}, {Sanchez Carrasco},
  {Sanz-Forcada}, {Sarkis}, {Sarmiento}, {Sch{\"a}fer}, {Schiller}, {Schmidt},
  {Schmitt}, {Sch{\"o}fer}, {Schweitzer}, {Shulyak}, {Solano}, {Stahl},
  {Storz}, {Tabernero}, {Tala}, {Tal-Or}, {Ulbrich}, {Veredas}, {Vico Linares},
  {Vilardell}, {Wagner}, {Winkler}, {Zapatero Osorio}, {Zechmeister},
  {Ammler-von Eiff}, {Anglada-Escud{\'e}}, {del Burgo}, {Garcia-Vargas},
  {Klutsch}, {Lizon}, {Lopez-Morales}, {Ofir}, {P{\'e}rez-Calpena}, {Perryman},
  {S{\'a}nchez-Blanco}, {Strachan}, {St{\"u}rmer}, {Su{\'a}rez}, {Trifonov},
  {Tulloch}, \& {Xu}}]{CARMENES1}
{Quirrenbach}, A., {Amado}, P.~J., {Caballero}, J.~A., {et~al.} 2016, in
  \procspie, Vol. 9908, Ground-based and Airborne Instrumentation for Astronomy
  VI, 990812

\bibitem[{{Reep} {et~al.}(2016){Reep}, {Warren}, {Crump}, \&
  {Sim{\~o}es}}]{Reep}
{Reep}, J.~W., {Warren}, H.~P., {Crump}, N.~A., \& {Sim{\~o}es}, P.~J.~A. 2016,
  \apj, 827, 145

\bibitem[{{Reid} {et~al.}(1995){Reid}, {Hawley}, \& {Gizis}}]{PMSU}
{Reid}, I.~N., {Hawley}, S.~L., \& {Gizis}, J.~E. 1995, \aj, 110, 1838

\bibitem[{{Reiners} {et~al.}(2014){Reiners}, {Sch{\"u}ssler}, \&
  {Passegger}}]{Reiners2014}
{Reiners}, A., {Sch{\"u}ssler}, M., \& {Passegger}, V.~M. 2014, \apj, 794, 144

\bibitem[{{Reiners} {et~al.}(2017){Reiners}, {Zechmeister}, {Caballero},
  {Ribas}, {Morales}, {Jeffers}, {Sch{\"o}fer}, {Tal-Or}, {Quirrenbach},
  {Amado}, {Kaminski}, {Seifert}, {Abril}, {Aceituno}, {Alonso-Floriano},
  {Ammler-von Eiff}, {Antona}, {Anglada-Escud{\'e}}, {Anwand-Heerwart},
  {Arroyo-Torres}, {Azzaro}, {Baroch}, {Barrado}, {Bauer}, {Becerril},
  {B{\'e}jar}, {Ben{\'{\i}}tez}, {Berdi{\~n}as}, {Bergond}, {Bl{\"u}mcke},
  {Brinkm{\"o}ller}, {del Burgo}, {Cano}, {C{\'a}rdenas V{\'a}zquez}, {Casal},
  {Cifuentes}, {Claret}, {Colom{\'e}}, {Cort{\'e}s-Contreras}, {Czesla},
  {D{\'{\i}}ez-Alonso}, {Dreizler}, {Feiz}, {Fern{\'a}ndez}, {Ferro},
  {Fuhrmeister}, {Galad{\'{\i}}-Enr{\'{\i}}quez}, {Garcia-Piquer},
  {Garc{\'{\i}}a Vargas}, {Gesa}, {G{\'o}mez}, {Galera}, {Gonz{\'a}lez
  Hern{\'a}ndez}, {Gonz{\'a}lez-Peinado}, {Gr{\"o}zinger}, {Grohnert},
  {Gu{\`a}rdia}, {Guenther}, {Guijarro}, {de Guindos}, {Guti{\'e}rrez-Soto},
  {Hagen}, {Hatzes}, {Hauschildt}, {Hedrosa}, {Helmling}, {Henning}, {Hermelo},
  {Hern{\'a}ndez Arab{\'{\i}}}, {Hern{\'a}ndez Casta{\~n}o}, {Hern{\'a}ndez
  Hernando}, {Herrero}, {Huber}, {Huke}, {Johnson}, {de Juan}, {Kim}, {Klein},
  {Kl{\"u}ter}, {Klutsch}, {K{\"u}rster}, {Lafarga}, {Lamert}, {Lamp{\'o}n},
  {Lara}, {Laun}, {Lemke}, {Lenzen}, {Launhardt}, {L{\'o}pez del Fresno},
  {L{\'o}pez-Gonz{\'a}lez}, {L{\'o}pez-Puertas}, {L{\'o}pez Salas},
  {L{\'o}pez-Santiago}, {Luque}, {Mag{\'a}n Madinabeitia}, {Mall}, {Mancini},
  {Mandel}, {Marfil}, {Mar{\'{\i}}n Molina}, {Maroto}, {Fern{\'a}ndez},
  {Mart{\'{\i}}n}, {Mart{\'{\i}}n-Ruiz}, {Marvin}, {Mathar}, {Mirabet},
  {Montes}, {Moreno-Raya}, {Moya}, {Mundt}, {Nagel}, {Naranjo}, {Nortmann},
  {Nowak}, {Ofir}, {Oreiro}, {Pall{\'e}}, {Panduro}, {Pascual}, {Passegger},
  {Pavlov}, {Pedraz}, {P{\'e}rez-Calpena}, {P{\'e}rez Medialdea}, {Perger},
  {Perryman}, {Pluto}, {Rabaza}, {Ram{\'o}n}, {Rebolo}, {Redondo}, {Reffert},
  {Reinhart}, {Rhode}, {Rix}, {Rodler}, {Rodr{\'{\i}}guez},
  {Rodr{\'{\i}}guez-L{\'o}pez}, {Rodr{\'{\i}}guez Trinidad}, {Rohloff},
  {Rosich}, {Sadegi}, {S{\'a}nchez-Blanco}, {S{\'a}nchez Carrasco},
  {S{\'a}nchez-L{\'o}pez}, {Sanz-Forcada}, {Sarkis}, {Sarmiento},
  {Sch{\"a}fer}, {Schmitt}, {Schiller}, {Schweitzer}, {Solano}, {Stahl},
  {Strachan}, {St{\"u}rmer}, {Su{\'a}rez}, {Tabernero}, {Tala}, {Trifonov},
  {Tulloch}, {Ulbrich}, {Veredas}, {Vico Linares}, {Vilardell}, {Wagner},
  {Winkler}, {Wolthoff}, {Xu}, {Yan}, \& {Zapatero Osorio}}]{Reiners2017}
{Reiners}, A., {Zechmeister}, M., {Caballero}, J.~A., {et~al.} 2017, ArXiv
  e-print [\eprint[arXiv]{1711.06576}]

\bibitem[{{Riaz} {et~al.}(2006){Riaz}, {Gizis}, \& {Harvin}}]{Ria06}
{Riaz}, B., {Gizis}, J.~E., \& {Harvin}, J. 2006, \aj, 132, 866

\bibitem[{{Robertson} {et~al.}(2016){Robertson}, {Bender}, {Mahadevan}, {Roy},
  \& {Ramsey}}]{Robertson}
{Robertson}, P., {Bender}, C., {Mahadevan}, S., {Roy}, A., \& {Ramsey}, L.~W.
  2016, \apj, 832, 112

\bibitem[{{Robrade} \& {Schmitt}(2005)}]{Robrade2005}
{Robrade}, J. \& {Schmitt}, J.~H.~M.~M. 2005, \aap, 435, 1073

\bibitem[{{Rousselot} {et~al.}(2000){Rousselot}, {Lidman}, {Cuby}, {Moreels},
  \& {Monnet}}]{Rousselot}
{Rousselot}, P., {Lidman}, C., {Cuby}, J.-G., {Moreels}, G., \& {Monnet}, G.
  2000, \aap, 354, 1134

\bibitem[{{Rubio da Costa} \& {Kleint}(2017)}]{rubiodacosta1}
{Rubio da Costa}, F. \& {Kleint}, L. 2017, \apj, 842, 82

\bibitem[{{Rubio da Costa} {et~al.}(2016){Rubio da Costa}, {Kleint},
  {Petrosian}, {Liu}, \& {Allred}}]{Rubiodacosta}
{Rubio da Costa}, F., {Kleint}, L., {Petrosian}, V., {Liu}, W., \& {Allred},
  J.~C. 2016, \apj, 827, 38

\bibitem[{{Schmidt} {et~al.}(2007){Schmidt}, {Cruz}, {Bongiorno}, {Liebert}, \&
  {Reid}}]{Schmidt-M7flare}
{Schmidt}, S.~J., {Cruz}, K.~L., {Bongiorno}, B.~J., {Liebert}, J., \& {Reid},
  I.~N. 2007, \aj, 133, 2258

\bibitem[{{Schmieder} {et~al.}(1987){Schmieder}, {Forbes}, {Malherbe}, \&
  {Machado}}]{Schmieder1987}
{Schmieder}, B., {Forbes}, T.~G., {Malherbe}, J.~M., \& {Machado}, M.~E. 1987,
  \apj, 317, 956

\bibitem[{{Scholz} {et~al.}(2005){Scholz}, {Meusinger}, \&
  {Jahrei{\ss}}}]{Sch05}
{Scholz}, R.-D., {Meusinger}, H., \& {Jahrei{\ss}}, H. 2005, \aap, 442, 211

\bibitem[{{Schrijver}(2001)}]{Schrijver2001}
{Schrijver}, C.~J. 2001, \solphys, 198, 325

\bibitem[{{Shkolnik} {et~al.}(2010){Shkolnik}, {Hebb}, {Liu}, {Reid}, \&
  {Collier Cameron}}]{Shk10}
{Shkolnik}, E.~L., {Hebb}, L., {Liu}, M.~C., {Reid}, I.~N., \& {Collier
  Cameron}, A. 2010, \apj, 716, 1522

\bibitem[{{Short} \& {Doyle}(1998)}]{Short1}
{Short}, C.~I. \& {Doyle}, J.~G. 1998, \aap, 336, 613

\bibitem[{{Su{\'a}rez Mascare{\~n}o} {et~al.}(2016){Su{\'a}rez Mascare{\~n}o},
  {Rebolo}, \& {Gonz{\'a}lez Hern{\'a}ndez}}]{SM16}
{Su{\'a}rez Mascare{\~n}o}, A., {Rebolo}, R., \& {Gonz{\'a}lez Hern{\'a}ndez},
  J.~I. 2016, \aap, 595, A12

\bibitem[{{Tang}(1983)}]{Tang1983}
{Tang}, F. 1983, \solphys, 83, 15

\bibitem[{{{\v S}vestka}(1972)}]{Svestka}
{{\v S}vestka}, Z. 1972, \araa, 10, 1

\bibitem[{{{\v S}vestka} {et~al.}(1962){{\v S}vestka}, {Kopeck{\'y}}, \&
  {Blaha}}]{Svestka1962}
{{\v S}vestka}, Z., {Kopeck{\'y}}, M., \& {Blaha}, M. 1962, Bulletin of the
  Astronomical Institutes of Czechoslovakia, 13, 37

\bibitem[{{Verwichte} {et~al.}(2017){Verwichte}, {Antolin}, {Rowlands},
  {Kohutova}, \& {Neukirch}}]{Verwichte2017b}
{Verwichte}, E., {Antolin}, P., {Rowlands}, G., {Kohutova}, P., \& {Neukirch},
  T. 2017, \aap, 598, A57

\bibitem[{{Walkowicz} \& {Hawley}(2009)}]{Walkowicz}
{Walkowicz}, L.~M. \& {Hawley}, S.~L. 2009, \aj, 137, 3297

\bibitem[{{West} {et~al.}(2015){West}, {Weisenburger}, {Irwin},
  {Berta-Thompson}, {Charbonneau}, {Dittmann}, \& {Pineda}}]{West2015}
{West}, A.~A., {Weisenburger}, K.~L., {Irwin}, J., {et~al.} 2015, \apj, 812, 3

\bibitem[{{Worden} {et~al.}(1984){Worden}, {Schneeberger}, {Giampapa},
  {Deluca}, \& {Cram}}]{Worden}
{Worden}, S.~P., {Schneeberger}, T.~J., {Giampapa}, M.~S., {Deluca}, E.~E., \&
  {Cram}, L.~E. 1984, \apj, 276, 270

\bibitem[{{Zarro} {et~al.}(1988){Zarro}, {Canfield}, {Metcalf}, \&
  {Strong}}]{Zarro1988}
{Zarro}, D.~M., {Canfield}, R.~C., {Metcalf}, T.~R., \& {Strong}, K.~T. 1988,
  \apj, 324, 582

\bibitem[{{Zechmeister} {et~al.}(2017){Zechmeister}, {Reiners}, {Amado},
  {Azzaro}, {Bauer}, {B{\'e}jar}, {Caballero}, {Guenther}, {Hagen}, {Jeffers},
  {Kaminski}, {K{\"u}rster}, {Launhardt}, {Montes}, {Morales}, {Quirrenbach},
  {Reffert}, {Ribas}, {Seifert}, {Tal-Or}, \& {Wolthoff}}]{Zechmeister2017}
{Zechmeister}, M., {Reiners}, A., {Amado}, P.~J., {et~al.} 2017, \aap, 609, A12

\end{thebibliography}

\newpage
\onecolumn
\footnotesize

\tablecaption{Detected asymmetries. }\label{asyms}
\tablehead{\hline \hline no. & Name & Epoch & $t_{\rm exp}$  &Gauss 1& & & &Gauss 2& & & &A$_{\rm broad}$/ &Remarks$^{c}$\\
&     & Julian date & & Area   &     $v_{\rm central}$ &$\sigma$& FWHM & Area  &    $v_{\rm central}$&  $\sigma$&FWHM&A$_{\rm total}$& \\
&     & -2\,400\,000[d] & [s]     & [\AA]  &  [km\,s$^{-1}$]& [\AA]  & [km\,s$^{-1}$]& [\AA] & [km\,s$^{-1}$]& [\AA]&[km\,s$^{-1}$] & &  \\ 
\noalign{\smallskip}
\hline
\noalign{\smallskip}
}
\begin{supertabular}{lllc|cccc|cccc|cl}
1&V388 Cas    &  57749.4020& 1589 &4.13    & 0.9& 0.54 & 58.1 &  1.10   &33.8& 2.06  &221.5 &0.21 &  IR\\ 
2&YZ Cet      &  57622.5718& 444   & 2.71&0.9& 0.46 & 49.5 &  0.24   & 99.5& 0.74  &79.6 &0.08 &  tech\\
3&Barta 161 12&  57622.6627 &1700 &0.86 &29.2  &0.74 & 79.6&0.16 &-37.4&0.5& 53.8 &0.16&tech\\
4&            &  57642.6235& 1800 &1.72 &25.1  &0.54 & 58.1 &0.74 &-46.1&2.20& 236.5& 0.30&IR\\
5&            &  57688.5189& 1802 &1.87 &16.4  &0.73 &78.5 & 3.47 &98.2&2.97& 319.3 &0.65&IR\\
6&            &  57703.4290& 1719 &2.41 &13.7  &0.80 &86.0 &0.35&-90.4&1.00& 107.5 &0.13&IR\\
7&            &  57763.3199& 1802 &0.42 &-25.6 &0.33 &35.5&1.26 &-1.4&1.61& 173.1&0.75&IR\\
8&TZ Ari      &  57766.2898& 504  &0.87    & 0.& 0.54 & 58.1 &  0.32    & -11.0& 1.51  &162.3 &0.42 & \\
9&GJ 166 C   &   57634.6243& 263  &0.97     & 2.7& 0.69 & 74.2 &  0.96   & 10.5& 2.11  &226.9 & 0.50 & tech\\
10&            &  57636.6648& 277  &1.14    & 0.9& 0.55 & 59.1 &  0.23   & -0.5& 1.54  &165.6 & 0.17 &  tech\\
11&            &  57705.5112& 222  &0.93    & -0.5& 0.58 & 62.3 &  0.69  & 9.6& 1.86  &200.0 & 0.43 & \\
12&            &  57712.5133& 551  &1.24    & -0.9& 0.58 & 62.3 &  0.75  &19.2& 1.93  &207.5 &0.38 &\\
13&            &  57799.3539& 492  &1.17    & -1.4& 0.49 & 52.7 &  0.52  &-10.0& 2.26  &243.0 & 0.31 &\\
14&V2689 Ori   &  57414.3922& 769  &0.28    & -1.4& 0.50 & 53.8 &  0.07  & 44.3& 1.61  &173.1 & 0.2 & tech\\
15&G 099-049   & 57694.6742& 362  & 0.63   & 0.& 0.54 & 58.1 &  0.31    &-8.2& 1.20  &129.0 & 0.32 & tech\\
16&            &  57699.6356& 702  &0.96    & 0.& 0.54 & 58.1 &  0.41    & 26.5& 1.09  &117.2 & 0.30 &\\
17&YZ CMi      &  57414.5293& 1000  &0.72    & 0.9& 0.58 & 62.3 &  0.84   &-11.0& 1.61  &173.1 & 0.54 &tech\\
18&            &  57415.4739& 1100 &1.11    & -0.5& 0.53 & 57.0 &  0.72  &27.4& 2.27  &244.1 & 0.39 &tech\\
19&            &  57421.3962& 650  &0.96    & -1.8& 0.62 & 66.6 &  0.65  & 3.7& 2.40  &258.0 & 0.40 &tech\\
20&            &  57441.3985& 1000  &0.99    & -6.3& 0.57 & 61.3 &  0.26  & 46.6& 0.80  &86.0 &0.21 & \\
21&            &  57510.3330& 800    &0.57    & -9.6& 0.80 & 86.0 &  0.92  &-33.8& 1.80  &193.5 & 0.62 &tech\\
22&            &  57673.6632& 239  &1.08    & -1.4& 0.56 & 60.2 &  0.55  & -31.5& 1.80  &193.5 & 0.34 &\\
23&            &  57692.7212& 161  &2.23    & 0.9& 0.60 & 64.5 &  0.20   &103.2& 0.84  &90.3 &0.08 &\\
24&            &  57704.5994& 267  &1.64    & 0.& 0.59 & 63.4 &  0.57    &39.7& 1.88  &202.1 & 0.26 &\\
25&            &  57760.4882& 262  &1.66    & -1.4& 0.55 & 59.1 &  1.08  &-36.5& 2.02  &217.2 & 0.39 &\\
26&            &  57762.6101& 234  &2.15    & -2.7& 0.64 & 68.8 &  0.62  &19.6& 2.10  &225.7 & 0.22 &tech\\
27&            &  57798.4688& 424  &2.27    & -0.9& 0.59 & 63.4 &  1.55  &0.5& 1.96  &210.7 & 0.41 &tech\\
28&GJ 362      &  57442.4760& 1000 &0.20    & 0.9& 0.47 & 50.5 &  0.12   &1.4& 1.55  &166.7 & 0.37 &tech\\
29&CN Leo      &  57694.7164& 511  &10.14    & 2.7& 0.60 & 64.5 &  0.69   &83.5& 0.51  &54.8 & 0.06 &tech\\
30&            &  57792.5648& 1602 &15.00    & 0.5& 0.55 & 59.1 &  2.58   &-1.8& 1.68  &180.6 & 0.15 &IR\\
31&WX UMa      &  57558.3800& 1800 &0.98    & 11.4& 0.56 & 60.2 &  0.98  &-8.2& 2.19  &235.4 & 0.50 &tech\\
32&            &  57763.6722& 1801 &2.43    & 5.0& 0.55 & 59.1 &  0.50   &-99.1& 1.26  &135.5 & 0.17 & \\
33&            &  57766.6724& 1802 &2.51    & 1.8& 0.80 & 86.0 &  0.43   &-55.7& 1.73  &186.0 & 0.14 & \\
34&            &  57787.5886& 1802 &2.02    & 5.0& 0.80 & 86.0 &  0.49   &72.1& 0.90  &96.8 &0.20 &tech\\
35&StKM 2-809  &  57449.6542& 1500 &1.21   & -0.9& 0.49 & 52.7 &  1.15  &-5.5& 2.23  &239.8 & 0.49 &tech\\
36&            &  57530.4266& 1501 &0.35    & -0.9& 0.35 & 37.6 &  0.94  &67.6& 1.94  &208.6 & 0.73 & NaD. tech\\
37&            &  57747.6647& 1802 &0.94    & 0.5& 0.73 & 78.5 &  1.25   &-45.7& 2.25  &241.9 & 0.57 &IR\\
38&GL Vir      &  57466.5384& 1801 &1.67    & -5.0& 0.57 & 61.2 &  0.90  &12.8& 2.42  &260.2 & 0.35 & \\
39&OT Ser      &  57555.4514& 1500 &0.15    & 0.5& 0.54 & 58.1 &  0.18   &13.2& 2.13  &229.0 & 0.54 &tech\\
40&            &  57608.3260& 600  &0.25    & -11.9& 0.80 & 86.0 &  0.07 &-100.4& 1.76  &189.2 & 0.22 &tech\\
41&            &  57752.7040& 373  &0.74    & 2.3& 0.76 & 81.7 &  1.13   &-8.7& 3.21  &345.1 & 0.60 & NaD, IR\\
42&            &  57755.7238& 526  &0.52    & 5.5& 0.64 & 68.8 &  0.13   &-40.6& 2.43  &261.3 & 0.20 &tech\\ 
43&vB 8        &  57584.4138& 1800 &8.51    & 2.3& 0.53 & 57.0 &  0.59   &86.7& 0.50  &53.8 & 0.06 & \\ 
44&            &  57623.3675& 1801 &0.85    & -26.5& 0.80 & 86.0 &  8.91 &-32.0& 1.78  &191.4 & 0.91& tech\\ 
45&TYC 3529$^{a}$&57692.3211&711 &0.17   & -3.7& 0.55 & 59.1 &  0.19  &-31.0& 1.62  &174.2 & 0.53 &\\ 
46&G 141-036   &  57631.4553& 1801 &5.61   & -2.3& 0.67 & 72.0 &  3.08  &11.0& 2.11  &226.9 & 0.35 &IR\\ 
47&V374 Peg    &  57554.6214& 1400 &0.80    &10.5 & 0.69& 74.2  & 0.31   &-34.7& 0.93 & 100.0& 0.28 &tech\\
48&         &  57558.6245& 1100 &0.69    &-6.8& 0.60 & 64.5 & 0.48    &-10.5& 1.27 & 136.5& 0.41 &tech\\      
49&         &  57588.5737& 1500 &1.01    &22.4& 0.50 & 53.8 & 0.14    &80.8& 0.92 & 98.9 & 0.12 & \\      
50&            &  57753.3254& 602  &0.96    &12.8& 0.67 & 72.0 & 1.26    &-5.5& 1.95 & 209.7& 0.57&tech\\
51&            &  57754.3237& 712  &3.92      &-16.9 & 0.64& 68.8  & 1.78  &3.65& 1.77 & 190.3& 0.31 & \\
52&         &  57762.3027& 789  &0.75    &-23.7& 0.63 & 67.7 & 0.59   &24.2& 1.28 & 137.6 & 0.44 &\\      
53&         &  57763.2727& 803  &1.29    &5.9& 0.55 & 59.1 & 0.10     &84.9& 0.50 & 53.8& 0.07 &\\      
54& EV Lac     &  57632.6289& 119  &3.08   &-0.9 & 0.66& 71.0  & 2.60   &-30.1 & 1.94 & 208.6& 0.45 &IR\\
55&            &  57633.4671& 111  &11.09   &0.9 & 0.80& 86.0  & 8.94    &35.2& 2.48 & 266.6 & 0.44 &NaD, tech\\
56&         &  57634.6314& 105  &0.88    & -0.5& 0.66 & 71.0 & 0.50   &-89.0& 1.45  & 155.9& 0.36 &tech\\      
57&         &  57646.4848& 153  &1.11    &1.8& 0.66 & 71.0 & 0.38     &56.1& 2.21 & 237.6 & 0.26 &\\      
58&         &  57646.4884& 463  &1.02    &2.3& 0.66 & 71.0 & 0.53     &-0.5& 3.11 & 334.4 & 0.34 &\\      
59&            &  57647.3730& 307  &1.44    &1.4& 0.61 & 65.6 & 1.55     &34.2& 3.05 & 327.9 & 0.52 &IR\\      
60&         &  57650.5369& 274  &5.26    & -0.9& 0.59 & 63.4 & 2.74   &77.2  & 1.66 & 178.5 &0.34 &tech\\      
61&         &  57684.3777& 302  &1.13    &-1.4& 0.58 & 62.4 & 0.80    &56.6& 3.40 & 365.1& 0.41 &IR\\      
62&         &  57704.4115& 159  &1.16    &-2.3& 0.55 & 59.1 & 0.14    &-44.7& 0.75 & 80.6 & 0.11 &IR\\      
63&         &  57752.3258& 154  &1.40    &-0.9& 0.52 & 55.9 & 0.43    &26.9& 1.84 & 197.8& 0.23 &\\      
64&         &  57754.2751& 139  &0.79    &-0.5& 0.59 & 63.4 & 0.53    &2.7& 1.83 & 196.8& 0.40 &tech\\      
65& GT Peg     &  57587.5889& 1500 &1.12   &-0.9& 0.54& 58.1  & 0.58    &77.2& 3.63 & 390.3& 0.34 &tech\\
66&            &  57695.3379& 584  &0.61   & -5.0& 0.52& 55.9  & 0.39   &-3.7& 2.94 & 316.1& 0.39 &IR\\
67&            &  57762.2768& 705  &2.88   & -3.2& 0.59& 63.4  & 1.33   &56.6& 1.69 & 181.7& 0.32 &IR\\
68& RX J235+38$^{b}$&57643.5635& 1800&2.09  & 0.& 0.55& 59.1  & 2.54     &-19.2& 2.20 & 236.5& 0.55 &tech\\
69&            &  57659.6019& 1201 &0.85   &2.3& 0.50& 55.9  & 0.33     &-25.1& 2.78 & 298.9& 0.28 &\\
70&         &  57673.4268& 1801 &0.72    & 3.7& 0.72 & 77.4 & 0.32    &21.9& 2.99 & 321.5& 0.31 &\\      
71&         &  57706.3457& 1801 &0.75    &-2.3& 0.64 & 68.8 & 0.54    &24.2& 2.38 & 255.9& 0.42 &\\      
72&            &  57738.3102& 1801 &1.62   &2.3& 0.54& 58.1  & 1.37     &12.8& 1.99 &214.0& 0.46 &tech\\
\noalign{\smallskip}
\hline

\end{supertabular}

Note. $^{a}$ Full designation is TYC 3529-1437-1. $^{b}$ Full designation is: RX J2354.8+3831.
$^{c}$ In column Remarks: NaD: spectrum also exhibited \ion{Na}{i} D wings. Tech: IR spectrum is not
available due to technical reasons. IR: spectrum also exhibited wings
in the IR Helium line at 10830 \AA.

\twocolumn
\appendix

\section{Detected asymmetries}\label{appendix}

\begin{figure}
\begin{center}
\includegraphics[width=0.5\textwidth, clip]{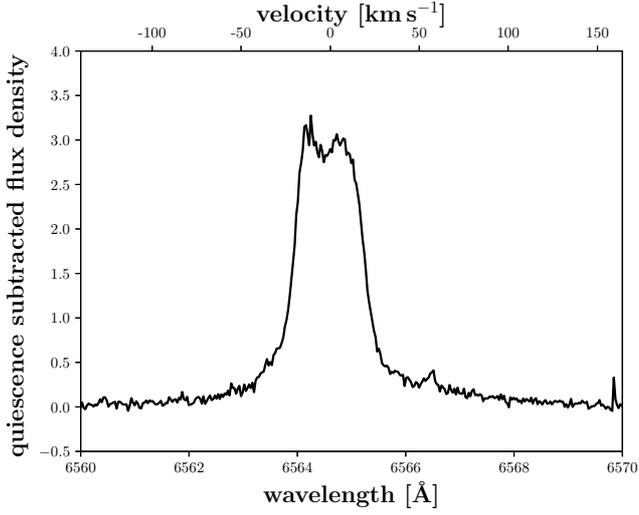}
\caption{\label{asymV388Cas} Spectrum with asymmetry for V388 Cas; the spectrum corresponds to entry no. 1 in Table \ref{asyms}.}
\end{center}
\end{figure}

\begin{figure}
\begin{center}
\includegraphics[width=0.5\textwidth, clip]{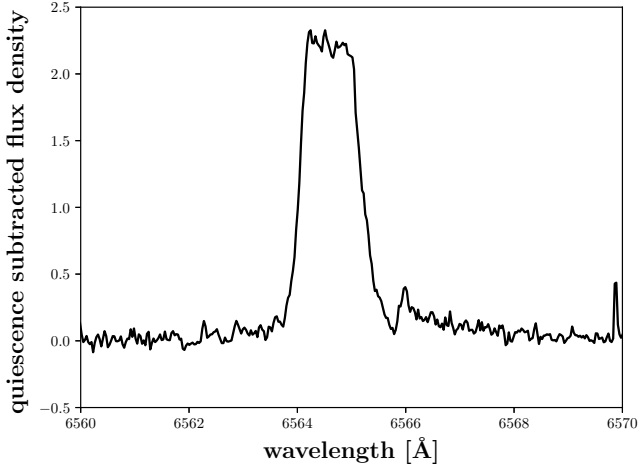}
\caption{\label{asymYZCet} Spectrum with asymmetry for YZ Cet; the spectrum corresponds to entry no. 2 in Table \ref{asyms}.}
\end{center}
\end{figure}

\begin{figure}
\begin{center}
\includegraphics[width=0.5\textwidth, clip]{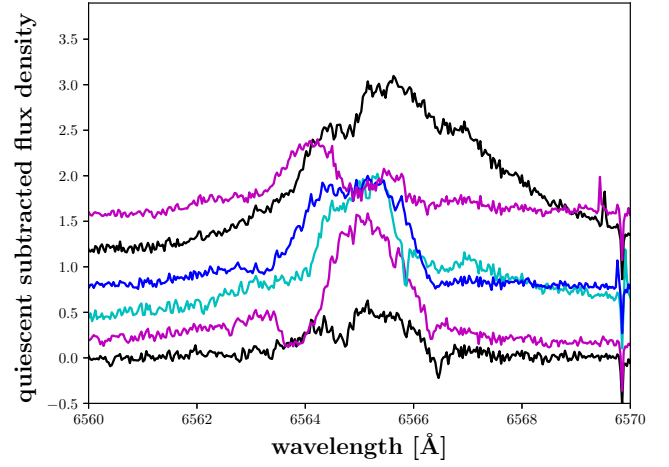}
\caption{\label{asymBarta} Spectra with asymmetries for Barta 161 12; from bottom to top the spectra correspond to entry nos. 3 -- 7 in Table \ref{asyms}.
The second spectrum from the top (black line) is shown in addition to the data 
in Table \ref{asyms} because it exhibits significant asymmetry but it
could not be fitted satisfactorily with two Gaussians.}
\end{center}
\end{figure}

\begin{figure}
\begin{center}
\includegraphics[width=0.5\textwidth, clip]{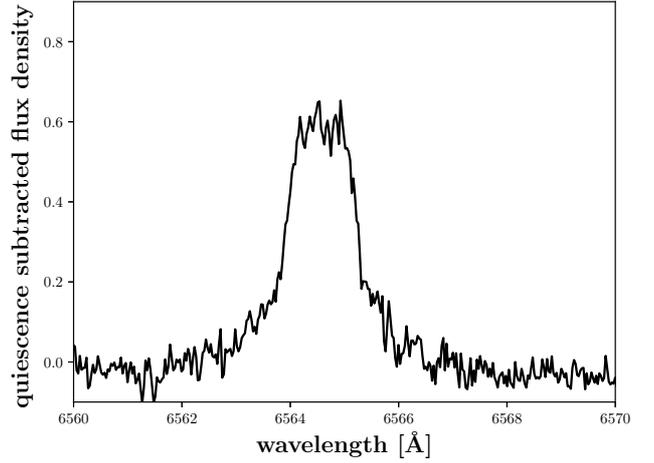}
\caption{\label{asymTZAri} Spectrum with asymmetry for TZ Ari; the spectrum corresponds to entry no. 8 in Table \ref{asyms}.}
\end{center}
\end{figure}

\begin{figure}
\begin{center}
\includegraphics[width=0.5\textwidth, clip]{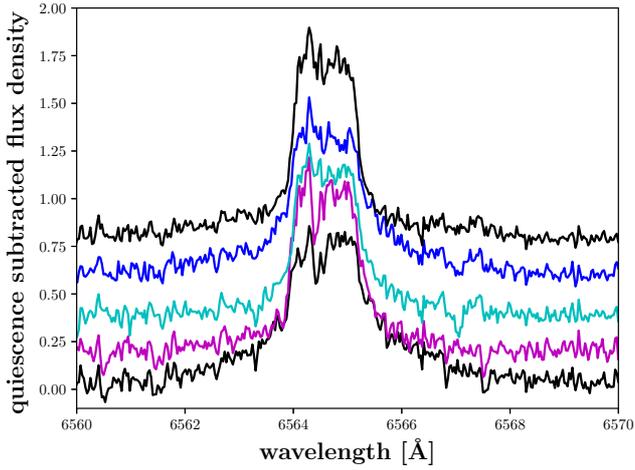}
\caption{\label{asymomi02EriC} Spectra with asymmetry for GJ 166 C; from bottom to top the spectra correspond to entry nos. 9 -- 13 in Table \ref{asyms}.}
\end{center}
\end{figure}

\begin{figure}
\begin{center}
\includegraphics[width=0.5\textwidth, clip]{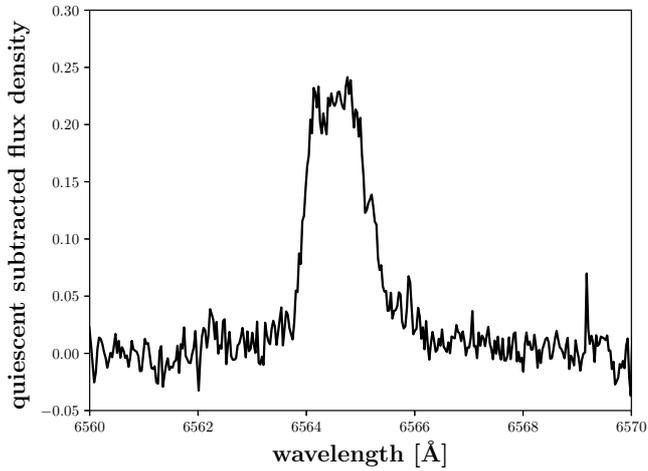}
\caption{\label{asymV2689Ori} Spectrum with asymmetry for V2689 Ori; the spectrum corresponds to entry no. 14 in Table \ref{asyms}.}
\end{center}
\end{figure}

\begin{figure}
\begin{center}
\includegraphics[width=0.5\textwidth, clip]{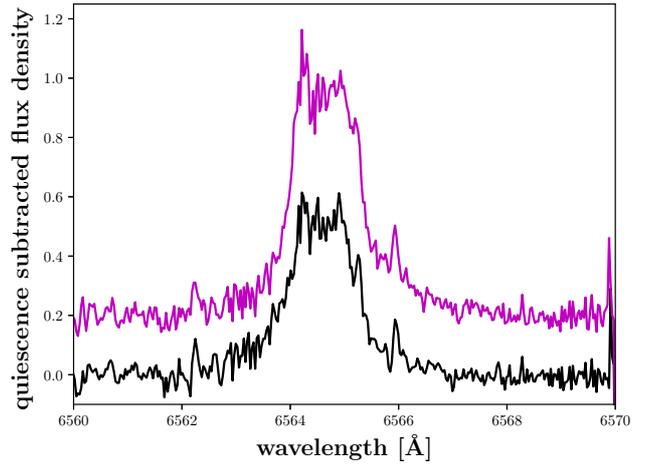}
\caption{\label{asymG099-049} Spectra with asymmetry for G099-049; the spectrum corresponds to entry no. 15 in Table \ref{asyms}.}
\end{center}
\end{figure}

\begin{figure}
\begin{center}
\includegraphics[width=0.5\textwidth, clip]{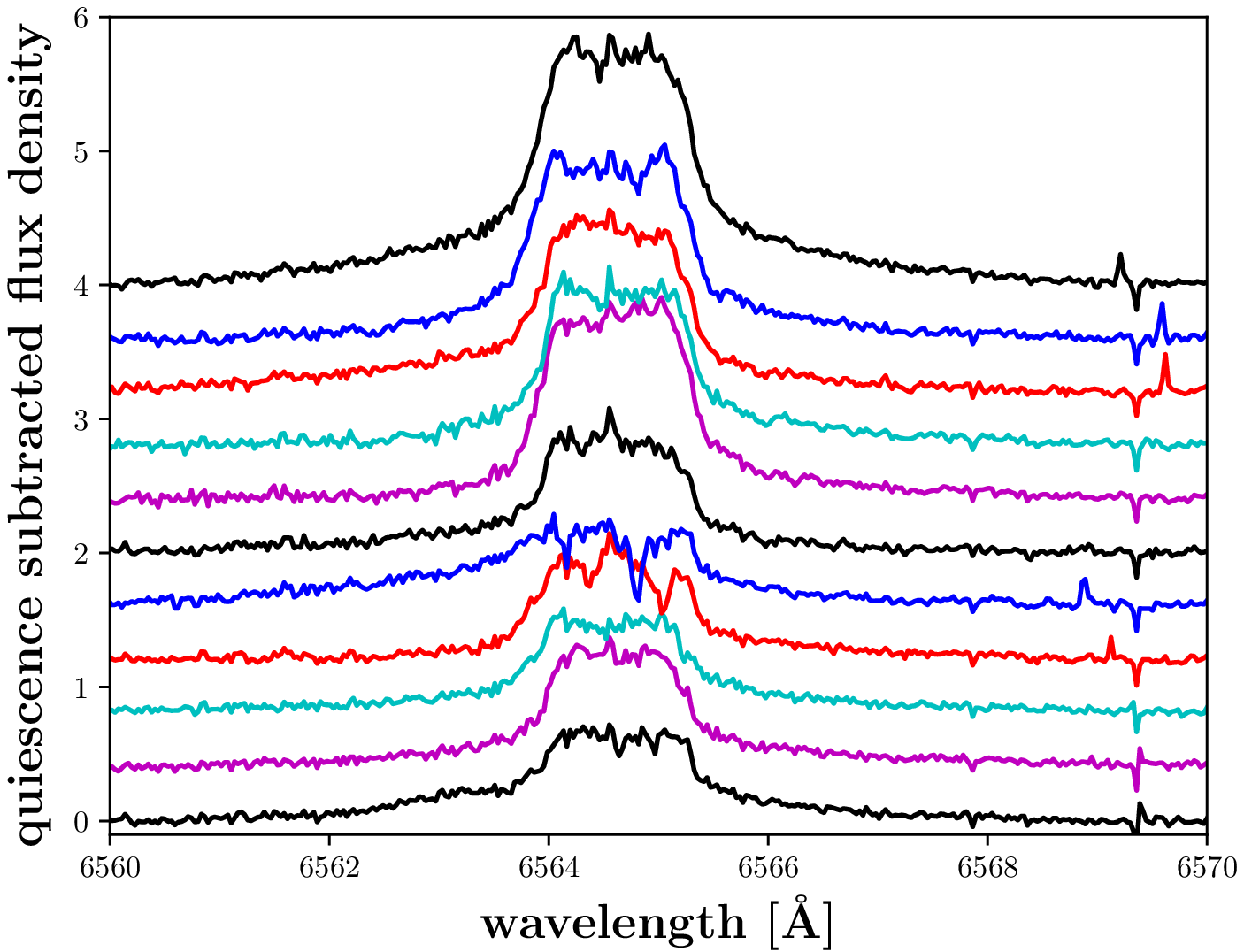}
\caption{\label{asymYZCMi} Spectra with asymmetry for YZ CMi; from bottom to top the spectra correspond to entry nos. 17 -- 27 in Table \ref{asyms}.}
\end{center}
\end{figure}

\begin{figure}
\begin{center}
\includegraphics[width=0.5\textwidth, clip]{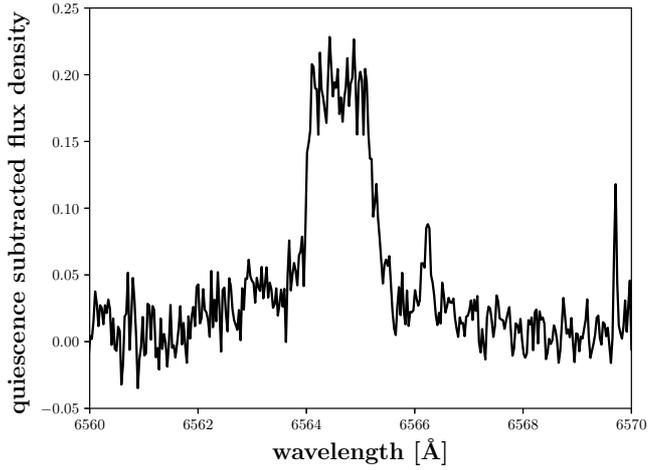}
\caption{\label{asymGJ362} Spectrum with asymmetry for GJ 362; the spectrum corresponds to entry no. 28 in Table \ref{asyms}.}
\end{center}
\end{figure}

\begin{figure}
\begin{center}
\includegraphics[width=0.5\textwidth, clip]{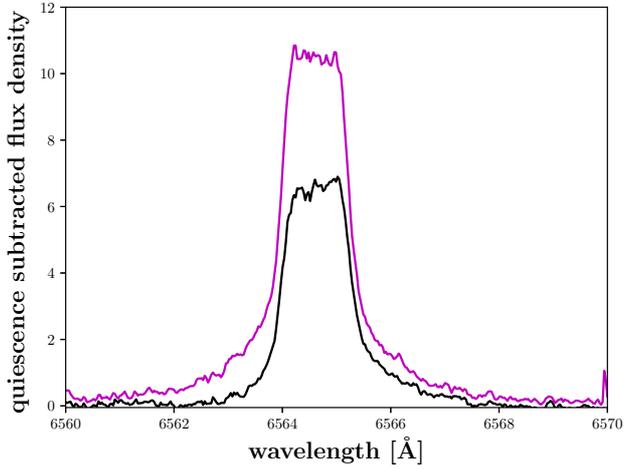}
\caption{\label{asymCNLeo} Spectra with asymmetry for CN Leo; from bottom to top the spectra correspond to entry nos. 29 -- 30 in Table \ref{asyms}.}
\end{center}
\end{figure}

\begin{figure}
\begin{center}
\includegraphics[width=0.5\textwidth, clip]{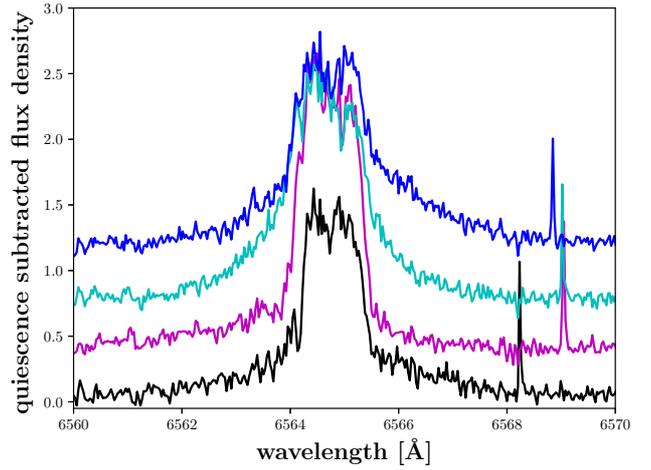}
\caption{\label{asymWXUMa} Spectra with asymmetry for WX UMa; from bottom to top the spectra correspond to entry nos. 31 -- 34 in Table \ref{asyms}.}
\end{center}
\end{figure}

\begin{figure}
\begin{center}
\includegraphics[width=0.5\textwidth, clip]{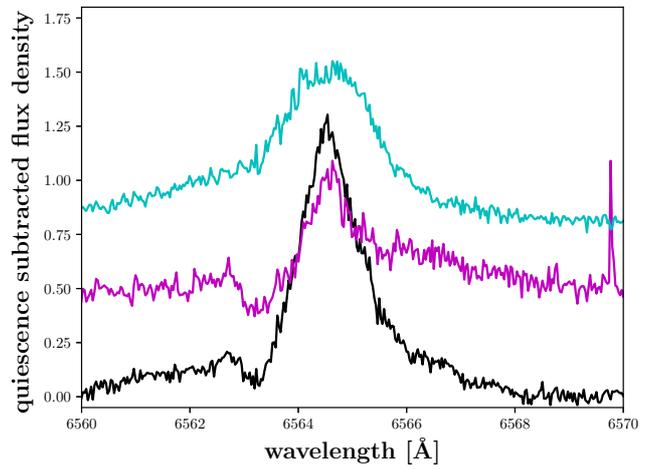}
\caption{\label{asymStKM2} Spectra with asymmetry for StKM 2; from bottom to top the spectra correspond to entry nos. 35 -- 37 in Table \ref{asyms}.}
\end{center}
\end{figure}

\begin{figure}
\begin{center}
\includegraphics[width=0.5\textwidth, clip]{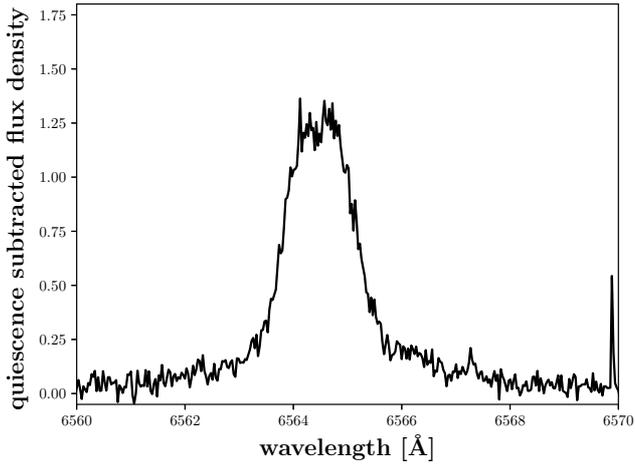}
\caption{\label{asymGLVir} Spectrum with asymmetry for GL Vir; the spectrum corresponds to entry no. 38 in Table \ref{asyms}.}
\end{center}
\end{figure}

\begin{figure}
\begin{center}
\includegraphics[width=0.5\textwidth, clip]{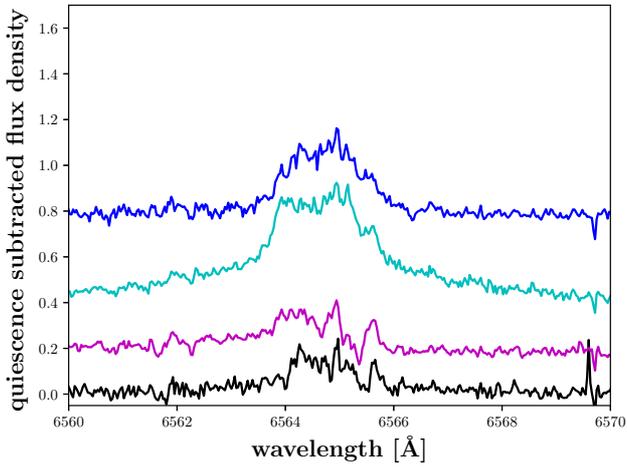}
\caption{\label{asymOTSer} Spectra with asymmetry for OT Ser; from bottom to top the spectra correspond to entry nos. 39 -- 42 in Table \ref{asyms}.}
\end{center}
\end{figure}

\begin{figure}
\begin{center}
\includegraphics[width=0.5\textwidth, clip]{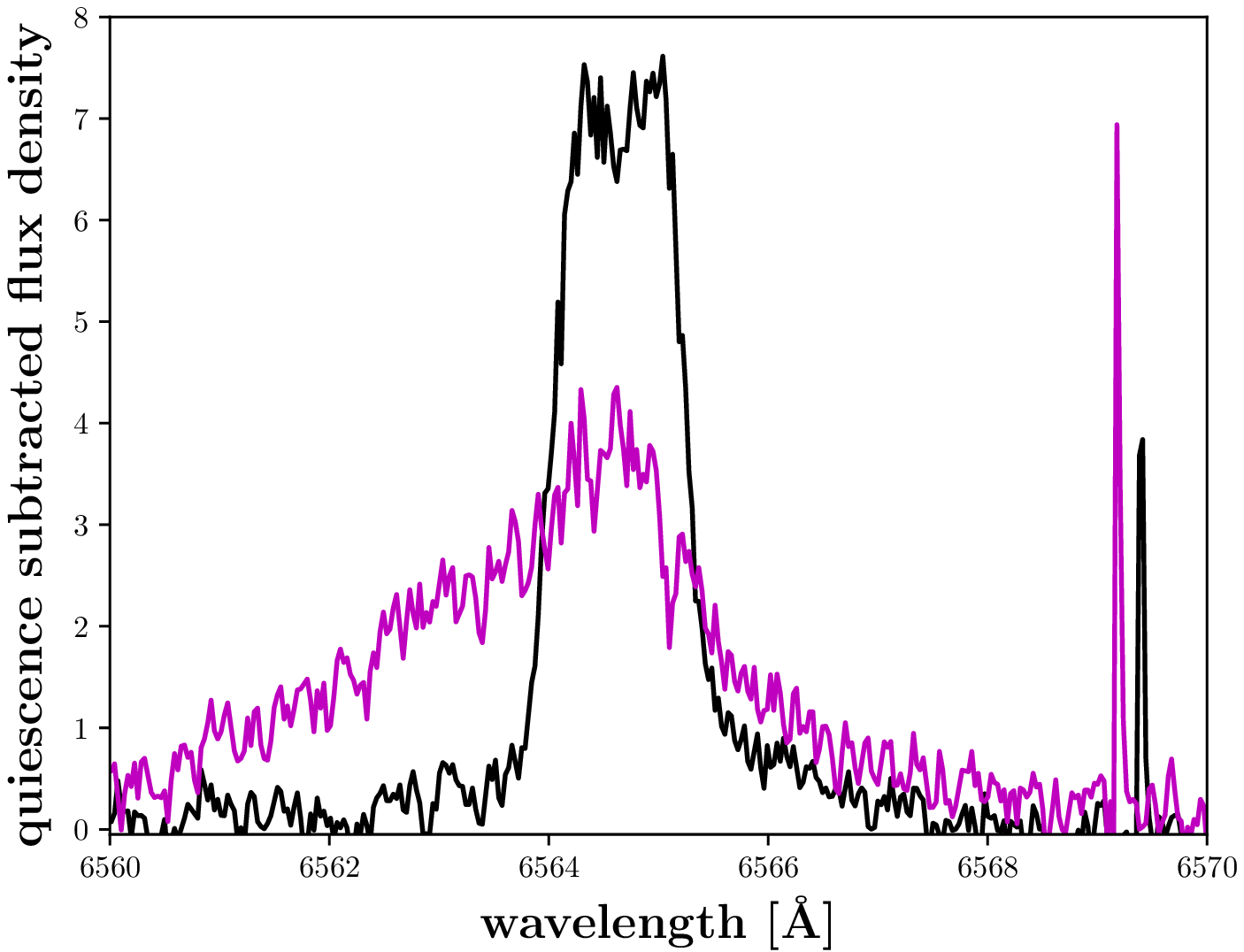}
\caption{\label{asymvB8} Spectra with asymmetry for vB 8; from bottom to top the spectra correspond to entry nos. 43 -- 44 in Table \ref{asyms}.}
\end{center}
\end{figure}

\begin{figure}
\begin{center}
\includegraphics[width=0.5\textwidth, clip]{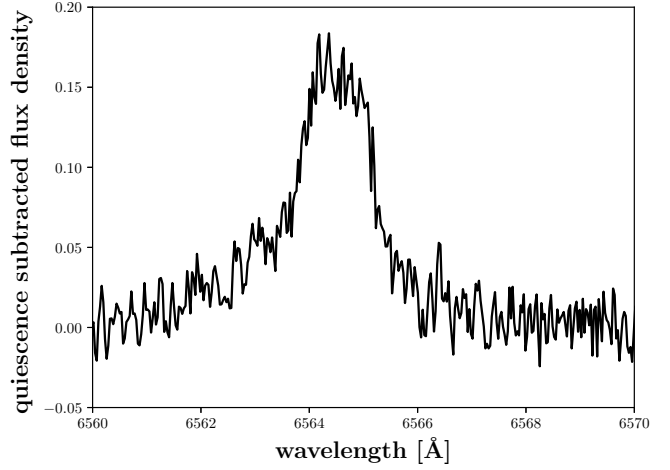}
\caption{\label{asymJ18174} Spectrum with asymmetry for J18174+483; the spectrum corresponds to entry no. 45 in Table \ref{asyms}.}
\end{center}
\end{figure}

\begin{figure}
\begin{center}
\includegraphics[width=0.5\textwidth, clip]{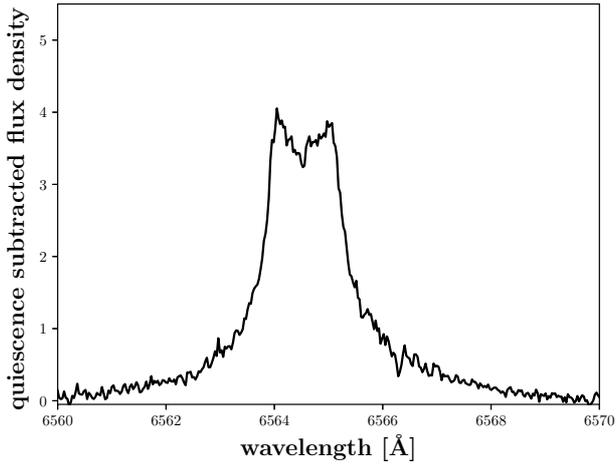}
\caption{\label{asymG141-036} Spectrum with asymmetry for G141-036; the spectrum corresponds to entry no. 46 in Table \ref{asyms}.}
\end{center}
\end{figure}

\begin{figure}
\begin{center}
\includegraphics[width=0.5\textwidth, clip]{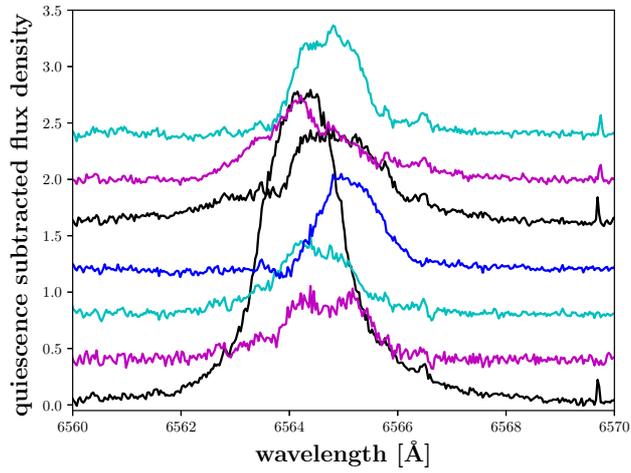}
\caption{\label{asymV374Peg} Spectra with asymmetry for V374 Peg; from bottom to top the spectra correspond to entry nos. 47 -- 53 in Table \ref{asyms}.}
\end{center}
\end{figure}

\begin{figure}
\begin{center}
\includegraphics[width=0.5\textwidth, clip]{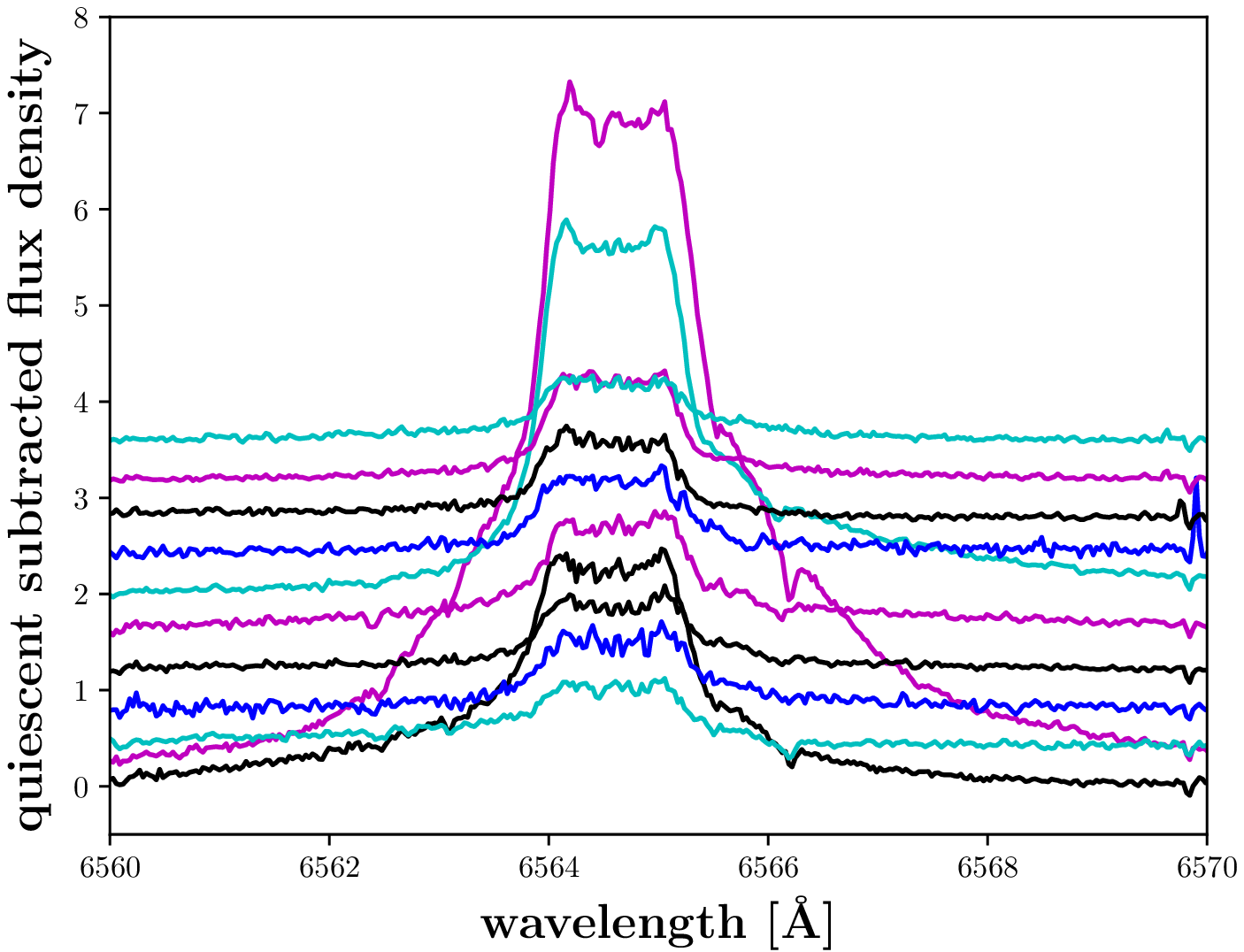}
\caption{\label{asymEVLac} Spectra with asymmetry for EV Lac; from bottom to top the spectra correspond to entry nos. 54 -- 64 in Table \ref{asyms}.}
\end{center}
\end{figure}

\begin{figure}
\begin{center}
\includegraphics[width=0.5\textwidth, clip]{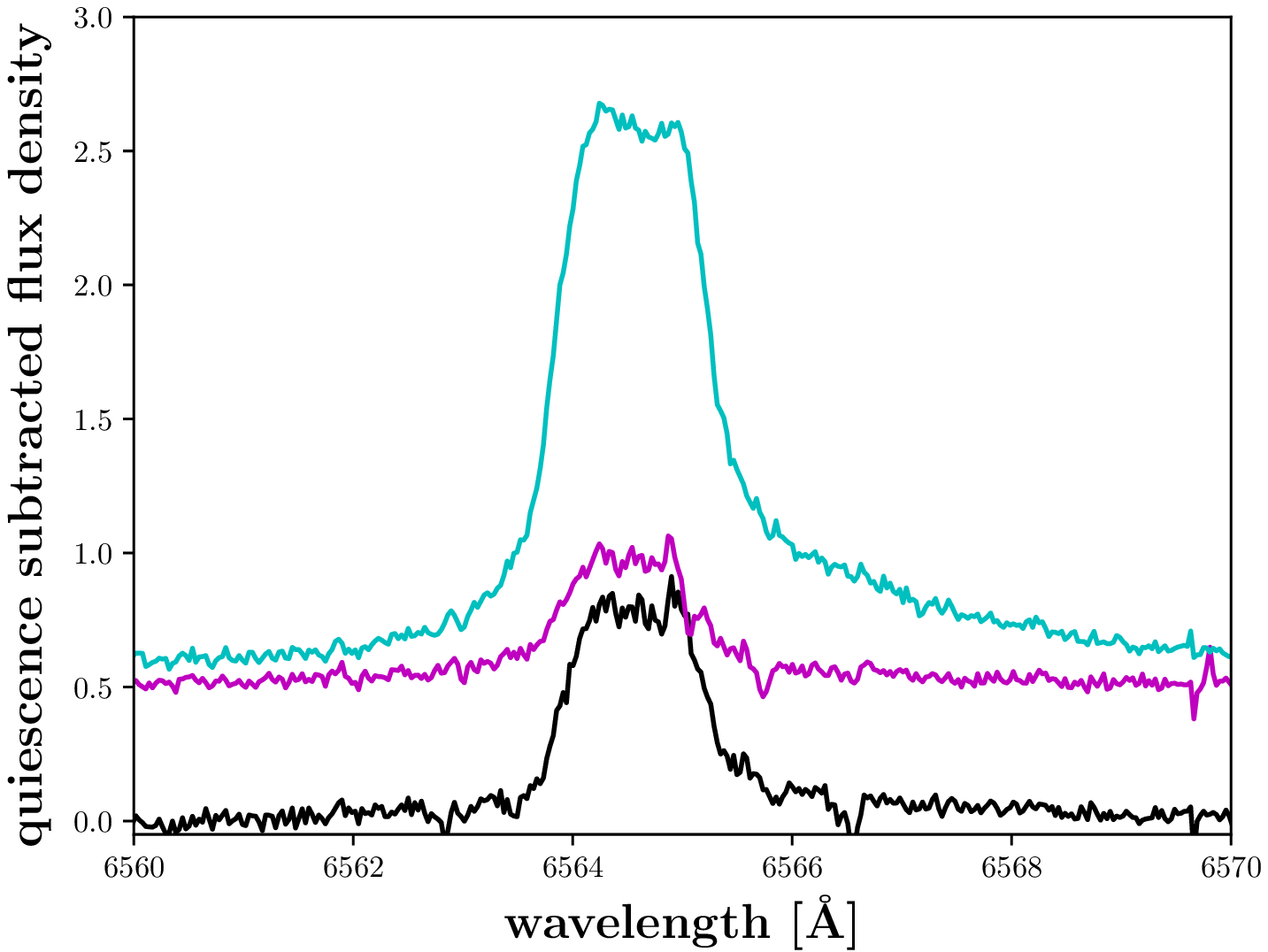}
\caption{\label{asymGTPeg} Spectra with asymmetry for GT Peg; from bottom to top the spectra correspond to entry nos. 65 -- 67 in Table \ref{asyms}.}
\end{center}
\end{figure}

\begin{figure}
\begin{center}
\includegraphics[width=0.5\textwidth, clip]{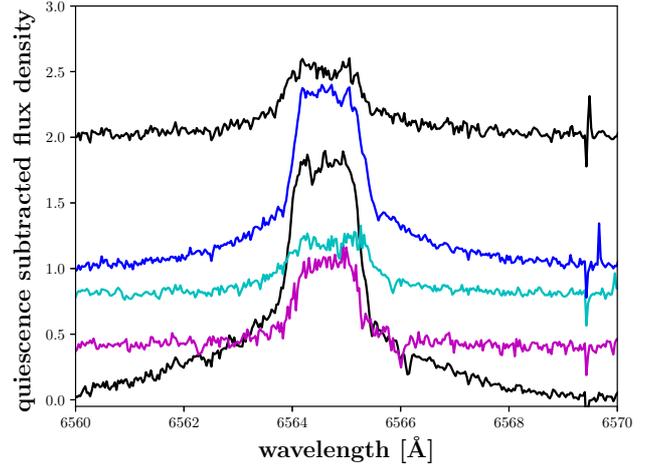}
\caption{\label{asymJ23548} Spectra with asymmetry for J23548+385; from bottom to top the spectra correspond to entry nos. 68 -- 72 in Table \ref{asyms}.}
\end{center}
\end{figure}

We present here all H$\alpha$ spectra with detected asymmetries in Figs.
\ref{asymV388Cas} to \ref{asymJ23548}. The quiescent spectrum is already
subtracted. Moreover, we plot a velocity scale in Fig. \ref{asymV388Cas}
  for reference.
If the star shows an asymmetry in more than one spectrum, the spectra are plotted with
an offset for better comparison and in cycling colours.

\clearpage

\section{Detected asymmetries in \ion{Na}{i} D and \ion{He}{i} D$_{3}$}
\label{appendixb}
\begin{figure*}
\begin{center}
\includegraphics[width=0.5\textwidth, clip]{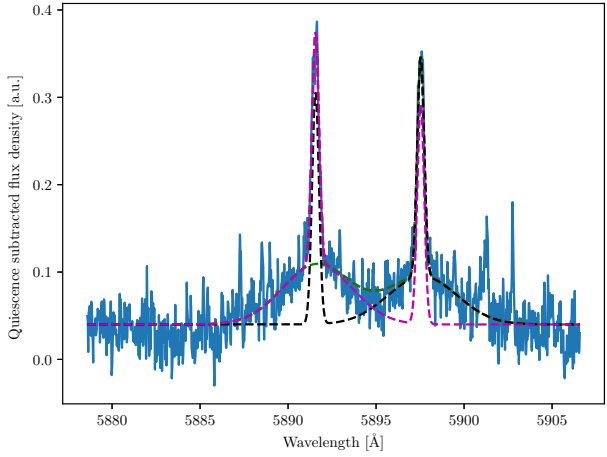}
\includegraphics[width=0.5\textwidth, clip]{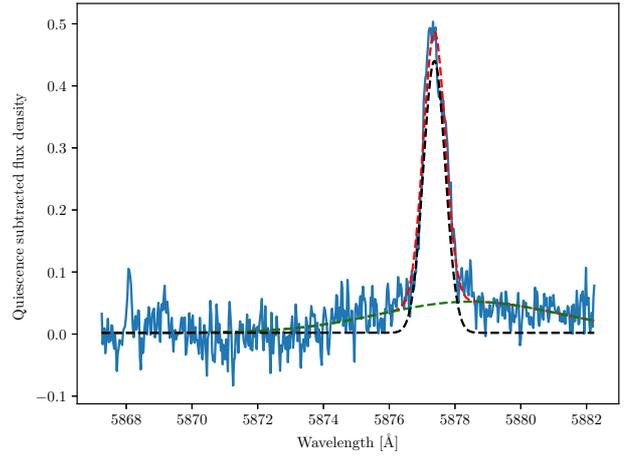}
\caption{\label{asymnad1} Spectrum with asymmetry for OT Ser.}
\end{center}
\end{figure*}

\begin{figure*}
\begin{center}
\includegraphics[width=0.5\textwidth, clip]{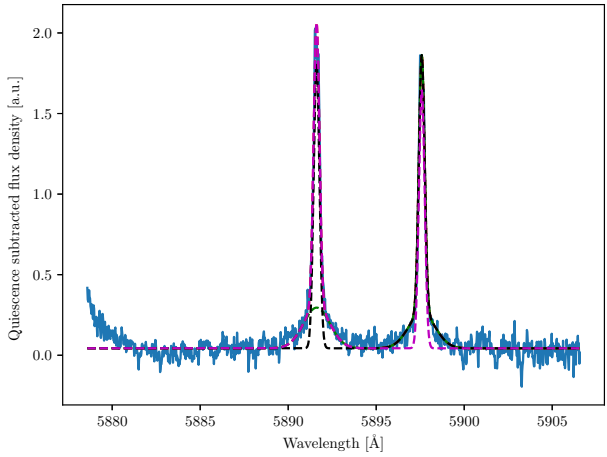}
\includegraphics[width=0.5\textwidth, clip]{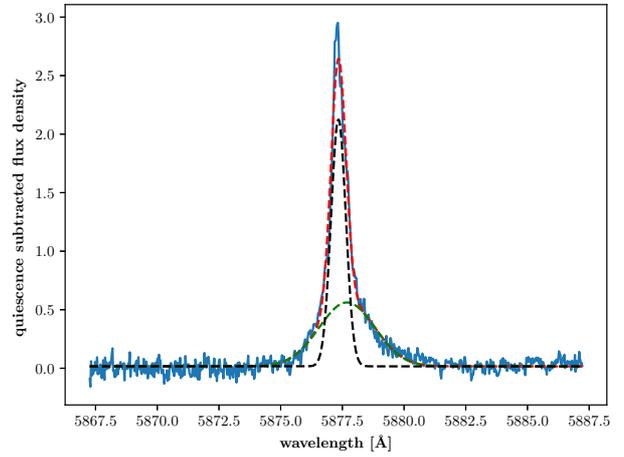}
\caption{\label{asymnad2} Spectrum with asymmetry for EV Lac.}
\end{center}
\end{figure*}

\begin{figure*}
\begin{center}
\includegraphics[width=0.5\textwidth, clip]{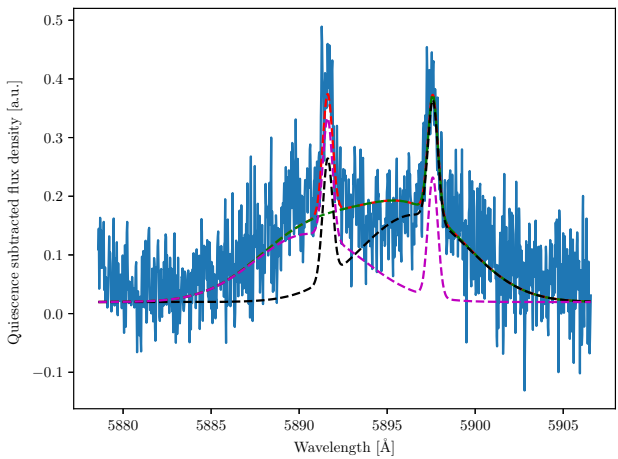}
\includegraphics[width=0.5\textwidth, clip]{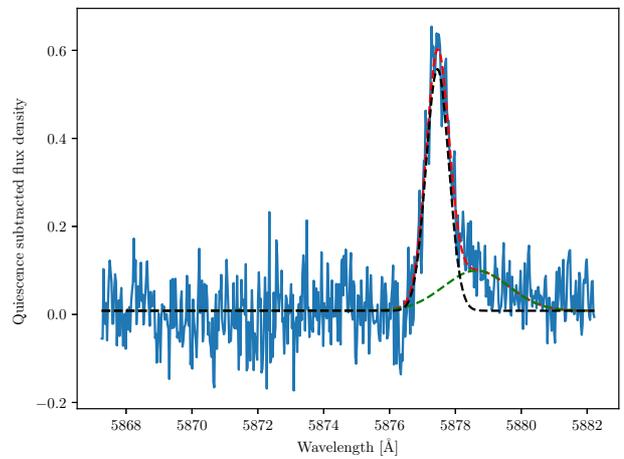}
\caption{\label{asymnad3} Spectrum with asymmetry for StKM2-809.}
\end{center}
\end{figure*}

We present here all \ion{Na}{i} D and \ion{He}{i} D$_{3}$ spectra with detected
asymmetries with the quiescent spectrum subtracted and the best fit overlaid
in Figs. \ref{asymnad1} to \ref{asymnad3}.

\section{Detected asymmetries in \ion{He}{i} 10833 \AA\, and corresponding excess fluxes in Pa$\beta$}
\label{appendixc}
\begin{figure*}
\begin{center}
\includegraphics[width=0.5\textwidth, clip]{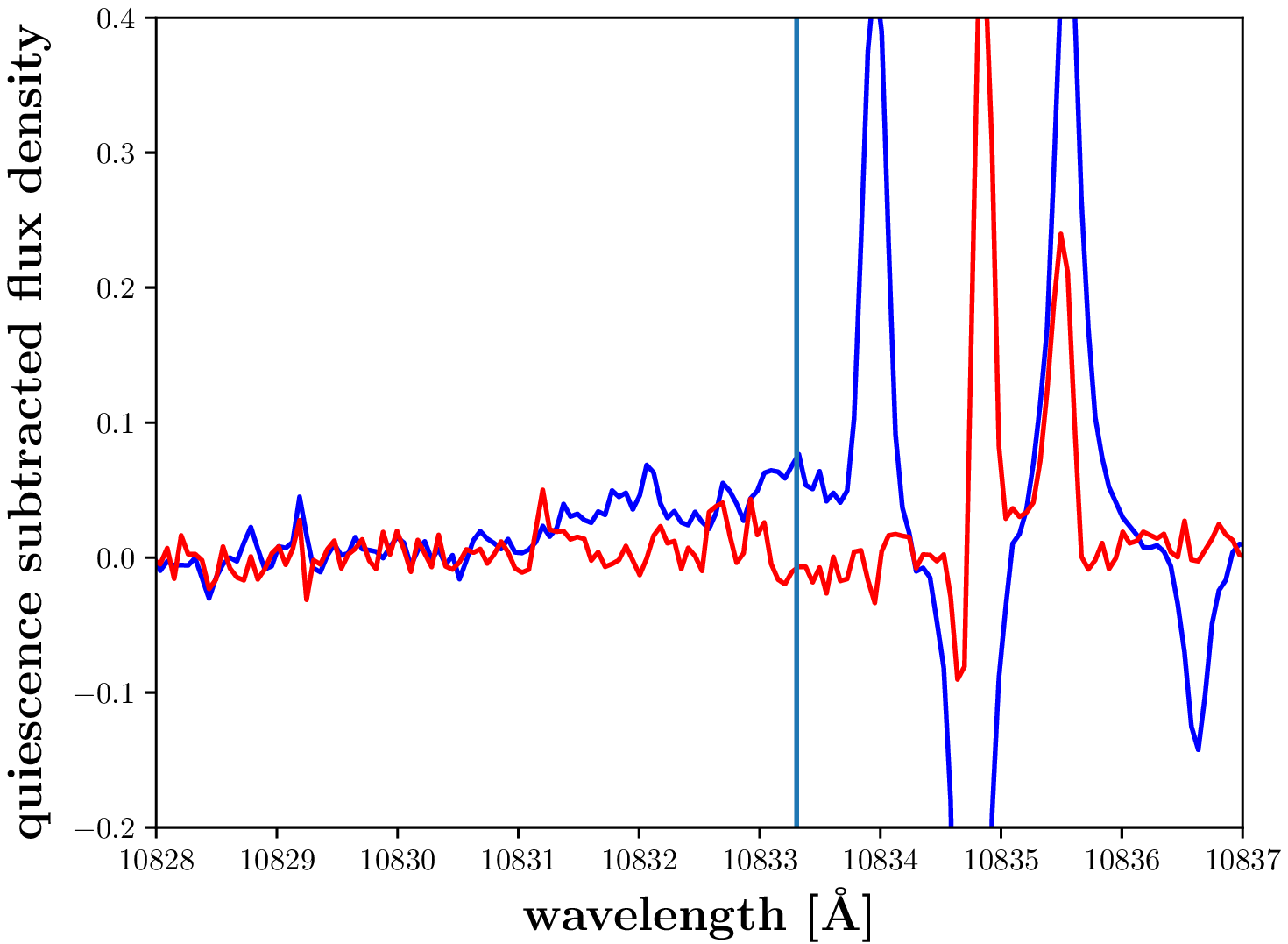}
\includegraphics[width=0.5\textwidth, clip]{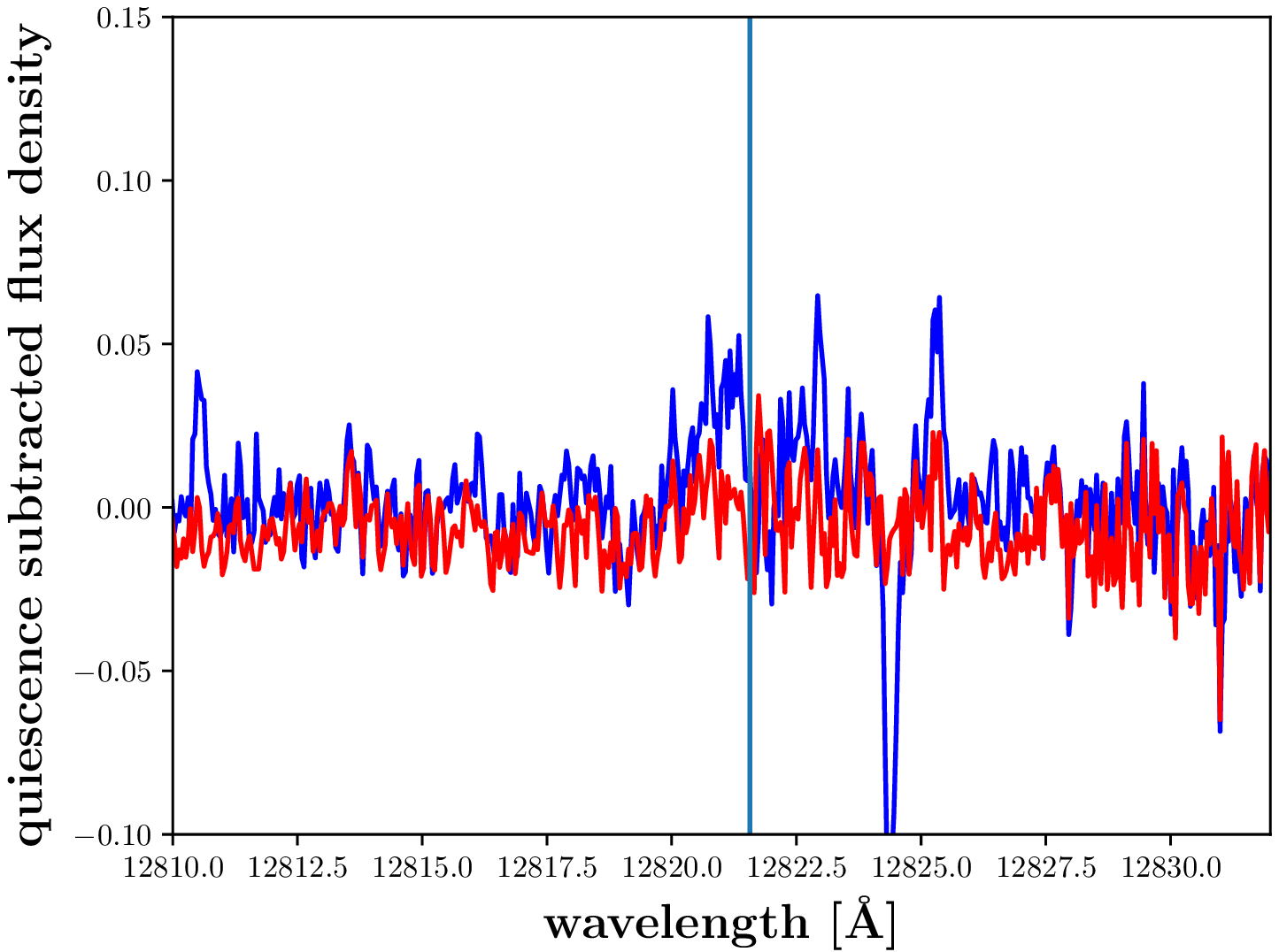}
\caption{\label{He10830V388} Spectrum with broad wing in \ion{He}{i} 10833 \AA\, for V388 Cas in
  blue, a quiescent spectrum for comparison in red.}
\end{center}
\end{figure*}

\begin{figure*}
\begin{center}
\includegraphics[width=0.5\textwidth, clip]{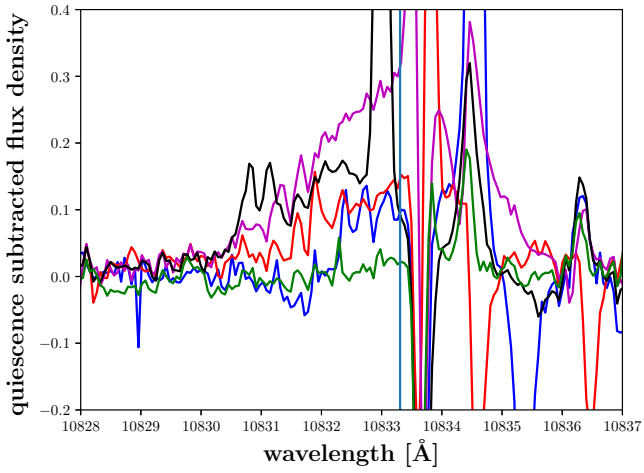}
\includegraphics[width=0.5\textwidth, clip]{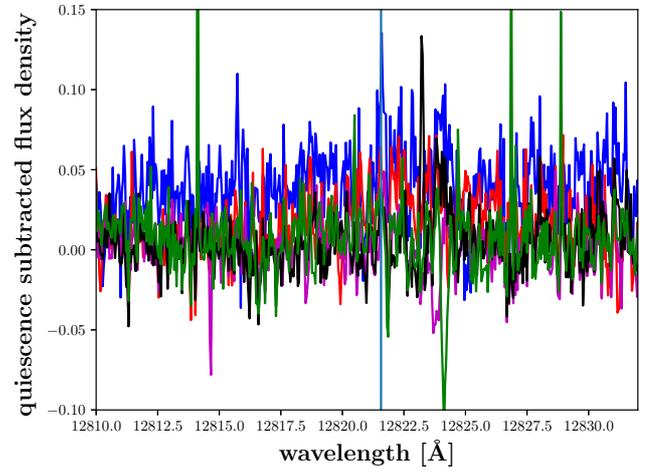}
\caption{\label{He10830Barta} Spectrum with broad wing in \ion{He}{i} 10833 \AA\, for Barta 161 12 in
  blue, black, red, and magenta, and a quiescent spectrum for comparison in green.}
\end{center}
\end{figure*}

\begin{figure*}
\begin{center}
\includegraphics[width=0.5\textwidth, clip]{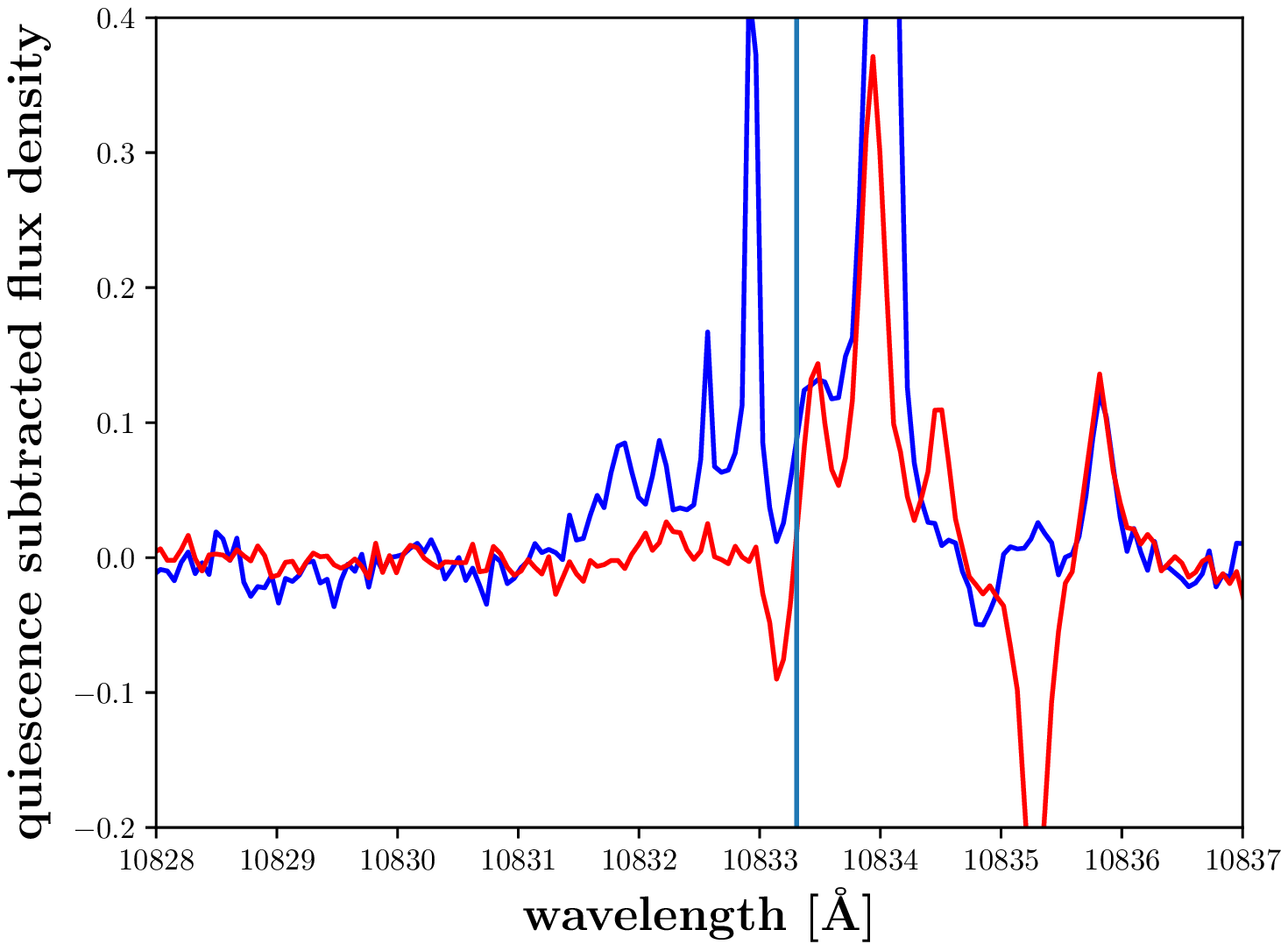}
\includegraphics[width=0.5\textwidth, clip]{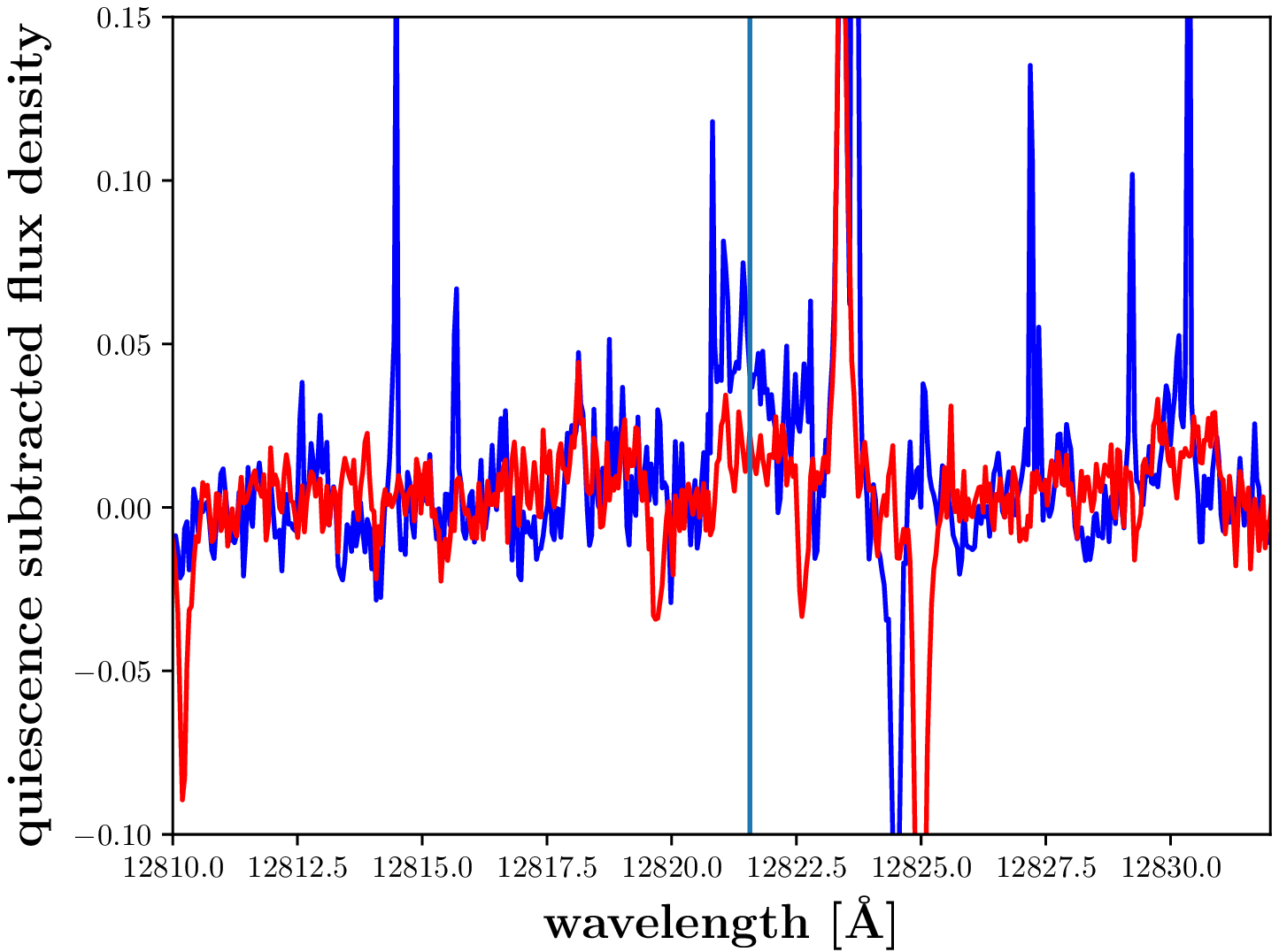}
\caption{\label{He10830cnleo} Spectrum with broad wing in \ion{He}{i} 10833 \AA\, for CN Leo in blue,
  a quiescent spectrum for comparison in red.}
\end{center}
\end{figure*}

\begin{figure*}
\begin{center}
\includegraphics[width=0.5\textwidth, clip]{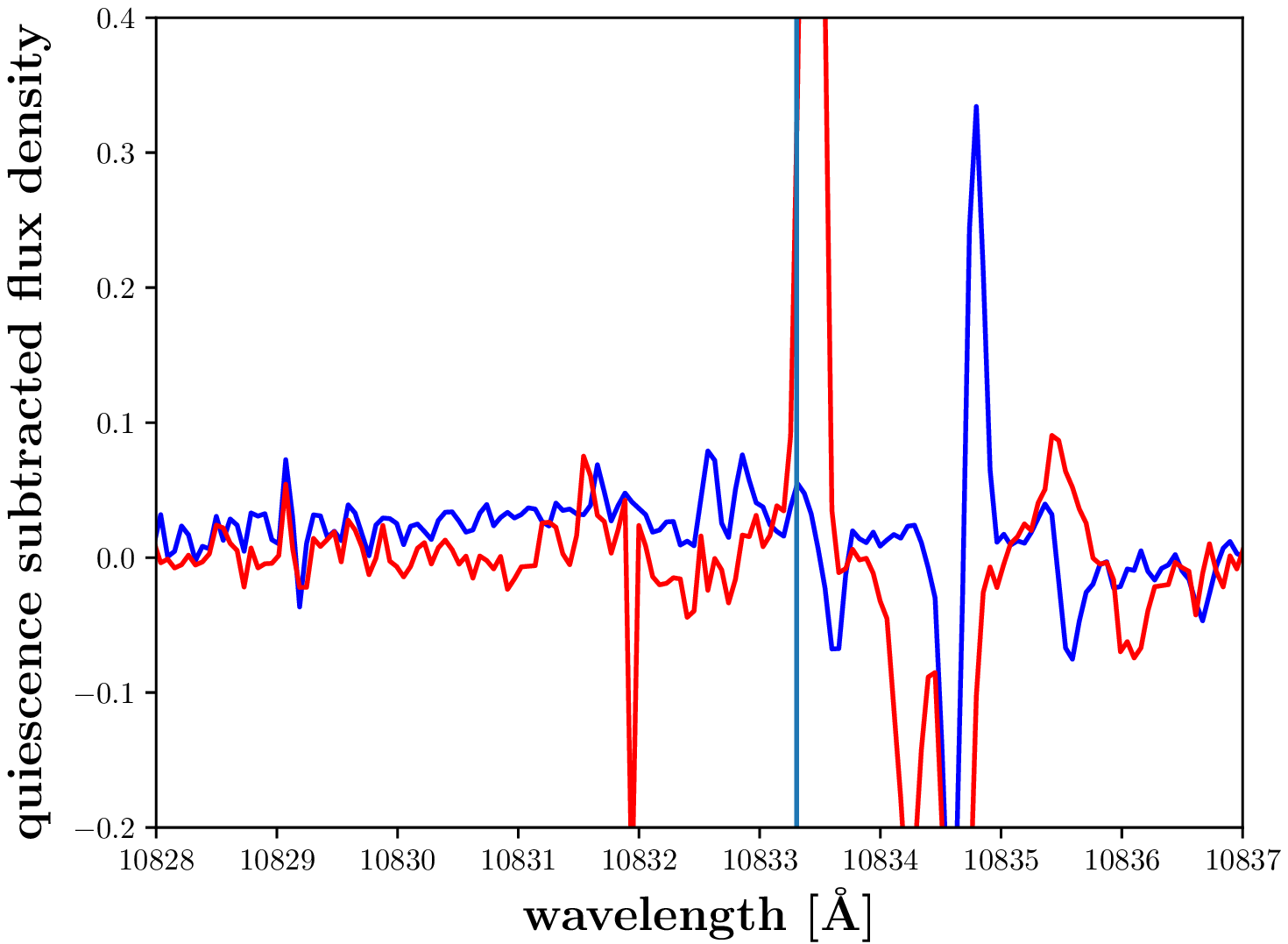}
\includegraphics[width=0.5\textwidth, clip]{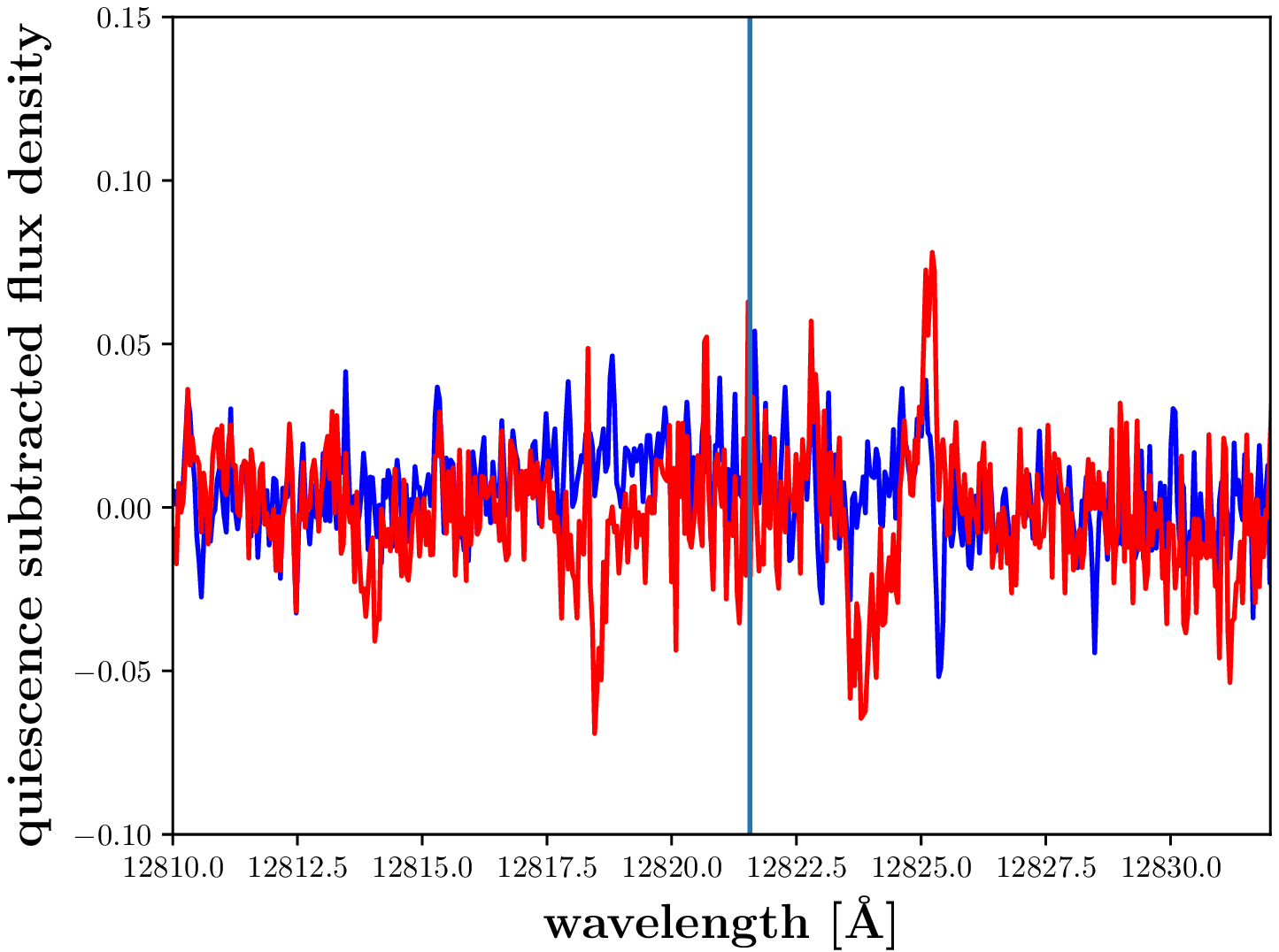}
\caption{\label{He10830StKM2} Spectrum with broad wing in \ion{He}{i} 10833 \AA\, for StKM2-809 in blue,
  a quiescent spectrum for comparison in red.}
\end{center}
\end{figure*}

\begin{figure*}
\begin{center}
\includegraphics[width=0.5\textwidth, clip]{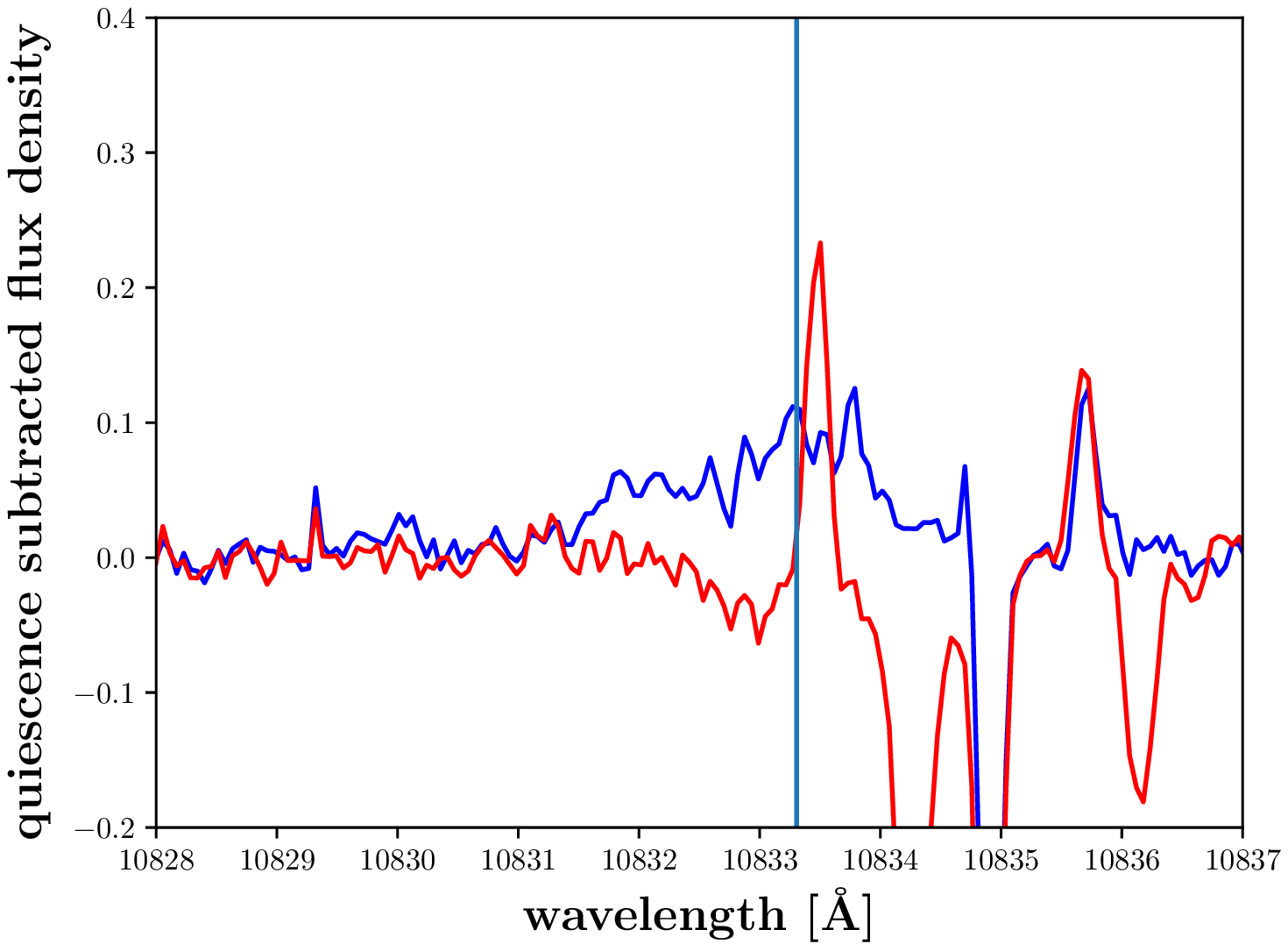}
\includegraphics[width=0.5\textwidth, clip]{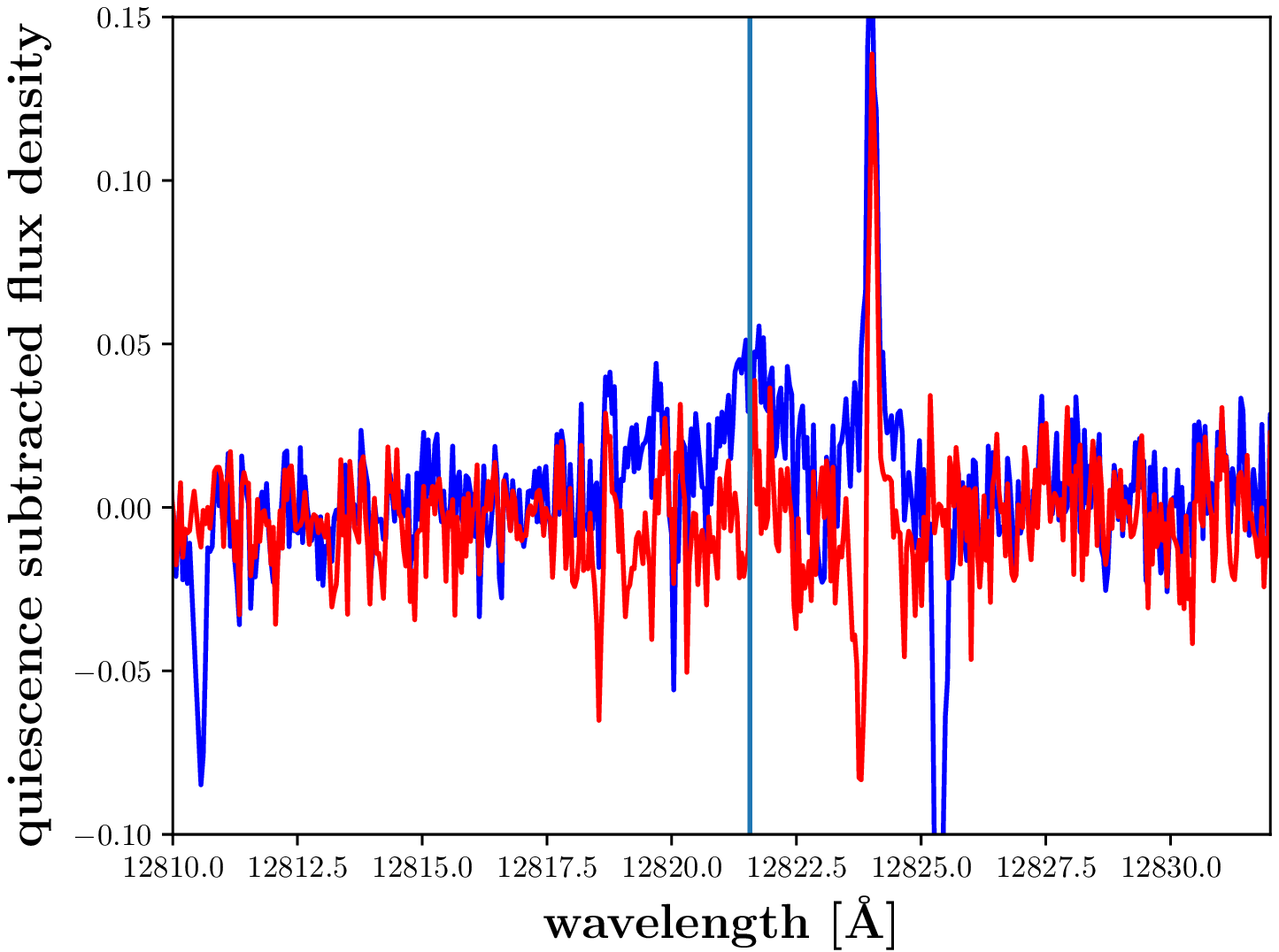}
\caption{\label{He10830otser} Spectrum with broad wing in \ion{He}{i} 10833 \AA\, for OT Ser in blue,
  a quiescent spectrum for comparison in red.}
\end{center}
\end{figure*}

\begin{figure*}
\begin{center}
\includegraphics[width=0.5\textwidth, clip]{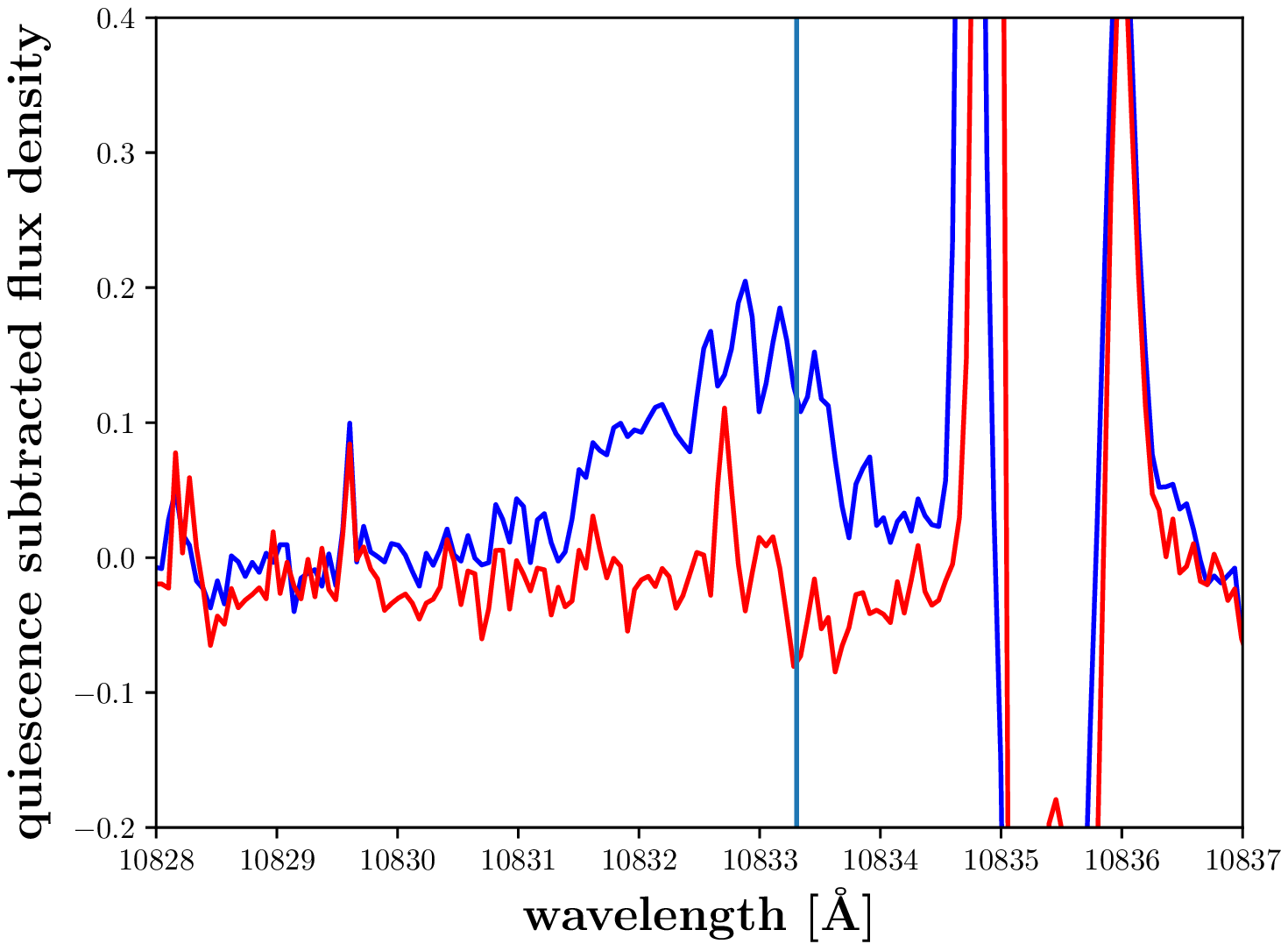}
\includegraphics[width=0.5\textwidth, clip]{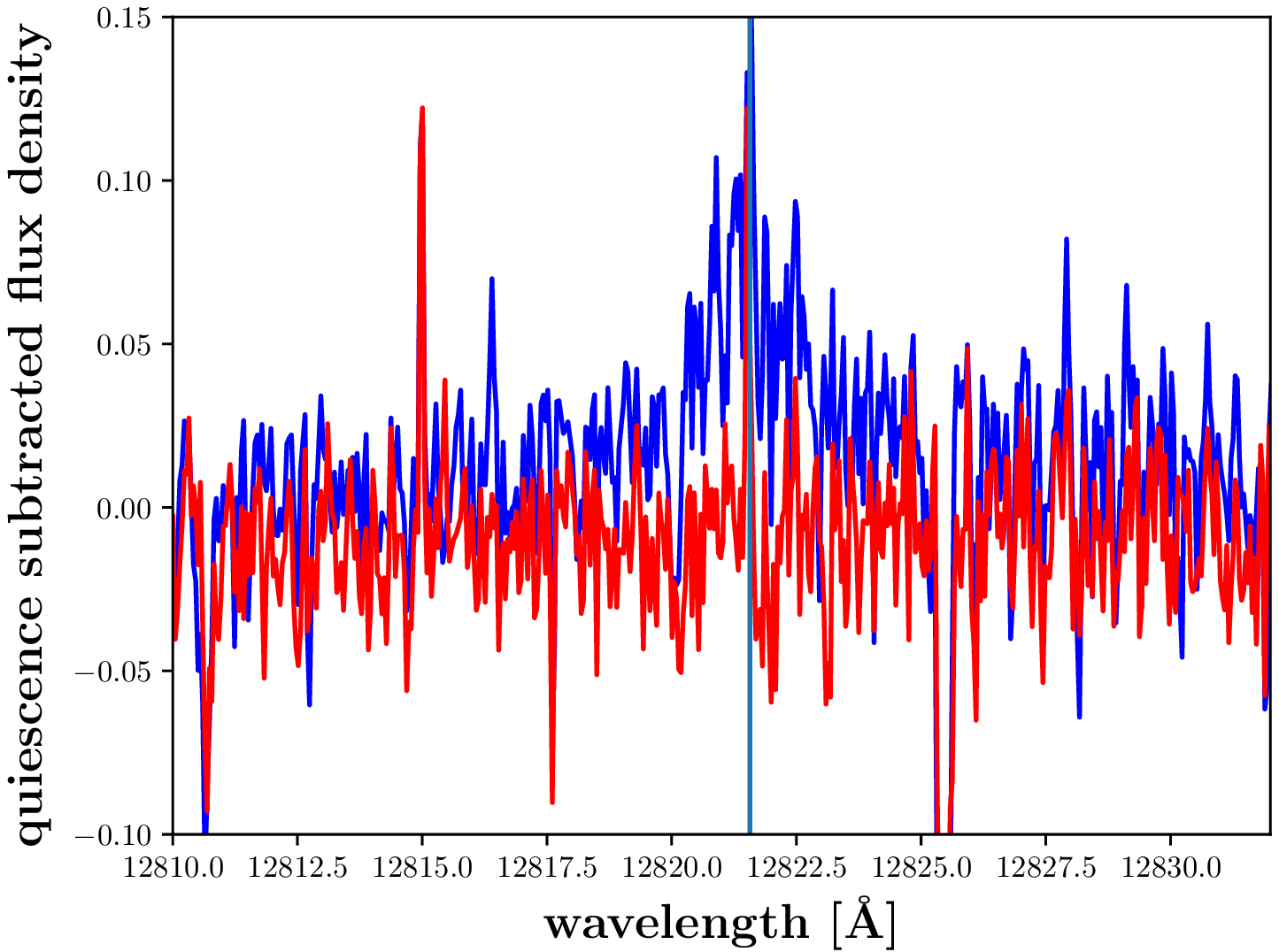}
\caption{\label{He10830G141} Spectrum with broad wing in \ion{He}{i} 10833 \AA\, for G141-036 in blue,
  a quiescent spectrum for comparison in red.}
\end{center}
\end{figure*}

\begin{figure*}
\begin{center}
\includegraphics[width=0.5\textwidth, clip]{evlache10830.eps}
\includegraphics[width=0.5\textwidth, clip]{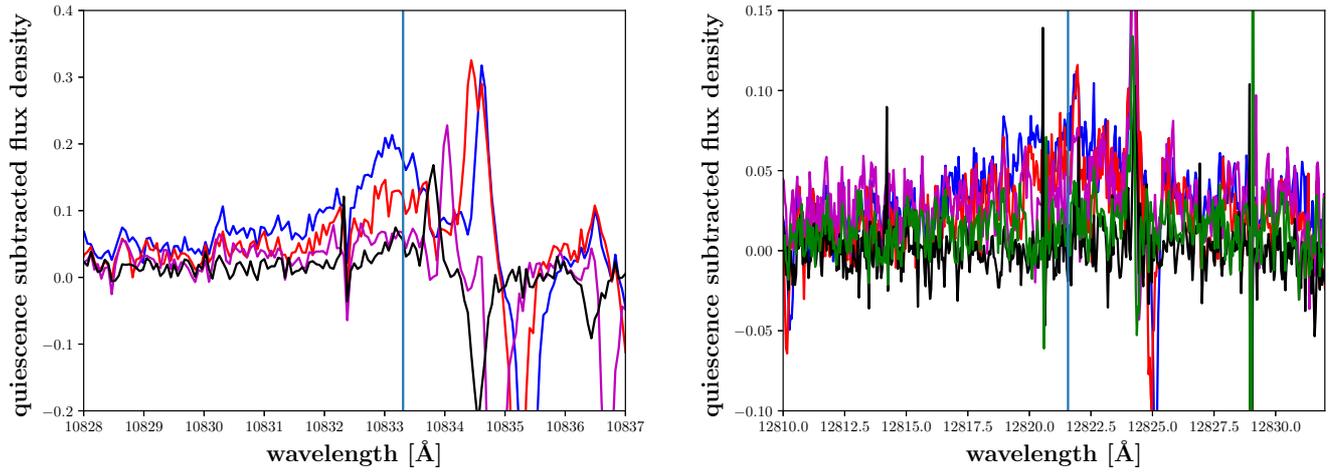}
\caption{\label{He10830evlac} Spectra with broad wing in \ion{He}{i} 10833 \AA\, for EV Lac, the
  green spectrum is the quiescent comparison spectrum.}
\end{center}
\end{figure*}

\begin{figure*}
\begin{center}
\includegraphics[width=0.5\textwidth, clip]{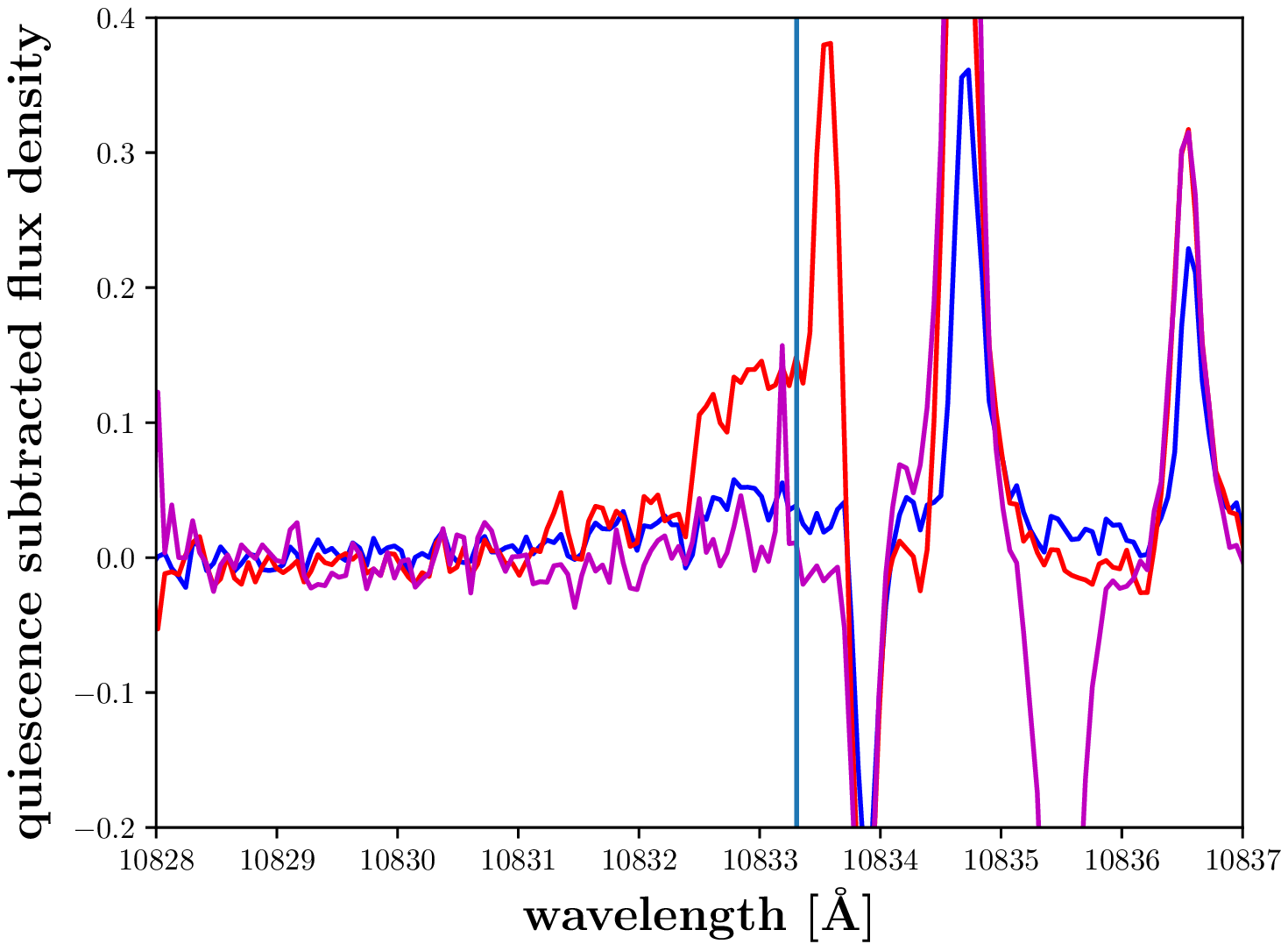}
\includegraphics[width=0.5\textwidth, clip]{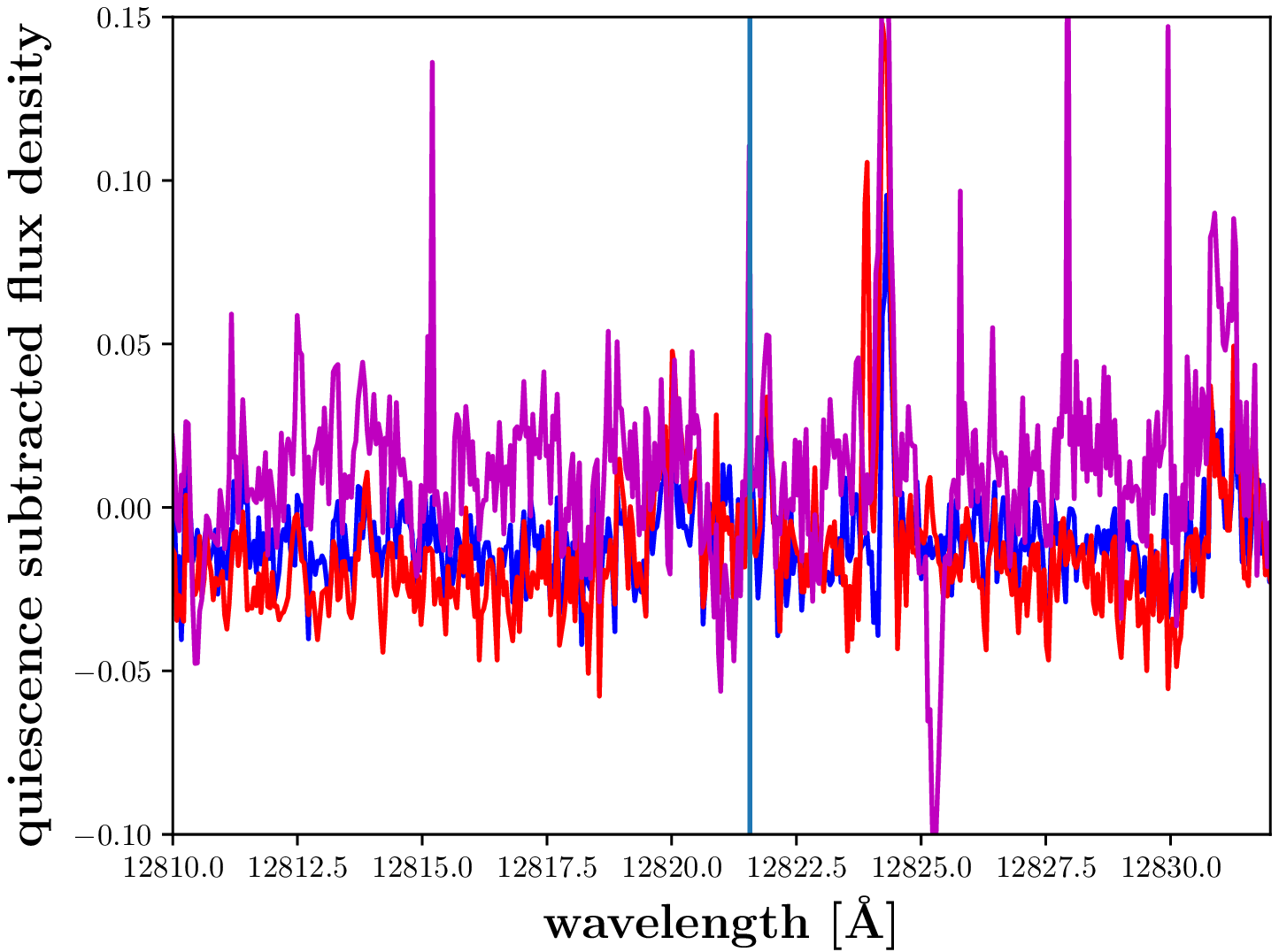}
\caption{\label{He10830GTPeg} Spectra with broad wing in \ion{He}{i} 10833 \AA\, for GT Peg, the
  magenta spectrum is the quiescent comparison spectrum.}
\end{center}
\end{figure*}

Here we present all quiescent flux density subtracted spectra with a \ion{He}{i} 10833 \AA\, broad wing in
Figs. \ref{He10830V388} to \ref{He10830GTPeg}. We
searched in all spectra exhibiting an H$\alpha$ asymmetry. The problem of the strong air glow and telluric
lines in the region is apparent. Besides the broad wing spectra we also always present a quiescent spectrum.
In the right panel we show the same  for the corresponding Pa$\beta$ spectra.
It is quite apparent that the line reacts in a much less sensitive manner. For both spectral lines
we mark the reference central wavelength with a blue vertical line.

\section{Examples of the H$\alpha$ profile fit for the weakest and strongest asymmetries}
\label{appendixd}

Here we present examples of our fits of the H$\alpha$ line profile. For each of the categories
(i) red asymmetry, (ii) blue asymmetry, and (iii) symmetric broadening, we show the strongest and the weakest
asymmetry and its fit.

\begin{figure}
\begin{center}
\includegraphics[width=0.5\textwidth, clip]{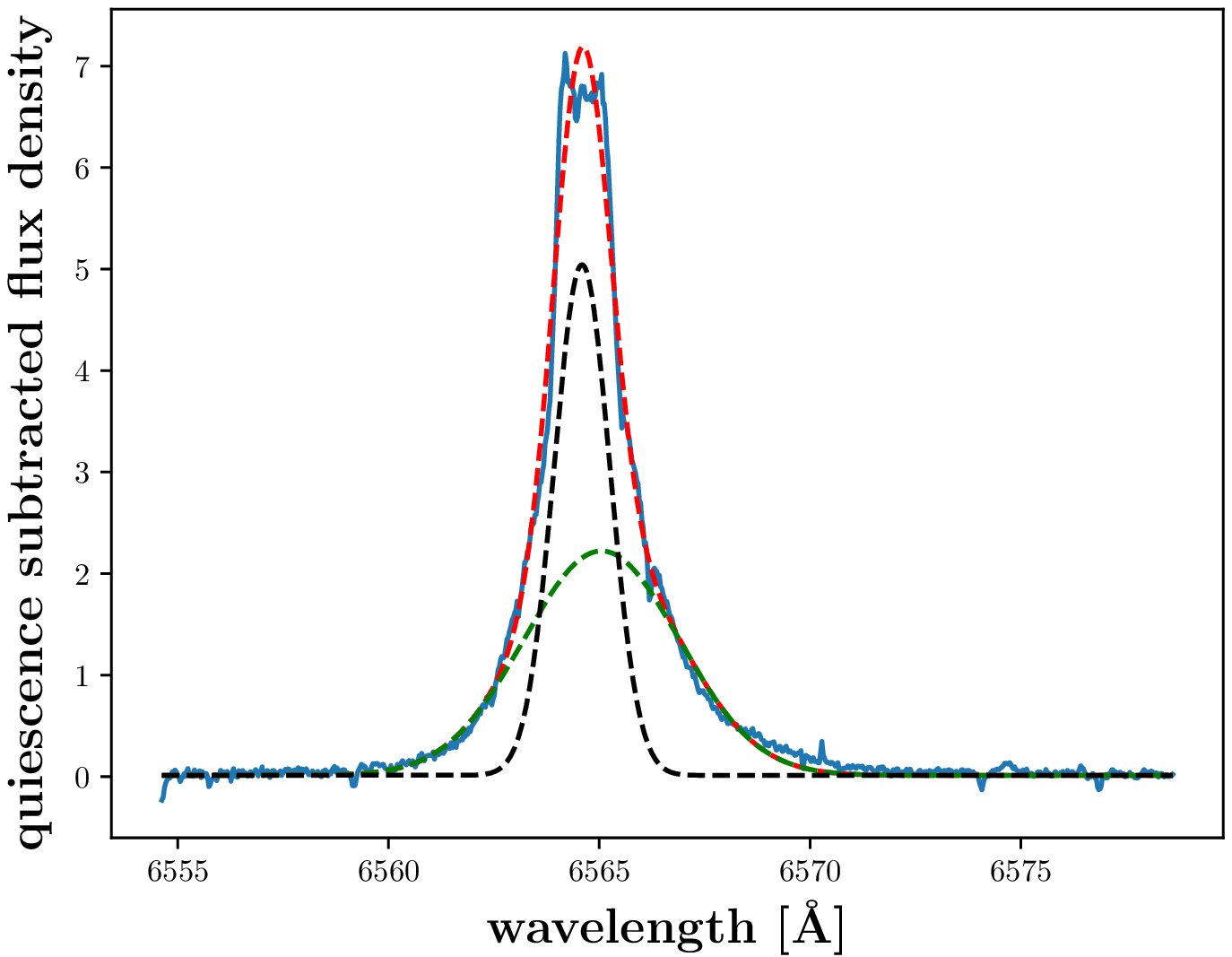}
\caption{\label{fit1} Strongest red asymmetry was found for EV Lac corresponding to asymmetry no. 55
  in Table \ref{asyms}.}
\end{center}
\end{figure}

\begin{figure}
\begin{center}
\includegraphics[width=0.5\textwidth, clip]{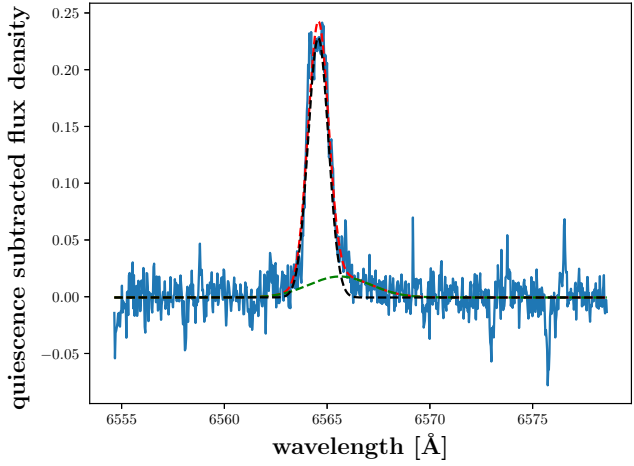}
\caption{\label{fit2} Weakest red asymmetry was found for V2689 Ori corresponding to asymmetry no. 14
  in Table \ref{asyms}.}
\end{center}
\end{figure}

\begin{figure}
\begin{center}
\includegraphics[width=0.5\textwidth, clip]{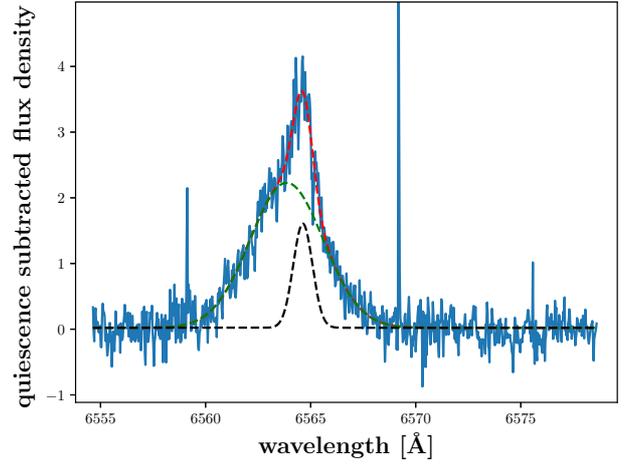}
\caption{\label{fit3} Strongest blue asymmetry was found for vB 8 corresponding to asymmetry no. 44
  in Table \ref{asyms}.}
\end{center}
\end{figure}

\clearpage

\begin{figure}
\begin{center}
\includegraphics[width=0.5\textwidth, clip]{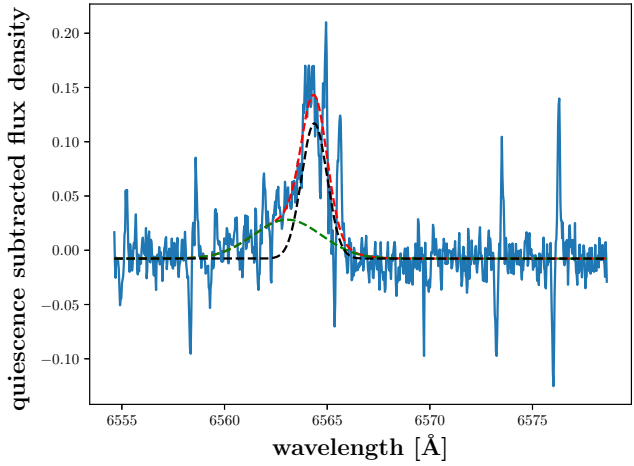}
\caption{\label{fit4} Weakest blue asymmetry was found for OT Ser corresponding to asymmetry no. 40
  in Table \ref{asyms}.}
\end{center}
\end{figure}

\begin{figure}
\begin{center}
\includegraphics[width=0.5\textwidth, clip]{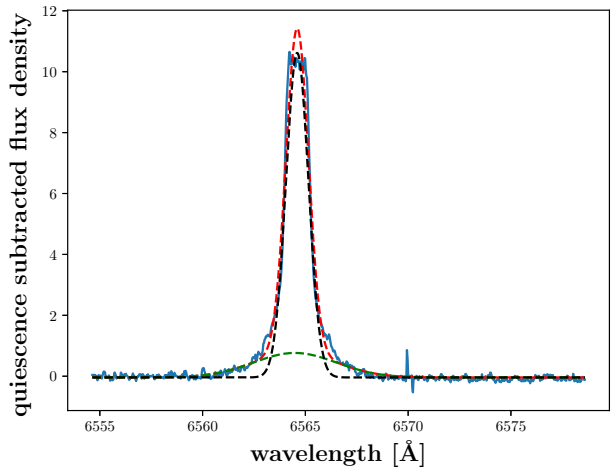}
\caption{\label{fit5} Strongest symmetric broadening was found for CN Leo corresponding to asymmetry no. 30
  in Table \ref{asyms}.}
\end{center}
\end{figure}

\begin{figure}
\begin{center}
\includegraphics[width=0.5\textwidth, clip]{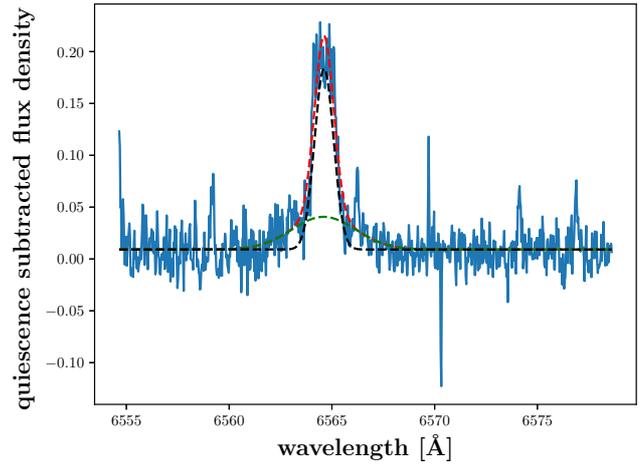}
\caption{\label{fit6} Weakest symmetric broadening was found for GJ 362 corresponding to asymmetry no. 28
  in Table \ref{asyms}.}
\end{center}
\end{figure}

\end{document}